\documentclass[ALICE,manyauthors]{cernphprep}

\usepackage[comma,square,numbers,sort&compress]{natbib}

\usepackage[T1]{fontenc}
\usepackage{orcidlink}
\usepackage{hyperref}
\usepackage{lineno}
\usepackage{xspace}
\usepackage{color}
\usepackage{bm}
\usepackage{setspace} %double spacing

\usepackage{floatrow}
\usepackage{rotating}
\usepackage{amssymb} % needed for math
\usepackage{amsmath} % needed for math
\usepackage{amsfonts}
\usepackage{notoccite}
\usepackage{upgreek}
\usepackage{hyperref}  % links im text
\usepackage{siunitx}
\usepackage{float} 
\usepackage{subfigure}
\usepackage{graphicx}
\usepackage{graphicx}

\makeatletter
\renewcommand{\p@subfigure}{\thefigure\space}
\makeatother

\begin{document}
%%%%%%%%%%%%%%%%%%%%%%%%%%%%%%%%%%%%%%%%%%%%%%%%%%
% These are some new commands that may be useful 
% for paper writing in general. If other newcommands
% are needed for your specific paper, please feel 
% free to add here. 
%
% The currently available commands are organized in: 
% 1) Systems
% 2) Quantities
% 3) Energies and units
% 4) Detectors
% 5) Particle species
% 6) Colors (AB)
%%%%%%%%%%%%%%%%%%%%%%%%%%%%%%%%%%%%%%%%%%%%%%%%%%

% 1) SYSTEMS 
\newcommand{\pp}           {pp\xspace}
\newcommand{\ppbar}        {\mbox{$\mathrm {p\overline{p}}$}\xspace}
\newcommand{\XeXe}         {\mbox{Xe--Xe}\xspace}
\newcommand{\PbPb}         {\mbox{Pb--Pb}\xspace}
\newcommand{\pPb}          {\mbox{p--Pb}\xspace}
\newcommand{\AuAu}         {\mbox{Au--Au}\xspace}
\newcommand{\dAu}          {\mbox{d--Au}\xspace}

% 2) QUANTITIES 
%\newcommand{\s}            {\ensuremath{\sqrt{s}}\xspace}
\newcommand{\snn}          {\ensuremath{\sqrt{s_{\mathrm{NN}}}}\xspace}
\newcommand{\pt}           {\ensuremath{p_{\rm T}}\xspace}
\newcommand{\meanpt}       {$\langle p_{\mathrm{T}}\rangle$\xspace}
\newcommand{\ycms}         {\ensuremath{y_{\rm CMS}}\xspace}
\newcommand{\ylab}         {\ensuremath{y_{\rm lab}}\xspace}
\newcommand{\etarange}[1]  {\mbox{$\left | \eta \right |~<~#1$}}
\newcommand{\yrange}[1]    {\mbox{$\left | y \right |~<~#1$}}
\newcommand{\dndy}         {\ensuremath{\mathrm{d}N_\mathrm{ch}/\mathrm{d}y}\xspace}
\newcommand{\dndeta}       {\ensuremath{\mathrm{d}N_\mathrm{ch}/\mathrm{d}\eta}\xspace}
\newcommand{\avdndeta}     {\ensuremath{\langle\dndeta\rangle}\xspace}
\newcommand{\dNdy}         {\ensuremath{\mathrm{d}N_\mathrm{ch}/\mathrm{d}y}\xspace}
\newcommand{\Npart}        {\ensuremath{N_\mathrm{part}}\xspace}
\newcommand{\Ncoll}        {\ensuremath{N_\mathrm{coll}}\xspace}
\newcommand{\dEdx}         {\ensuremath{\textrm{d}E/\textrm{d}x}\xspace}
\newcommand{\RpPb}         {\ensuremath{R_{\rm pPb}}\xspace}

% 3) ENERGIES, UNITS
\newcommand{\nineH}        {$\sqrt{s}~=~0.9$~Te\kern-.1emV\xspace}
\newcommand{\seven}        {$\sqrt{s}~=~7$~Te\kern-.1emV\xspace}
\newcommand{\twoH}         {$\sqrt{s}~=~0.2$~Te\kern-.1emV\xspace}
\newcommand{\twosevensix}  {$\sqrt{s}~=~2.76$~Te\kern-.1emV\xspace}
\newcommand{\five}         {$\sqrt{s}~=~5.02$~Te\kern-.1emV\xspace}
\newcommand{\twosevensixnn}{$\sqrt{s_{\mathrm{NN}}}~=~2.76$~Te\kern-.1emV\xspace}
\newcommand{\fivenn}       {$\sqrt{s_{\mathrm{NN}}}~=~5.02$~Te\kern-.1emV\xspace}
\newcommand{\LT}           {L{\'e}vy-Tsallis\xspace}
\newcommand{\GeVc}         {Ge\kern-.1emV/$c$\xspace}
\newcommand{\MeVc}         {Me\kern-.1emV/$c$\xspace}
\newcommand{\GeVmass}      {Ge\kern-.2emV/$c^2$\xspace}
\newcommand{\MeVmass}      {Me\kern-.2emV/$c^2$\xspace}
\newcommand{\lumi}         {\ensuremath{\mathcal{L}}\xspace}

% 4) DETECTORS 
\newcommand{\ITS}          {\rm{ITS}\xspace}
\newcommand{\TOF}          {\rm{TOF}\xspace}
\newcommand{\ZDC}          {\rm{ZDC}\xspace}
\newcommand{\ZDCs}         {\rm{ZDCs}\xspace}
\newcommand{\ZNA}          {\rm{ZNA}\xspace}
\newcommand{\ZNC}          {\rm{ZNC}\xspace}
\newcommand{\SPD}          {\rm{SPD}\xspace}
\newcommand{\SDD}          {\rm{SDD}\xspace}
\newcommand{\SSD}          {\rm{SSD}\xspace}
\newcommand{\TPC}          {\rm{TPC}\xspace}
\newcommand{\TRD}          {\rm{TRD}\xspace}
\newcommand{\VZERO}        {\rm{V0}\xspace}
\newcommand{\VZEROA}       {\rm{V0A}\xspace}
\newcommand{\VZEROC}       {\rm{V0C}\xspace}
\newcommand{\Vdecay} 	   {\ensuremath{V^{0}}\xspace}

% 5) PARTICLE SPECIES 
\newcommand{\ee}           {\ensuremath{e^{+}e^{-}}} 
\newcommand{\pip}          {\ensuremath{\pi^{+}}\xspace}
\newcommand{\pim}          {\ensuremath{\pi^{-}}\xspace}
\newcommand{\kap}          {\ensuremath{\rm{K}^{+}}\xspace}
\newcommand{\kam}          {\ensuremath{\rm{K}^{-}}\xspace}
\newcommand{\pbar}         {\ensuremath{\rm\overline{p}}\xspace}
\newcommand{\kzero}        {\ensuremath{{\rm K}^{0}_{\rm{S}}}\xspace}
\newcommand{\lmb}          {\ensuremath{\Lambda}\xspace}
\newcommand{\almb}         {\ensuremath{\overline{\Lambda}}\xspace}
\newcommand{\Om}           {\ensuremath{\Omega^-}\xspace}
\newcommand{\Mo}           {\ensuremath{\overline{\Omega}^+}\xspace}
\newcommand{\X}            {\ensuremath{\Xi^-}\xspace}
\newcommand{\Ix}           {\ensuremath{\overline{\Xi}^+}\xspace}
\newcommand{\Xis}          {\ensuremath{\Xi^{\pm}}\xspace}
\newcommand{\Oms}          {\ensuremath{\Omega^{\pm}}\xspace}
\newcommand{\pion}          {\ensuremath{\pi\xspace}}

% 6) COLORS:
\definecolor{dgreen}{cmyk}{1.,0.,1.,0.4} % dark green
\definecolor{orange}{cmyk}{0.,0.353,1.,0.} % orange
\newcommand{\orange}[1]{\textcolor{orange}{#1}}
\newcommand{\blue}[1]{{\color{blue}{#1}}}
\newcommand{\green}[1]{{\color{green}{#1}}}
\newcommand{\dgreen}[1]{{\textcolor{dgreen}{#1}}}
\newcommand{\red}[1]{{\color{red}{#1}}}
\newcommand{\magenta}[1]{{\color{magenta}{#1}}}
\def \new {\blue}
\def \old {\orange}
\def \ask {\magenta}
\def \tmp {\green}

\newcommand{\mt}{m_\text{T}}
\newcommand{\ptText}{$p_\text{T}$}
\newcommand{\mtText}{$m_\text{T}$}
\newcommand{\Epsilon}{\mathcal{E} }
\newcommand{\piPlus}{\uppi^+}
\newcommand{\piMinus}{\uppi^-}
\newcommand{\piPlusMinus}{\uppi^\pm}

\newcommand{\ppiPlus}{\text{p--}\uppi^+}
\newcommand{\ppiMinus}{\text{p--}\uppi^-}
\newcommand{\ppiPlusMinus}{\text{p--}\uppi^\pm}

\newcommand{\pppiPlus}{\text{p--p--}\uppi^+}
    \newcommand{\pppiMinus}{\text{p--p--}\uppi^-}
\newcommand{\pppiPlusMinus}{\text{p--p--}\uppi^\pm}

\newcommand{\snnFemto}{\sqrt{s} =13~\text{ TeV}}

%%%%%%%%%%%%%%%  Title page %%%%%%%%%%%%%%%%%%%%%%%%
\begin{titlepage}
% The dates below correspond to CERN approval
%, please don't touch: EB chairs will take care
\PHyear{2025}       % required, will be obtained from CERN
\PHnumber{034}      % required, will be obtained from CERN
\PHdate{24 February}  % required, will be obtained from CERN
%%%%%%%%%%%%%%%%%%%%%%%%%%%%%%%%%%%%%%%%%%%%%%%%%%%%

%%% Put your own title + short title here:
\title{Investigating the $\mathbf{\ppiPlusMinus}$ and  $\mathbf{\pppiPlusMinus}$ dynamics \\ with femtoscopy in pp collisions at $\mathbf{\sqrt{\textit{s}}=13}$~TeV}
\ShortTitle{$\ppiPlusMinus$ and $\pppiPlusMinus$ femtoscopy in pp collisions at $\snnFemto$ }   % appears on right page headers

%%% Do not change the next lines
\Collaboration{ALICE Collaboration\thanks{See Appendix~\ref{app:collab} for the list of collaboration members}}
\ShortAuthor{ALICE Collaboration} % appears on the right page headers; do not change

\begin{abstract}

 The interaction between pions and nucleons plays a crucial role in hadron physics. It represents a fundamental building block of the low-energy QCD dynamics and is subject to several resonance excitations. This work studies the $\ppiPlusMinus$ dynamics using femtoscopic correlations in high-multiplicity pp collisions at $\sqrt{s} = 13$~TeV measured by ALICE at the LHC. As the final-state interaction between protons and pions is well constrained by scattering experiments and the study of pionic hydrogen, the results give access to information on the particle-emitting source in pp collisions using the femtoscopy methods. The scaling of the source size of primordial protons and pions against their pair transverse mass is extracted. The results are compared with the source sizes studied with p--p, $\text{p--K}^+$, and $\uppi^\pm$--$\uppi^\pm$ pairs by ALICE in the same collision system and are found to be in agreement for the different particle pairs. This reinforces recent findings by ALICE of a common emission source for all hadron-pairs in pp collisions at LHC energies. 
Furthermore, the $\pppiPlusMinus$ systems are studied using three-particle femtoscopy in pp collisions at $\sqrt{s} = 13$~TeV. The presence of three-body effects is analyzed utilizing the cumulant expansion method. In this formalism, the known two-body interactions are subtracted in order to isolate the three-body effects. For both, $\pppiPlus$ and $\pppiMinus$, a non-zero cumulant is found, indicating effects beyond pairwise interactions. These results give information on the coupling of the pion to multiple nucleons.

\end{abstract}
\end{titlepage}

\setcounter{page}{2} %please do not remove this line

%%%%%%%%%%%%%%%%%%%%%%%%%%%%%%%%
% begin the main text
%%%%%%%%%%%%%%%%%%%%%%%%%%%%%%%%

%%%%%%%%%%%%%%%%%%%%%%%%%%%%%%%%%%%%%%%
\section{Introduction} 
\label{s:Introduction}

The residual strong interaction between pions and nucleons~\cite{WEINBERG19913,Moorhouse:1969va, Ericson:1988gk} is rather well known since a comprehensive set of scattering data are available~\cite{Arndt:1995bj,PDG}, and chiral effective field theories anchored to this data provide an exhaustive description of this interaction~\cite{Fettes:1998ud,GASPARYAN2010126}. 
The pion is key to understanding low-energy quantum
chromodynamics (QCD) in a generalized context since it plays the role of an exchange boson for the nuclear interaction~\cite{Lee:2002eq}. Indeed, the spontaneous breaking of the chiral symmetry of the QCD Lagrangian leads to the emergence of massless Nambu–Goldstone bosons. Due to the additional, explicit breaking of chiral symmetry by the small quark masses, these bosons acquire a mass, with the pion being the lightest among them.
Consequently, the pion plays a central role in mediating long-range interactions governed by chiral dynamics. Investigating pion--nucleon interactions offers valuable insights into the underlying symmetries and dynamics of QCD, especially within the low-energy domain where conventional perturbative methods are not applicable.

The pion--nucleon dynamics is characterized by a rich spectrum of short-lived resonances~\cite{Meissner:1999vr, PDG} which are visible in scattering data but also studied at intermediate ($\sim$GeV) and high energy ($\geq $100 GeV) \mbox{hadron--hadron} inelastic collisions. 
Indeed, pion--nucleon scattering has the potential to induce the excitation of nucleons to higher-energy states, such as the $\Delta(1232)$  resonance, which is particularly prominent due to its relatively low mass compared to other baryonic resonance states.  The behavior of such processes is characterized by complex coupled-channel dynamics. Overall, such resonances  require partial-wave analysis methods to correctly interpret pion--nucleon data and disentangle resonant and non-resonant parts of the interaction~\cite{Cutkosky:1979fy,Arndt:2006bf, Andronic:2018qqt} .

While the two-body pion--nucleon interaction constitutes a solid reference for low-energy QCD, extending our understanding to pion--multinucleon systems presents exciting challenges and opportunities for both experimental and theoretical research in low-energy QCD. Scattering experiments with pions and deuterons have been carried out, which allowed constraining the $\text{N--N--}\uppi$ interactions~\cite{Brueckner:1953zza,Baru:2012iv}. Analyses of pionic deuterium also helped to constrain the isospin dependent part of the pion--nucleon interaction~\cite{Strauch:2010vu,DORING2005673}. On the other hand, $\text{p--p--}\uppi$ systems have never been investigated so far. 
The study of systems containing pions together with several nucleons allowed one to study the in-medium  properties of the pion. The analysis of the X-ray spectra of pionic atoms, in which a negatively charged pion substitutes an atomic electron, enabled the extraction of the modified decay constant and mass of the pion viewed as evidence of partial restoration of chiral symmetry~\cite{Suzuki:2002ae,piAF:2022gvw}. Furthermore, measurements of pion flow in heavy-ion collisions at intermediate energies of about 1 GeV per nucleon have been used to infer the equation-of-state of dense baryonic matter~\cite{FOPI:2006ifg, Li:2005gfa}.
Recent works that investigate the properties of hypothetical axions in dense nuclear matter also connect axion properties to the behavior of the pions in the same environment~\cite{Balkin:2020dsr}.

The femtoscopy technique applied to hadron--hadron collisions measured at the LHC could provide a novel access to the pion--(multi)nucleon interactions. This method has already shown sensitivity to the residual final-state interaction (FSI) among hadron pairs and triplets with different quark content. In particular, the ALICE Collaboration measured several interactions with a precision that, in the case of unstable hadrons, challenges the scarce available scattering data. If one considers only u and d quarks, $\uppi$--$\uppi$ and p--p correlations~\cite{ALICE:2023sjd, ALICE:2018ysd, ppSourceErratum} have been analysed. If one includes s and c  quarks, $\text{p--}\text{K}^\pm$, $\text{K}^\pm$--$\text{K}^\pm$, $\text{K}_\text{S}^0$--$\text{K}_\text{S}^0$, $\text{K}_\text{S}^0$--$\text{K}^\pm$, $\Lambda$--p, $\Sigma$--p, $\Xi^-$--p, $\Omega^-$--p, $\Lambda$--$\Xi^-$, and $\text{D}$--p, $\text{D}/\text{D}^*$--$\uppi$, $\text{D}/\text{D}^*$--$\text{K}$~\cite{ALICE:2019gcn, ALICE:2021njx, ALICE:2019buq, ALICE:2019hdt, ALICE:2020mfd, ALICE:2022uso, ALICE:2022enj, ALICE:2024bhk, ALICE:2023zbh, ALICE:2021ovd} have been already measured.
In the case of interactions that have already been precisely determined in scattering experiments, the analysis of two-particle correlations allows the extraction of a parameterization of the particle-emitting source. The latter characterizes the space--time coordinates of the produced hadrons in any inelastic collision. The emitting source describes the probability density distribution of the inter-particle distance at which hadrons are created following a collision. The analysis of p--p, $\text{p--}\Lambda$, $\text{p--}\text{K}^+$, and $\uppi$--$\uppi$ correlations carried out by the ALICE Collaboration~\cite{ppSourceErratum,ALICE:2023sjd,Mihaylov:2023pyl} observed a common scaling behavior of the hadron-pair source in high-multiplicity pp collisions at $\sqrt{s}=13$~TeV, whose size monotonically decreases as a function of the pair transverse mass $m_\text{T} = \sqrt{\overline{m}^2 + k_\text{T} ^2}$. In this definition, $\overline{m}$ is the average mass of the two particles in the pair, while $k_\text{T} = |\mathbf{p}_{\text{T},1} + \mathbf{p}_{\text{T},2}|/2$ is defined with the transverse momenta $\mathbf{p}_{\text{T}}$ of the particles 1 and 2 in the laboratory frame. A common scaling is expected to appear for pairs of identical particle, as it has been already discussed in the context of heavy-ion collisions~\cite{Lisa:2005dd, Kisiel:2009eh}. However, the origin and underlying mechanism of this behavior remains, especially for pairs of non-identical particles, an open question. 

In this paper, proton--pion correlations are analyzed to further test whether the observation of a common hadron-pair source holds true also for this specific hadron--hadron combination which covers a very broad $m_\text{T}$ range, in comparison to the more limited $m_\text{T}$ range of the pairs considered so far. Moreover, $\ppiPlusMinus$ correlations provide access to the resonance production and modification of the resonance properties due to rescattering effects in the pion-rich environment created in hadron--hadron collisions at the LHC. Such effects have been studied in heavy-ion collisions for fixed-target experiments at center-of-mass energies of about 1 GeV~\cite{Reichert:2019lny,Reichert:2022uha,Adamczewski-Musch:2020edy} and at colliders~\cite{ALICE:2023edr}, including ultrarelativistic d--Au collisions at RHIC~\cite{STAR:2008twt}. Understanding the aforementioned resonance properties and their modifications is crucial for transport models such as UrQMD~\cite{Bass:1998ca, Bleicher:1999xi} or SMASH~\cite{Petersen:2018jag}. Indeed, short-lived (c$\tau \approx\,$ 1 fm) resonances such as $\Delta(1232)^{0,++}$ can decay but also regenerate by pion--nucleon scattering before the decoupling of the hadrons produced in the collision. The decay and subsequent secondary formation of the $\Delta$ resonances can affect the measured mass and width of these resonances. This effect occurs as rescattering processes, involving other pions produced within the same collision, modify momenta, leading to a downward shift in resonance masses with respect to their nominal values due to phase space considerations.  
A detailed understanding of such processes is very important to correctly interpret the two-body $\ppiPlusMinus$ correlation but also to evaluate the effect of rescattering in ultrarelativistic hadron collisions.

Leaning on the reference provided by the two-body $\ppiPlusMinus$ correlations, one can extend the study to $\pppiPlusMinus$ triplets. Similar studies have already been carried out by the ALICE Collaboration through measurements of p--p--p, $\overline{\text{p}}$--p--p, p--p--$\Lambda$~\cite{ALICE:ppp-ppL}, and p--p--$\mathrm{K}^{\pm}$ systems~\cite{ALICE:ppK} in pp collisions at $\sqrt{s}$~=~13~TeV. The results have shown that the measured correlation functions provide access to the dynamics of the three-body systems.
In parallel with the analysis of the triplet correlations, methods to interpret the data have recently been developed. Cumulant expansion~\cite{Kubo,ALICE:3pi} and an application of the latter within the projector method~\cite{DelGrande:2021mju} have been tested on the data to demonstrate that it is possible to see the evidence of genuine three-body effects by means of three-particle correlations, provided that all mutual two-particle interactions are known and correctly accounted for. Also, fully-fledged theoretical calculations of three-body correlation functions and comparisons with measured correlations demonstrated the sensitivity to the short-range part of the strong interaction in such observables~\cite{KievskyThreeBody}.
These new methods also provide a promising avenue to investigate systems composed of pions and many nucleons and, hence, access to many-body interactions.

In this paper, measurements of the $\ppiPlusMinus$ and $\pppiPlusMinus$ femtoscopic correlations, obtained in high-multiplicity pp collisions at $\snnFemto$ recorded by ALICE, are presented.  The article is organized as follows. In Sec.~\ref{s:DataAnalysis}, the ALICE detector is introduced, and the applied event and charged track selections are explained. Section~\ref{Sec. Study of the ppi system} introduces two-body femtoscopy (Sec.~\ref{Sec. Two-Particle Correlation}), the modeling of the $\ppiPlusMinus$ correlation function (Sec.~\ref{Sec. Two-Body Modeling}) as well as the obtained results (Sec.~\ref{Sec. Results Two-Body}). Lastly, three-body femtoscopy and the cumulant formalism are presented in Sec.~\ref{Sec. Three-particle correlation function}. The results of the $\pppiPlusMinus$ three-particle systems are shown in Sec.~\ref{Sec. Three-Body results}. A summary of the results is presented in Sec.~\ref{Sec. Summary}.

%%%%%%%%%%%%%%%%%%%%%%%%%%%%%%%%%%%%%%%
\section{Event and track selection} 
\label{s:DataAnalysis}

Both $\ppiPlusMinus$ and $\pppiPlusMinus$ femtoscopic correlations are measured using identical event and track selection criteria. The analyzed data consist of high-multiplicity (HM) pp collisions at center-of-mass energy $\snnFemto$ recorded by ALICE~\cite{ALICE:2008ngc, ALICE:2014sbx, ALICE:2022wpn} during the LHC Run 2 data campaign in the years 2016--2018. The events are triggered based on the measured amplitude in the V0 detector, which consists of two arrays of plastic scintillators covering the pseudorapidity ranges $2.8 < \eta < 5.1$ and $-3.7 < \eta < -1.7$~\cite{ALICE:2013axi}. Only events classified as INEL $>$ 0 are considered, i.e., at least one charged track is registered within the pseudorapidity
range $|\eta| < 1$. From the HM-triggered data sample, the events with the 0.17\% largest charged-track multiplicity (concerning all inelastic collisions) are selected for the analysis. For the selected HM collisions, on average 30 charged tracks are found in $|\eta| < 0.5$~\cite{ALICE:2020mfd}. The position of the primary vertex (PV) is determined from the charged-particle tracks reconstructed in the Inner Tracking System (ITS)~\cite{ALICE:2008ngc,ALICE:2010tia} and the Time Projection Chamber (TPC)~\cite{Alme:2010ke}, which cover $|\eta| < 0.9$. The PV position along the beam axis (denoted as the $z$-direction) is required to be $|\text{PV}_z|<10$~cm of the nominal interaction point of ALICE to guarantee a uniform coverage of the tracking detectors in $|\eta| < 0.8$. Events with multiple PVs (so-called pile-up events) that stem from multiple collisions in the same or nearby bunch crossings are removed using the procedure described in Refs.~\cite{ALICE:2018ysd, ALICE:2019hdt, ppSourceErratum}. This rejection of pile-up events removes about 3\% of the selected events. In addition, a selection on the transverse sphericity~\cite{ALEPH:2003obs, ALICE:2019bdw, ALICE:2012cor, ALICE:2023zbh} is employed, which classifies the shape of an event. Events with $S_\text{T}\rightarrow 0$ are ``pencil-like'' and hence dominated by two back-to-back jets, while events with $S_\text{T}\rightarrow 1$ are spherical. Only events with $S_\text{T} > 0.7$ (corresponding to about 53\% of the selected event sample) are chosen in order to reduce background from mini-jet contributions which are jet-like structures originating from hard parton--parton scatterings~\cite{ALICE:2019gcn, ALICE:2021cpv, ALICE:2021cyj, ALICE:2023wjz}. Employing the discussed event selections, a total of about $5.4\times10^8$ events are available for both analyses. The interactions of particle and antiparticle systems are assumed to be the same. Hence pairs (triplets) of particles and the corresponding antiparticles are combined, i.e. $\ppiPlus \equiv \ppiPlus \bigoplus \overline{\text{p}}\text{--}\uppi^-$, $\ppiMinus \equiv \ppiMinus \bigoplus \overline{\text{p}}\text{--}\uppi^+$, $\pppiPlus \equiv \pppiPlus \bigoplus \overline{\text{p}}\text{--}\overline{\text{p}}\text{--}\uppi^-$ and  $\pppiMinus \equiv \pppiMinus \bigoplus \overline{\text{p}}\text{--}\overline{\text{p}}\text{--}\mathrm{\uppi}^+$. This assumption is explicitly tested by comparing the correlation functions between particles and antiparticles, which are found to be in agreement within their statistical uncertainties.

The proton and pion track selections are discussed below. The presented selections apply equally to both particles and antiparticles. The systematic uncertainties are evaluated by varying the kinematic selections in the track reconstruction as well as selections on the track-quality and particle identification (PID). The corresponding values of these systematic variations are presented in the text below within the parentheses next to the corresponding selection.  

The proton track selection follows Refs.~\cite{ppSourceErratum,ALICE:2018ysd, ALICE:ppp-ppL}. Proton candidate tracks are reconstructed using only the tracking information provided by the TPC in the region $|\eta| < 0.8~(0.77,~0.85)$ and within a transverse momentum range of $ 0.5~(0.4,~0.6) < p_\text{T} < 4.05 \text{ GeV}/c$. To guarantee a good track quality and momentum resolution, a minimum of 80 (70, 90) out of the 159 available space-point hits within the TPC is required. In order to further assure the quality of the proton candidate tracks, at least 70 out of the total of 159
rows of the TPC readout pads are required to be crossed. Additionally, clusters in at least 83\% of these crossed rows have to be found~\cite{ALICE:2018ysd}. For tracks with $p_\text{T} < 0.75 \text{ GeV}/c$, the PID is performed using the specific energy loss $\text{d}E/\text{d}x$ of the TPC only. Proton candidates are required to have a $\text{d}E/\text{d}x$ compatible within three standard deviations ($n_{\sigma,\text{TPC}} < 3~(2.5,~3.5)$) with the expected energy loss of a proton in the TPC gas. For candidates with larger transverse momenta, additional information from the Time-Of-Flight (TOF) detector~\cite{Akindinov:2013tea} is used for PID. For these tracks, protons must have a combined standard deviation $n_{\sigma, \text{comb}} = \sqrt{n_{\sigma,\text{TPC}}^2 + n_{\sigma,\text{TOF}}^2} < 3~(2.5,~3.5)$. The distance-of-closest approach (DCA) of the track to the primary vertex is selected to be less than 0.1~cm in the transverse plane and less than 0.2~cm along the beam axis. The purity of the selected proton sample is determined with Monte Carlo (MC) simulations using the  PYTHIA 8.2~\cite{Sjostrand:2014zea} (Monash 2013 Tune~\cite{Skands:2014pea}) event generator together with the GEANT3 software package~\cite{GEANT3} for particle transport. A HM selection is used to mimic the V0 HM trigger and the same event and track selection criteria as for the experimental data are employed. The purity is estimated by comparing the \ptText{} spectra of selected true protons to the overall selected proton candidate samples in the MC simulation. The resulting \ptText-averaged purity of protons is 98\%. The amount of primary protons, as well as those stemming from weak decays and interaction with the detector material is determined with MC template fits of the DCA distributions in the transverse plane to the experimental data. Within the selected proton sample, 86\% are primaries (\ptText-averaged). The contribution of material is found to be negligible. 

Pion-track candidates are reconstructed with combined information of the ITS and TPC and are considered in the transverse momentum range of $0.14~(0.12,~0.15) < p_\text{T} < 4.0  \text{ GeV}/c$ and $|\eta| < 0.8~(0.77,~0.85)$. Details on the track selection can be found in Ref.~\cite{ALICE:2023sjd}. Similar to the selection of proton candidates, a minimum of 80 (70, 90) out of the 159 available TPC space points is required to ensure good track quality. Pion PID for $p_\text{T} < 0.5 \text{ GeV}/c$ is performed using the specific energy loss of the TPC and requiring a signal compatible within 3 (2.5, 3.5) standard deviations $n_{\sigma,\text{TPC}}$ from the pion hypothesis. For larger transverse momenta, TOF information is used requiring a combined TPC+TOF standard deviation $n_{\sigma, \text{comb}} < 3~(2.5,~3.5)$. To suppress contributions from the feed-down of weak decays and interactions with the detector material, the DCA in the longitudinal and transverse directions are required to be less than 0.3~cm for pions. The purities and amount of primary pions are determined with MC simulations using the same procedure as described for the protons. The selected pion sample has a \ptText-averaged purity of 99\% while the fraction of primary pions is 95\%. The determined primary fraction of pions is multiplied by a factor of 0.878 to account for the fact that 12.2\% of primary pions come from strongly decaying particles with  $c\tau>5\text{ fm}$ such as the $\omega$ meson. As shown in Ref.~\cite{ALICE:2023sjd}, the decay of these particles can be treated as feed-down from long-lived particles in the modeling via the $\lambda$-parameter formalism (see Sec.~\ref{Sec. Two-Particle Correlation}).

For p and $\uppi^+$ tracks that are close in phase space, the reconstruction-related issues of track merging or splitting can occur and they have been investigated using Monte Carlo simulations. In order to avoid any bias due to these effects, an elliptic cut, the so-called close-pair rejection (CPR), is employed. For $\ppiPlus$ pairs, the selection is
\begin{equation}
    \frac{(\Delta \eta)^2}{(0.017)^2} + \frac{(\Delta \varphi^*)^2}{(0.04)^2} < 1 \,, \label{Eq. CPR PPion}
\end{equation}
where $\varphi^{*}$ is the average of the track azimuthal angle evaluated at 9 radii in the TPC and corrected for the change induced by the magnetic field.  A similar selection is
 not applied for $\ppiMinus$ as there are no visible effects of pair splitting or merging in the $\Delta \eta-\Delta\varphi^*$ distributions due to the opposite charge.  For the three-particle analysis of the $\pppiPlusMinus$ systems, a CPR between the two protons in the triplet has to be applied in addition to the CPR between protons and positively charged pions. For proton--proton pairs, the CPR is chosen as ${(\Delta \eta)^2 + (\Delta \varphi^*)^2 < (0.017)^2}$~\cite{ALICE:ppp-ppL}. The CPR for $\ppiPlusMinus$ pairs is varied by $\Delta\eta \pm 0.002$ and $\Delta\varphi^* \pm 0.005$ for the evaluation of the systematic uncertainties, while the one of p--p is varied to $\Delta\eta = 0.019$ and $\Delta\varphi^* = 0.019$ following Ref.~\cite{ALICE:ppp-ppL}.

The total systematic uncertainties of the experimental data are evaluated by simultaneous variation of the selection criteria for proton and pion candidates outlined above. The default and systematic variations of the selection criteria are randomly combined into sets, of which in total 44 are generated. As such, the procedure takes correlations between the systematic variations into account. A specific set of random variation is only accepted if the yield of $\ppiPlusMinus$ pairs ($\pppiPlusMinus$ triplets) does not vary by more than 15\% compared to the default selection in the kinematic region $k^* < 0.2 \text{ GeV}/c$ ($Q_3 < 0.5 \text{ GeV}/c$). This is done to ensure that the observed deviation between the systematic variation and the default selection does not originate from statistical fluctuations. The definition of these kinematic variables $k^*$ and $Q_3$~\cite{ALICE:3pi} will be introduced in Secs.~\ref{Sec. Two-Particle Correlation} and~\ref{Sec. Three-particle correlation function}, respectively.

%%%%%%%%%%%%%%%%%%%%%%%%%%%%%%%%%%%%%%%
\section{Study of the \texorpdfstring{$\mathbf{\ppiPlusMinus}$}{proton--pion} system}\label{Sec. Study of the ppi system}

\subsection{Two-particle correlation function}
\label{Sec. Two-Particle Correlation}

 The femtoscopy technique is used to study the space--time properties of particle emission in high-energy collisions as well as the final-state interactions of the emitted hadrons by analyzing momentum correlations between particles. This section focuses on introducing two-particle femtoscopy while its application to three-particle systems related to the study of the $\pppiPlusMinus$ correlations is introduced in Sec.~\ref{Sec. Three-particle correlation function}. Concerning the former case, the main observable of interest is the two-particle correlation function $C(k^*)$~\cite{Lisa:2005dd, Fabbietti:2020bfg}. The latter is defined as 
  \begin{equation}
     C(k^*) = \mathcal{N} \frac{N_{\text{same}}(k^*)}{N_{\text{mixed}}(k^*)} \,. \label{Eq. CF}
 \end{equation}
 The relative momentum between the two particles $k^* = |\mathbf{p}_1^*-\mathbf{p}_2^*|/2$ is defined in terms of the single particle momenta $\mathbf{p}_{i}^*$ evaluated in the pair rest frame. The $N_{\text{same}}(k^*)$ is the distribution of relative momenta between the two particles within the same collision event. This distribution might contain the signal related to the final-state interaction (FSI) as well as the characteristics of the underlying phase space, possible effects of quantum statistics, and non-femtoscopic background correlations such as mini-jet contributions. To account for the phase space, a mixed-event technique is employed where the relative momenta of two particles from two different events are calculated. The resulting mixed-event distribution $N_{\text{mixed}}(k^*)$ thus does not contain any FSI but emulates the underlying phase space. To avoid biases due to acceptance of the detector system as well as differences in the underlying phase space, the mixing procedure is performed only between events with similar $z$ position of the primary vertex ($\text{PV}_z$) and multiplicity~\cite{ppSourceErratum}. The interval width for $\text{PV}_z$ is 2~cm, and the difference in event multiplicity must be less than four particles. Additionally, the CPR between protons and pions presented in Sec.~\ref{s:DataAnalysis} is applied to both $N_{\text{same}}(k^*)$ and $N_{\text{mixed}}(k^*)$ to not introduce any bias. The normalization factor $\mathcal{N}$ is chosen such that the experimental correlation function equals unity in the $k^*$ range where no FSI signal is expected. The $\ppiPlusMinus$ correlation functions are measured in multiple ranges of the pair transverse mass. The considered \mtText{} ranges are $[0.54, 0.75)~\text{ GeV}/c^2$, $[0.75, 0.95)~\text{ GeV}/c^2$, $[0.95, 1.2)~\text{ GeV}/c^2$, $[1.2, 1.5)~\text{ GeV}/c^2$, $[1.5, 2.0)~\text{ GeV}/c^2$, and $[2.0, 2.5)~\text{ GeV}/c^2$. The \mtText-dependent ranges, where the mixed-event distributions of the $\ppiPlusMinus$ systems are normalized to the respective same-event distributions, are given in Table~\ref{Tab. Two-Body CF norm ranges}. The obtained correlation functions of $\ppiPlus$ and $\ppiMinus$ pairs are presented in Appendix~\ref{App. Two-Body CFs} for the $k^*$ range between 0 and 1.5~$\text{ GeV}/c$. In addition, Tab.~\ref{Tab. Two-Body CF norm ranges} provides the average \mtText{} for pairs with a $k^*<0.35~\text{ MeV}/c$.
 
 In the case of an attractive interaction between the two particles, the correlation is enhanced above unity at low $k^*$ (typically for $k^* < 0.2$~GeV$/c$). In contrast, a repulsive interaction leads to a depletion below unity in the correlation function at small relative momenta.  Additionally, the correlation function might be modified due to effects from quantum statistics, e.g.~Pauli exclusion principle for two identical spin 1/2 particles~\cite{ppSourceErratum}, or due to correlations related to the hadronization process (so-called mini-jet background). The latter is discussed in more detail later. In the absence of FSI and other correlations such as quantum statistics, mini-jet contributions between particle pairs or long-range correlations coming from energy-momentum conservation~\cite{PhysRevC.78.064903}, the same-event distribution is purely determined by the available phase space, and hence, the correlation function will be compatible with unity.
 \begin{table}[h]
    \centering
    \begin{tabular}{| c | c  | c | c |}
    \hline 
          & & \multicolumn{2}{c |}{Normalization range in GeV$/c$} \\
         \hline
        \mtText{} range in GeV$/c^2$ & $\langle$\mtText{}$\rangle$ in GeV$/c^2$ & $\ppiPlus$ &  $\ppiMinus$  \\
        \hline 
        \hline
        $[0.54, 0.75)$ & $0.68 \pm 0.05$ & 0.3 -- 0.38 & 0.35  -- 0.43\\
        $[0.75, 0.95)$ & $ 0.86 \pm 0.06$ & 0.35 -- 0.45 & 0.44 -- 0.52\\
        $[0.95, 1.2)$ &  $1.07 \pm 0.07$ & 0.42 -- 0.52 & 0.5--0.6\\
        $[1.2, 1.5)$ & $ 1.33\pm 0.08$ & 0.52 -- 0.62 & 0.6 -- 0.7\\
        $[1.5, 2.0)$ & $1.69 \pm 0.14$ & 0.6 -- 0.8 & 0.67 -- 0.83\\
        $[2.0, 2.5)$ & $2.14 \pm 0.11$ & 0.7 -- 0.95 & 0.8 -- 1.0\\
        \hline
    \end{tabular}
    \caption{Normalization ranges and average \mtText{} of the two-body correlation functions of $\ppiPlusMinus$ pairs for all \mtText{} ranges.}
    \label{Tab. Two-Body CF norm ranges}
\end{table}

 The genuine two-particle correlation function is defined from the theoretical perspective via the Koonin--Pratt equation~\cite{Lisa:2005dd, Fabbietti:2020bfg}
 \begin{equation}
     C(k^*) = \int S(r^*) |\Psi(\mathbf{r^*},\mathbf{k^*})|^2 \text{d}^3 r^* \,, \label{Eq. Koonin-Pratt}
 \end{equation}
where $r^*$ describes the relative distance between the two particles in the pair rest frame and $\Psi(\mathbf{r^*},\mathbf{k^*})$ is the two-particle wave function. 

The source function $S(r^*)$ is proportional to the probability of producing two particles at a relative distance $r^*$. The source of $\ppiPlusMinus$ pairs is modeled according to the Resonance Source Model (RSM)~\cite{ppSourceErratum}. The RSM differentiates between two classes of primary particles: primordial, which are produced directly by the initial collision, and particles stemming from the strong decay of short-lived resonances (with a lifetime $c\tau < 5$~fm). The latter contribution is relevant as the decay products can still undergo FSI and, hence, influence the observed correlation function by changing the emission source shape and thus effectively increasing the source size.

The emission of primordial pairs, the so-called core source, is modeled with a Gaussian distribution 

\begin{equation}
    S(r^*) = \frac{1}{(4\uppi r_{\text{core}}^2)^{3/2}} \exp\left( - \frac{r^{*2}}{4r_{\text{core}}^2} \right)\,.
\end{equation}
The size of the core source is given by the Gaussian width $r_\text{core}$.
The effects of the short-lived, strongly-decaying resonances are taken into account with MC simulations in which resonances are propagated and made to decay exponentially according to their respective lifetime. The EPOS event generator~\cite{EPOS} is used to model the propagation kinematics and angular distributions of the decay particles. The RSM source is generated by determining the relative distances $r^*$ for all proton--pion pairs, i.e. primordial--primordial, primordial--decay product, etc. The decay of short-lived resonances leads to a tail in the source distribution and thus effectively enlarges the source. In the implementation of the RSM, individual resonances decaying into protons and pions are not considered, and effective resonances with an average mass and average lifetime are used instead. The properties of these effective resonances and the amount of primordial protons and pions are determined from predictions of the statistical hadronization model using the Thermal-FIST software package~\cite{ThermalFist}. Details on this can be found in Refs.~\cite{ppSourceErratum, ALICE:2023sjd}. Considering only primordial particles and the decay products of resonances with $c\tau<5\text{ fm}$, these calculations provide a primordial proton yield of 35\% and an effective resonance with $\left<m_{\text{res}}^{\text{eff}}\right>_{\text{p}} = 1.36 \text{ GeV}/c^2$ and life-time $\left<\tau_{\text{res}}^{\text{eff}}\right>_{\text{p}} = 1.65 \text{ fm}/c$~\cite{ppSourceErratum}. Consequently, 65\% of the protons from the particle-emitting source stem from the decay of short-lived resonances. For pions, the respective primordial yield is 39\% and the effective resonance parameters are $\left<m_{\text{res}}^{\text{eff}}\right>_{\uppi} = 1.12 \text{ GeV}/c^2$ and $\left<\tau_{\text{res}}^{\text{eff}}\right>_{\uppi} = 1.5 \text{ fm}/c$~\cite{ALICE:2023sjd}. In the later fitting of the correlation functions, $\left<m_{\text{res}}^{\text{eff}}\right>_{\text{p},\uppi}$ and $\left<\tau_{\text{res}}^{\text{eff}}\right>_{\text{p},\uppi}$ are varied by $\pm 10\%$ to take systematic effects of the thermal models and the averaging process into account. As demonstrated in Ref.~\cite{ALICE:2023sjd}, the RSM approach (Gaussian core for primordial hadrons and tail due to short-lived resonances) effectively results in a source which is similar to Cauchy/exponential type source. 

Resonances with $c\tau>5\text{ fm}$ can be considered as feed-down and hence be absorbed in the $\lambda$ parameter formalism (see Ref.~\cite{ALICE:2023sjd}) explained below. The selected protons and pions, which are used to obtain the measured correlation function $C(k^*)$ (Eq.~\eqref{Eq. CF}), contain impurities, i.e. particles misidentified as protons (pions), as well as a mix of primary particles (i.e.~primordial or produced in decays of short-lived resonances) and particles originating from secondary weak decays ($c\tau > 5$~fm) as well as from interaction with the detector material. The impact of impurities and non-primary particles on the correlation function is taken into account via the $\lambda$-parameter formalism~\cite{ALICE:2018ysd}. The $\lambda$ parameters are calculated as $\lambda_{\text{ij}} = \nolinebreak P_\text{i}F_\text{i}P_\text{j}F_\text{j}$, where $P_{\text{i(j)}}$ is the purity of particle i(j) and  $F_{\text{i(j)}}$ is the fraction of particle i(j) coming from a particular channel, e.g. primary particles. Hence, the parameter $\lambda_{\text{ij}}$ is related to the probability of finding a particle pair with purities $P_{\text{i}}$ $P_{\text{j}}$ stemming from channels with respective fractions $F_{\text{i}}$ $F_{\text{j}}$. The  correlation function $C_\text{Interaction}(k^*)$ describing the interactions can be decomposed via the $\lambda$ parameters into 
\begin{equation}
    C_\text{Interaction}(k^*) = 1 + \lambda_{\text{gen}} \; (C_{\text{gen}}(k^*) - 1 ) + \sum_{\text{ij}} \lambda_{\text{ij}} \; (C_{\text{ij}}(k^*) - 1 ) \,.
\end{equation}
An example for a non-genuine part of the correlation function is the case, where the proton stems from the weak decay of $\Lambda$ ($i = \text{p}_\Lambda$) paired with a primordial pion ($j=\pi$). By definition, the sum of all $\lambda$ parameters equals unity. The decomposition is used in the modeling of the experimental correlation function $C_\text{exp}(k^*)$. The term $C_{\text{gen}}$ represents the genuine correlation function scaled by the genuine correlation strength parameter $\lambda_{\text{gen}}$. The genuine $\lambda_{\text{gen}}$ relates to the correctly identified primary protons and pions. The correlations $C_{\text{ij}}(k^*)$ are from feed-down of weakly-decaying particles, impurities, and material contributions, respectively scaled by the corresponding $\lambda_{\text{ij}}$. The $\lambda$ parameters are calculated by averaging the \ptText-dependent purities and fractions of protons and pions over the two-dimensional distribution of the transverse momenta of the proton and the pion. The resulting \mtText{} differential parameter $\lambda_{\text{gen}}$ are equivalent between $\ppiPlus$ and $\ppiMinus$ and are presented in Table~\ref{Tab. Lambda Gen Values}. In the later fitting procedure, the $\lambda$ parameters are varied by $\pm$10\% to estimate the systematic uncertainty.

\begin{table}[h]
    \centering
    \begin{tabular}{| c | c |}
    \hline 
        \mtText{} range in GeV$/c^2$ & $\lambda_{\text{gen}}$ \\
        \hline 
        \hline
        $[0.54, 0.75)$ & 0.677\\
        $[0.75, 0.95)$ & 0.687\\
        $[0.95, 1.2)$ & 0.689\\
        $[1.2, 1.5)$ & 0.688\\
        $[1.5, 2.0)$ & 0.661\\
        $[2.0, 2.5)$ & 0.616\\
        \hline
    \end{tabular}
    \caption{Parameter $\lambda_{\text{gen}}$ of $\ppiPlusMinus$ pairs for all used \mtText{} ranges.}
    \label{Tab. Lambda Gen Values}
\end{table}

\subsection{Modeling of two-particle correlation function of \texorpdfstring{$\mathbf{\ppiPlusMinus}$}{proton--pion}}
\label{Sec. Two-Body Modeling}
The experimental correlation functions of $\ppiPlus$ pairs are fitted with 
\begin{equation}
C_\text{$\ppiPlus$}(k^*) = \mathcal{N}_\text{Model} \times  C_\text{Background}(k^*) \times C_\text{Interaction}(k^*) + R_{\Delta^{++}} \,. \label{Eq. ProtonPion Exp Fit}
\end{equation}
 The function $C_\text{Background}(k^*)$ describes background contributions from correlated $\ppiPlus$ pairs originating from mini-jets. The latter are associated with hard processes of the initial partons and were already observed in the analysis of other meson--baryon and baryon--antibaryon systems~\cite{ALICE:2019gcn, ALICE:2021cpv, ALICE:2021cyj, ALICE:2023wjz}. The background function $C_\text{Background}(k^*)$ is modeled using  PYTHIA 8.2 (Monash 2013 Tune)~\cite{Sjostrand:2014zea, Skands:2014pea} MC simulations employing the same procedure as in Refs.~\cite{ALICE:2021cyj, ALICE:2023wjz}. In this approach, the simulated $\ppiPlus$ pairs are separated into two samples: pairs originating from a common parton (ancestor) and pairs not sharing such a parton in their evolution history (non-common ancestor). For both cases, a correlation function is obtained using the MC. The background function $C_\text{Background}(k^*)$ is then modeled as
 \begin{equation}
     C_\text{Background}(k^*) = w_\text{C} C_\text{C}(k^*) + (1-w_\text{C}) C_\text{NC}(k^*) + a_0 + a_1k^* + a_2k^{*2} + a_3k^{*3}\,, \label{Eq. ProtonPion Background}
 \end{equation}
 where $C_\text{C}(k^*)$ and $C_\text{NC}(k^*)$ are the common and non-common ancestor correlation functions, respectively. The weight $w_\text{C}$ is a free fit parameter and determines the relative amount of $\ppiPlus$ pairs, which share a common ancestor. All common and non-common ancestor correlation functions for $\ppiPlus$ pairs can be found in Appendix~\ref{App. Two-Body CFs}. The third-order polynomial is used to adjust the long-range correlations of the common and non-common ancestor correlation functions in the large $k^*$ region ($\gtrsim 0.5~\text{ GeV}/c$) as these are not properly described by the MC~\cite{ALICE:2021cyj, ALICE:2023wjz}. The parameters $a_{0,1,2,3}$ of the third-order polynomial are determined by prefitting the data in the large $k^*$ regime with $N_\text{prefit}\times C_\text{Background}$, where $N_\text{prefit}$ is the overall normalization of the prefit. The \mtText-dependent prefit ranges can be found in Table~\ref{Tab. ProtonPion PrefitRanges} in Appendix~\ref{App. Details Two-Body Modelling}. It was checked that the polynomial does not bias the FSI signal in the low $k^*$ region as it is flat in this region. 
 
 The interaction correlation function $ C_\text{Interaction}(k^*) = \left[ \lambda_{\text{gen}}C_{\text{gen}}(r_\text{core}, k^*) + (1-\lambda_{\text{gen}}) \right ]$ is multiplying the previously described background function. The function $C_{\text{gen}}(r_\text{core}, k^*)$ incorporates the correlation signal induced by the genuine interaction between p and $\uppi^+$ given by Coulomb and strong forces (excluding the $\Delta$ resonance). The strong force is described with a double-Gaussian local potential, which is tuned to reproduce scattering parameters (see Appendix~\ref{App. Details Two-Body Modelling}). The potential is evaluated with the CATS framework~\cite{Mihaylov:2018rva}, which numerically solves the Schrödinger equation to obtain the wave function needed to calculate the genuine correlation function using Eq.~\eqref{Eq. Koonin-Pratt}. This genuine correlation is corrected to take the finite momentum resolution of the detector into account (details in Appendix~\ref{App. Two-Body Smearing}). The shape of the genuine correlation depends on the source core radius $r_\text{core}$, which is treated as a free fit parameter. The term $C_{\text{gen}}(r_\text{core}, k^*)$ is scaled by the genuine $\lambda_{\text{gen}}$ parameter described in Sec.~\ref{Sec. Two-Particle Correlation}. Contributions from feed-down and misidentifications are assumed to be flat, resulting in the addition of $(1-\lambda_{\text{gen}})$.  
 
 The decay of $\Delta^{++}$ resonances into $\ppiPlus$ pairs are accounted for with $R_{\Delta^{++}} = \mathcal{N}_{\Delta^{++}} \times PS(p_{\text{T},\Delta^{++}}, T) \times \text{Sill}(M_{\Delta^{++}}, \Gamma_{\Delta^{++}})$. The function $\text{Sill}(M_{\Delta^{++}}, \Gamma_{\Delta^{++}})$ is the Sill distribution~\cite{GiacosaSill}, which depends on the resonance mass $M_{\Delta^{++}}$ and width $\Gamma_{\Delta^{++}}$. The Sill distribution is similar to a Breit--Wigner distribution. However, it considers threshold effects. The resonance width is a free parameter, while the mass is fixed to 1215 MeV$/c^2$~\cite{DeltaMass, Adamczewski-Musch:2020edy}. The term $PS(p_{\text{T},\Delta^{++}},T)$ is a Boltzmann-like phase space factor
\begin{equation}
    PS(p_{\text{T},\Delta^{++}}, T) \propto \frac{M_{\text{p}\uppi^+}}{\sqrt{M_{\text{p}\uppi^+}^2+p_{\text{T},\Delta^{++}}^2}}\exp\left[ - \frac{\sqrt{M_{\text{p}\uppi^+}^2+p_{\text{T},\Delta^{++}}^2}}{T} \right]\,, \label{Eq. PS factor}
\end{equation}
acting as a weight for the emission of the resonance with certain transverse momentum $p_{\text{T},\Delta^{++}}$. The invariant mass $M_{\text{p}\uppi^+}$ of the $\ppiPlus$ pair is directly related to the pair relative momentum $k^*$ via the two-body decay kinematics in the rest frame of the resonance. The transverse momentum $p_{\text{T}}$ can be estimated via the average \mtText{} of the proton--pion pair in the $k^*$ region from 0 to 350 MeV$/c$ where the resonance is placed. The parameter $T$ in Eq.~\eqref{Eq. PS factor} is treated as a free parameter to account for modifications of the resonance spectral shape due to hadronic rescattering effects and resonance regenerations ($\Delta \leftrightarrow$ p + $\uppi$), as discussed in Refs.~\cite{ Reichert:2019lny, Reichert:2022uha}.
The parameter $T$ is called ``$\Delta$ spectral temperature $T$`` throughout this paper. The scaling factor for the resonance $\mathcal{N}_{\Delta^{++}}$ is treated as a free fit parameter. Using the model presented in Eq.~\eqref{Eq. ProtonPion Exp Fit}, the \mtText-differential correlation functions of $\ppiPlus$ pairs are fitted in $k^*\in[0,0.45]~\text{ GeV}/c$. The upper limit of the fit is later varied by $\pm$10\%.  Overall, the fit is performed with 6 free parameters: the overall normalization $\mathcal{N}_\text{Model}$, the weight between common and non-common ancestors $\omega_\text{C}$, the source size $r_\text{core}$ as well as the quantities related to the $\Delta^{++}$ resonance ($\mathcal{N}_{\Delta^{++}}$, $T_{\Delta^{++}}$ and $\Gamma_{\Delta^{++}}$).

The modeling of the experimental correlation function of $\ppiMinus$ pairs is given by
\begin{equation}
C_\text{$\ppiMinus$}(k^*) = \mathcal{N}_\text{Model} \times  C_\text{Background}(k^*) \times C_\text{Interaction}(k^*) + R_{\Delta^{0}} +\mathcal{N}_\Lambda \rm{Gauss}(M_\Lambda, \Gamma_\Lambda) \,. \label{Eq. ProtonAntiPion Exp Fit}
\end{equation}
For the $\ppiMinus$ system, the background is fitted as
 \begin{equation}
     C_\text{Background} = 1 + N_\text{Background} [ w_\text{C} C_\text{C}(k^*) + (1-w_\text{C}) C_\text{NC}(k^*) -1 ] + \text{Sill}_1 + \text{Sill}_2\,.
 \end{equation}
 In this way, the shape of the background is fully determined by $w_\text{C}$, while its overall scaling is determined by $N_\text{Background}$. Both parameters, $w_\text{C}$ and $N_\text{Background}$, are free fit parameters. The common and non-common ancestor correlation functions for $\ppiMinus$ pairs can be found for all \mtText{} intervals in Appendix~\ref{App. Two-Body CFs}. The background is modeled in a more simplistic way compared to the $\ppiPlus$ as the $\ppiMinus$ system contains overall more structures in the correlations function and is thus more complex. For $k^* > 0.4$~GeV$/c$, two structures are present in the correlation function which are related to the decay of higher-mass resonances into $\ppiMinus$ pairs. In order not to bias the background shape due to these structures, two Sill distributions, denoted as $\text{Sill}_1$ and $\text{Sill}_2$, are included in the modeling of the background. The effective masses (widths) are $M_1=1500$~MeV$/c^2$ ($\Gamma_1 = 65$~MeV) and $M_2=1660$~MeV$/c^2$ ($\Gamma_2 = 70$~MeV), respectively.  The term $C_\text{Interaction}$ describes the FSI of the  $p-\uppi^-$ pair. As in the case of $\ppiPlus$, a flat feed-down contribution is assumed. The genuine interaction is modeled by using a double-Gaussian local potential tuned to the available scattering parameters (see Appendix~\ref{App. Details Two-Body Modelling}) and corrected for the finite momentum resolution of the detector (Appendix~\ref{App. Two-Body Smearing}). The function $R_{\Delta^{0}}$ takes into account the decay of the $\Delta^0(1232)$ resonance into a $\ppiMinus$ pair. It is modeled in the same way as $R_{\Delta^{++}}$, the mass of the $\Delta^0(1232)$ is constrained to 1215~MeV$/c^2$. As in the case of the $\Delta^{++}(1232)$, the decay width $\Gamma_{\Delta^0}$, the $\Delta^{++}$ spectral temperature $T$, and the overall scaling $\mathcal{N}_{\Delta^{0}}$ are free fit parameters. The decay $\Lambda \rightarrow \text{p} + \uppi^-$ is modeled with a Gaussian distribution $\text{Gauss}(M_\Lambda, \Gamma_\Lambda)$, which is scaled by $\mathcal{N}_\Lambda$. As in the previous case, the data of $\ppiMinus$ correlations are fitted using Eq.~\eqref{Eq. ProtonAntiPion Exp Fit} in the range of $k^*\in[0,0.45]~\text{ GeV}/c$. A variation of $\pm$10\% is considered for the upper limit of the fit.  The fit is consists of 11 free parameters: the overall normalization $\mathcal{N}_\text{Model}$, the weight between common and non-common ancestors $\omega_\text{C}$, the scaling of the background $N_\text{Background}$, the scaling of the two additional resonance bumps at larger $k^*$ ($\mathcal{N}_{\text{Sill}_1}$ and $\mathcal{N}_{\text{Sill}_2}$), the source size $r_\text{core}$, the quantities related to the $\Delta^{0}$ resonance ($\mathcal{N}_{\Delta^{0}}$, $T_{\Delta^{0}}$ and $\Gamma_{\Delta^{0}}$) and the quantities for the $\Lambda$ ($\mathcal{N}_\Lambda$ and $\Gamma_\Lambda$).

\subsection{Results}
\label{Sec. Results Two-Body}

\begin{figure}[!h]
  \centering
  \subfigure[]{
    \includegraphics[trim={0.2cm 0cm 1.5cm 0cm},clip,width=0.48\textwidth]{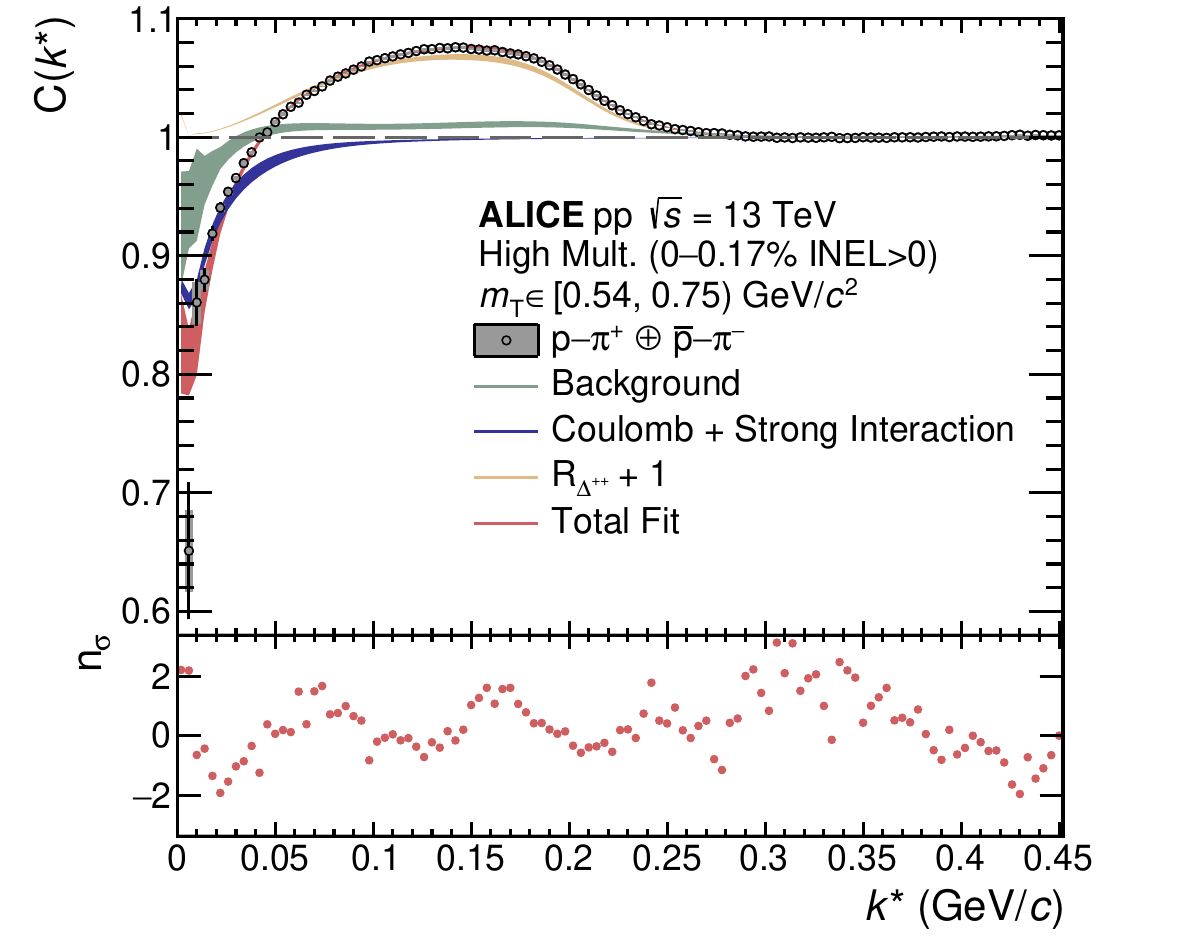}
    \label{Fig. CF ProtonPion mT 1}
  }
  \subfigure[]{
    \includegraphics[trim={0.2cm 0cm 1.5cm 0cm},clip,width=0.48\textwidth]{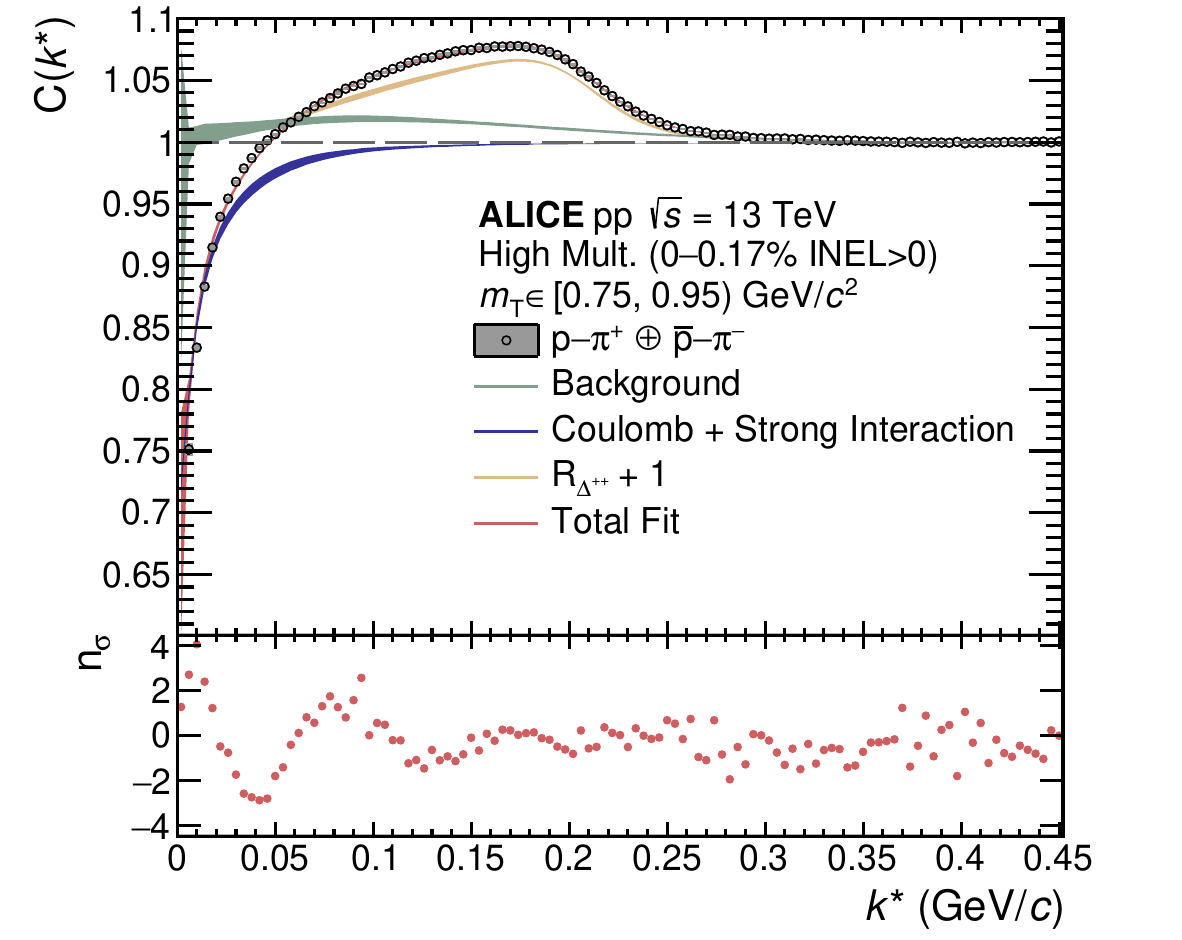}
    \label{Fig. CF ProtonPion mT 2}
  }\\[-0.8em]
    \subfigure[]{
    \includegraphics[trim={0.2cm 0cm 1.5cm 0cm},clip,width=0.48\textwidth]{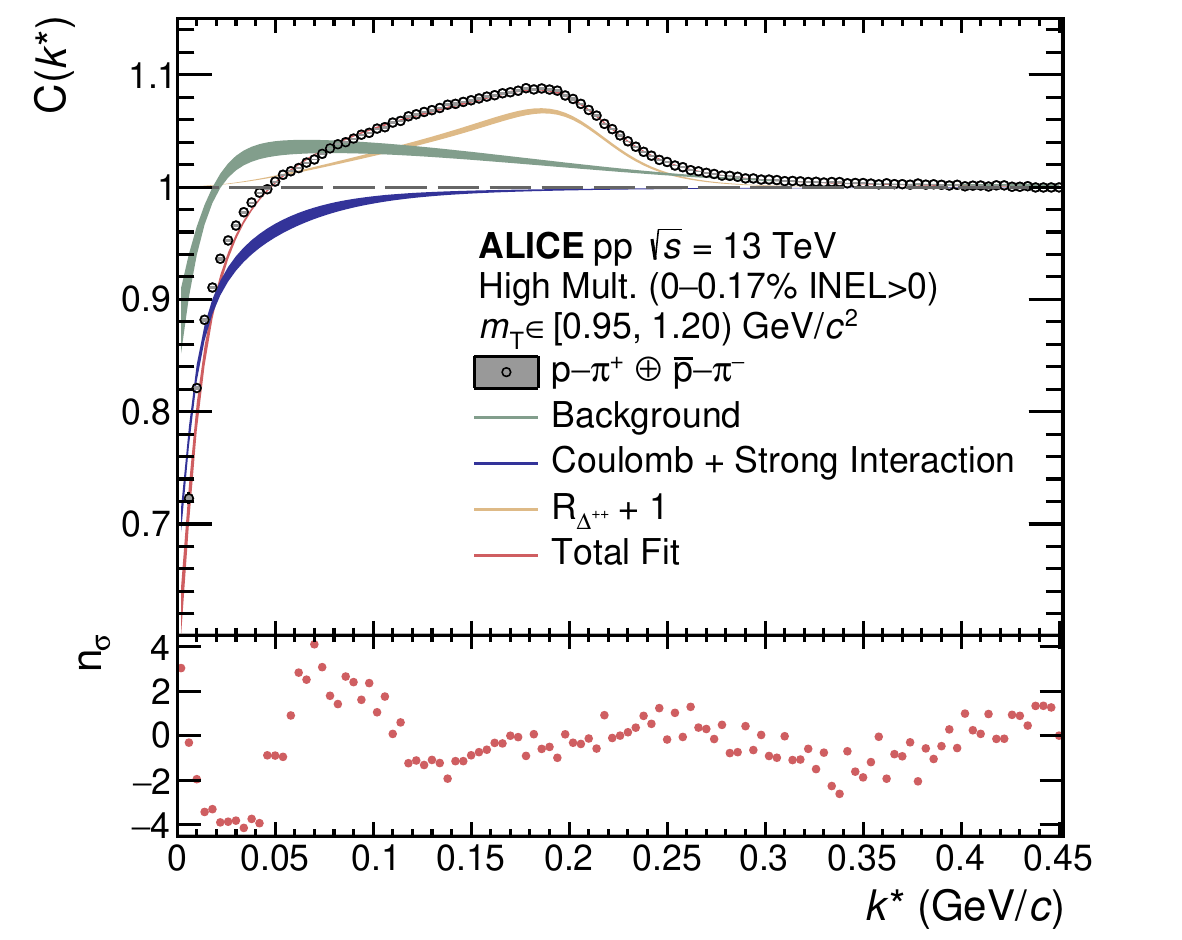}
    \label{Fig. CF ProtonPion mT 3}
  }
  \subfigure[]{
    \includegraphics[trim={0.2cm 0cm 1.5cm 0cm},clip,width=0.48\textwidth]{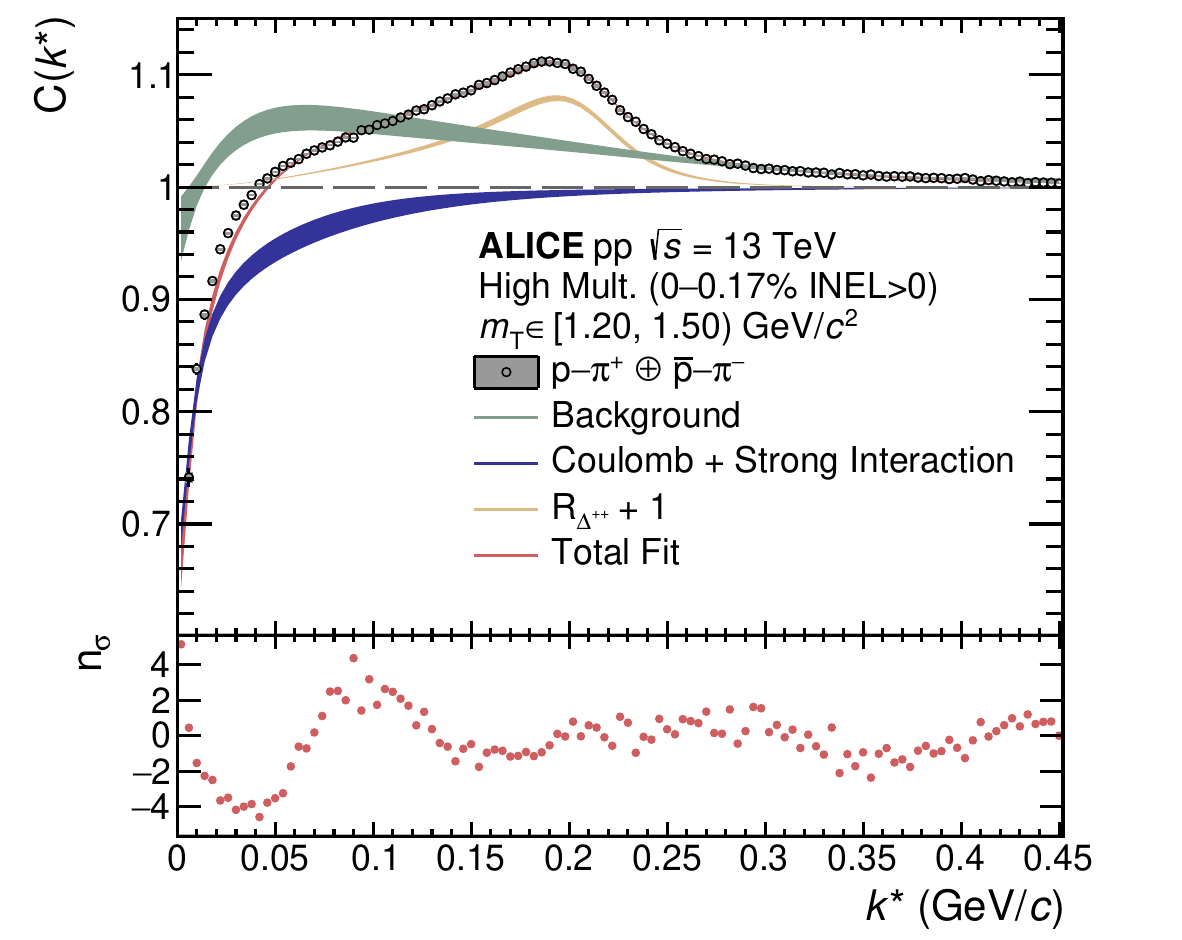}
    \label{Fig. CF ProtonPion mT 4}
  }\\[-0.8em]
    \subfigure[]{
    \includegraphics[trim={0.2cm 0cm 1.5cm 0cm},clip,width=0.48\textwidth]{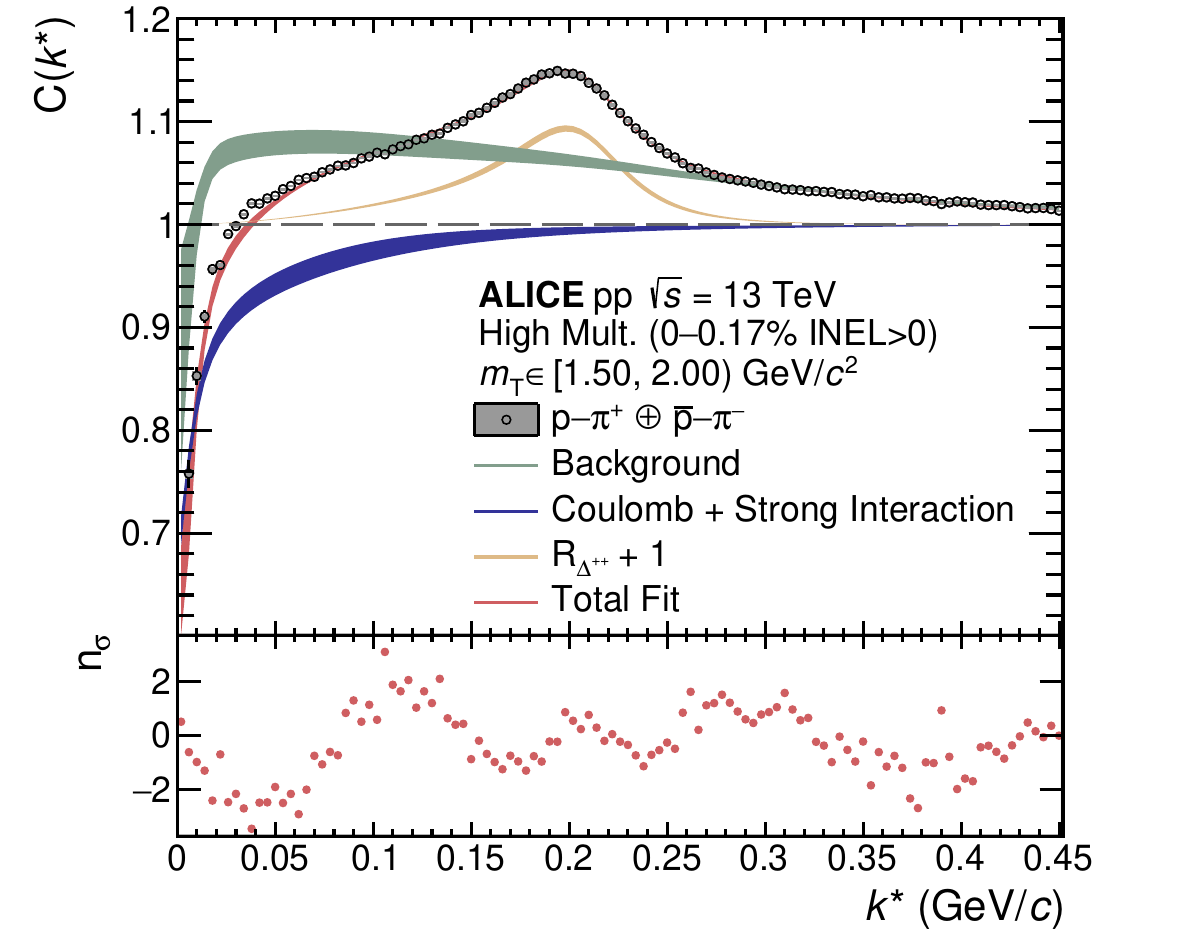}
    \label{Fig. CF ProtonPion mT 5}
  }
  \subfigure[]{
    \includegraphics[trim={0.2cm 0cm 1.5cm 0cm},clip,width=0.48\textwidth]{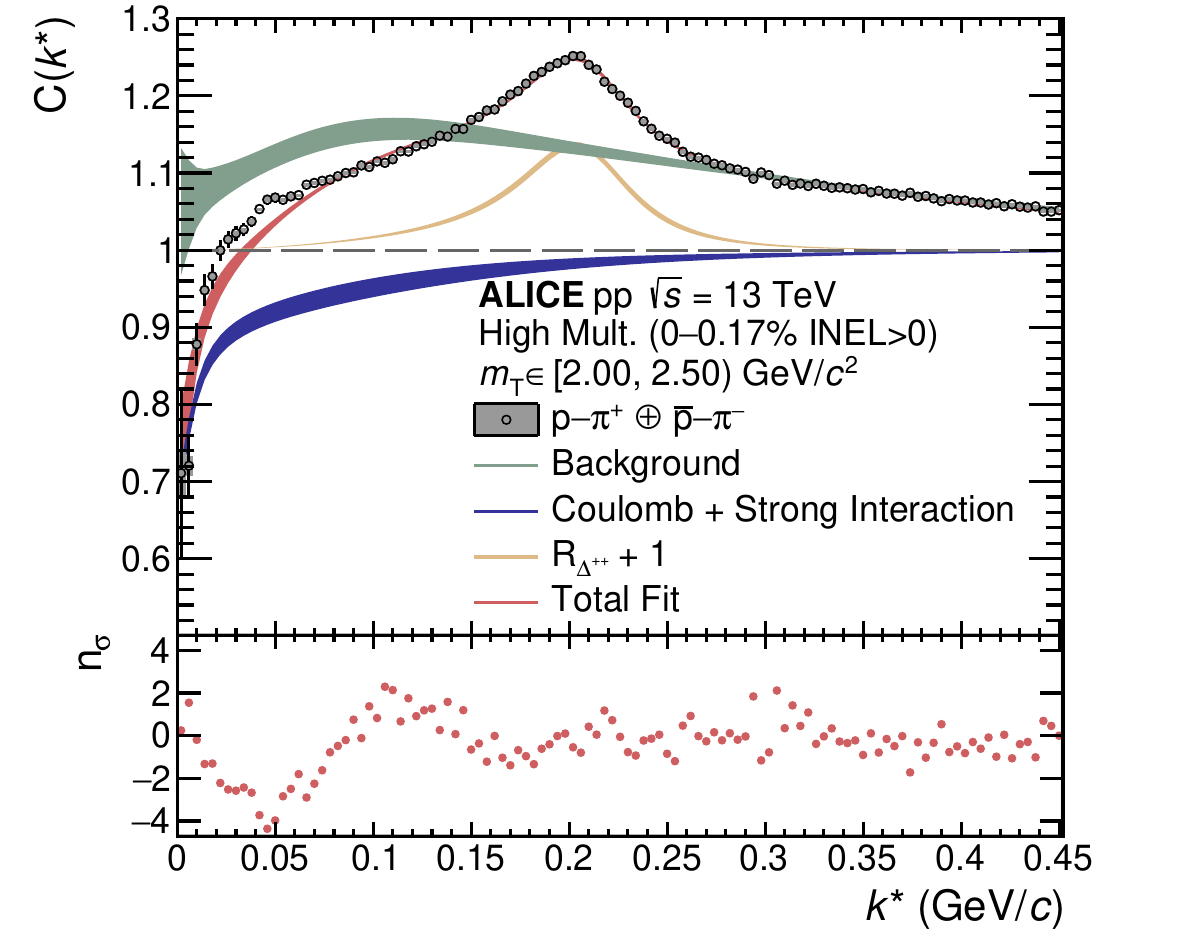}
    \label{Fig. CF ProtonPion mT 6}
  }\\[-0.8em]
  \caption{Upper panel: The experimental correlation function of $\ppiPlus$ pairs (black) as a function of the pair relative momentum $k^*$ in several intervals of the pair \mtText: (a) $[0.54, 0.75)\text{ GeV}/c^2$, (b) $[0.75, 0.95)\text{ GeV}/c^2$, (c) $[0.95, 1.2)\text{ GeV}/c^2$, (d) $[1.2, 1.5)\text{ GeV}/c^2$, (e) $[1.5, 2.0)\text{ GeV}/c^2$, and (f) $[2.0, 2.5)\text{ GeV}/c^2$. The lines (boxes) show the statistical (systematic) uncertainties of the experimental data. The red bands show the fit result according to Eq.~\eqref{Eq. ProtonPion Exp Fit}. The single contributions of correlated background (green), final-state interaction (blue), and $\Delta^{++}(1232)$ (yellow) are presented by the respective colored bands. The width of the bands represents the uncertainty from the fitting procedure.  Lower panel: point-by-point $n_\sigma$ between the overall fit and the experimental data.}
  \label{Fig. CF ProtonPion}
\end{figure}

The results for the correlation functions of $\ppiPlus$ pairs for all considered \mtText{} intervals are presented in Fig.~\ref{Fig. CF ProtonPion}. The correlation function shows a significant structure that peaks in the $k^*$ range of $(0.15-\nolinebreak 0.2)$~GeV$/c$ 
 in all \mtText{} ranges. This corresponds to the $\Delta^{++}(1232)$ resonance, decaying into a $\ppiPlus$ pair. The correlated background due to mini-jet contributions is visible as a tail at larger relative momenta, prominently seen for $k^* > 0.3~\text{ GeV}/c$. This background of correlated $\ppiPlus$ pairs becomes significantly more pronounced for larger pair transverse masses. This is expected, as the mini-jet contribution is related to the initial hard parton scatterings, which generally involves particles with higher \ptText, leading to larger \mtText.

\begin{figure}[h]
\centering
\includegraphics[trim={0cm 0cm 0cm 0cm},clip, width=0.7\textwidth]{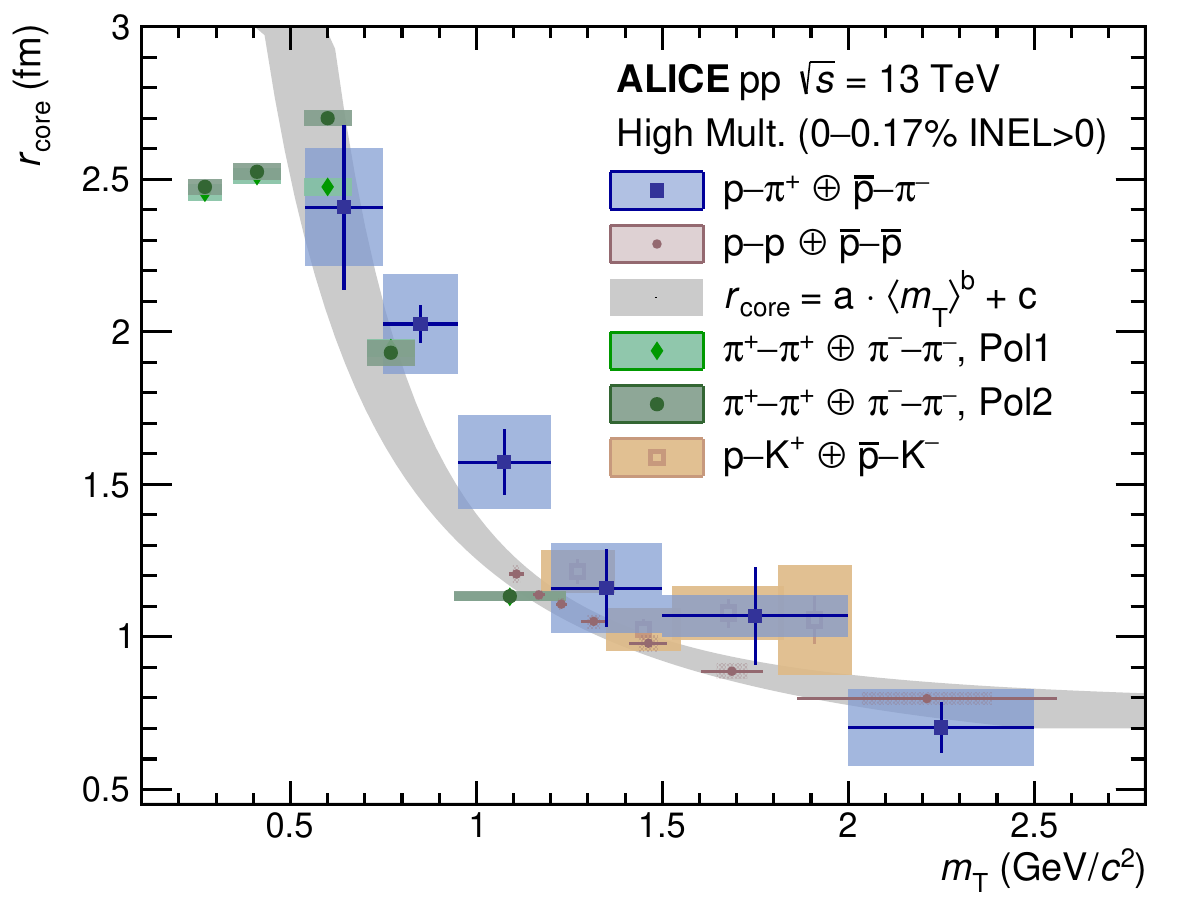}
\caption{Extracted source size $r_\text{core}$ of primordial $\ppiPlus$ pairs as a function of the pair transverse mass. The results are compared with the $r_\text{core}$ extracted from p--p pairs~\cite{ppSourceErratum} as well as $\uppi^\pm$--$\uppi^\pm$ and $\text{p--}\text{K}^+$ pairs~\cite{ALICE:2023sjd} in HM pp collisions at $\sqrt{s}=13$~TeV by ALICE. The lines (boxes) show the statistical (systematic) uncertainties. The gray band represents the extrapolated scaling of p--p pairs~\cite{ppSourceErratum}.}
\label{Fig. ProtonPion Source Size}
\end{figure}

The correlation functions are fitted with the formula given in Eq.~\eqref{Eq. ProtonPion Exp Fit}. The overall fit result is shown in Fig.~\ref{Fig. CF ProtonPion} in the top panel with the red band. The individual contributions of correlated background (green), the $\Delta^{++}(1232)$ (yellow), as well as the FSI between proton and $\uppi^+$ (blue) are presented as well. The width of these bands presents the 1$\sigma$ uncertainty. These uncertainties were obtained by repeating the fit procedure by randomly selected aforementioned variations in the modeling as well as variations of the experimental data according to the respective uncertainties. As reflected in the behavior of the experimental data, the contribution of the mini-jet background becomes more significant with the growing transverse mass. The compatibility between the total fit and the experimental data is evaluated with the p-value of the $\chi^2$ distribution, which is subsequently converted into a number of Gaussian standard deviations $n_\sigma$. The point-by-point $n_\sigma$ as a function of $k^*$ is always shown in the bottom panel of the respective figure. Overall, a good agreement between the fit and the experimental data is found, with a point-by-point $\text{n}_\sigma \lesssim 2$. However, starting from the third \mtText{} interval and higher, a small deviation between the fit and the experimental data can be seen at $k^* \approx \nolinebreak 40$~$\text{ MeV}/c$. This difference appears in the region where the background determined by MC simulations changes slope. This indicates difficulties of the PYTHIA 8.2 (Monash 2013 Tune)~\cite{Sjostrand:2014zea, Skands:2014pea} simulations in describing such correlated background in the regime of large \mtText{}. The experimental data can, therefore, be used as valuable input to MC generators to study such correlations in the future. However, the range of this deviation ($k^* \approx 20-60\text{ MeV}/c$) is small compared to signal of the FSI (0-300$\text{ MeV}/c$ in $k^*$). Hence, the final results are not influenced by this deviating background. The contribution of the $\Delta^{++}(1232)$ shows distortion from the typical Breit--Wigner-like shape, especially for the correlation functions at low \mtText.
\begin{figure}[h]
  \centering

    \subfigure[]{
    \includegraphics[trim={0.1cm 0 0.4cm 0cm},clip,width=0.48\textwidth]{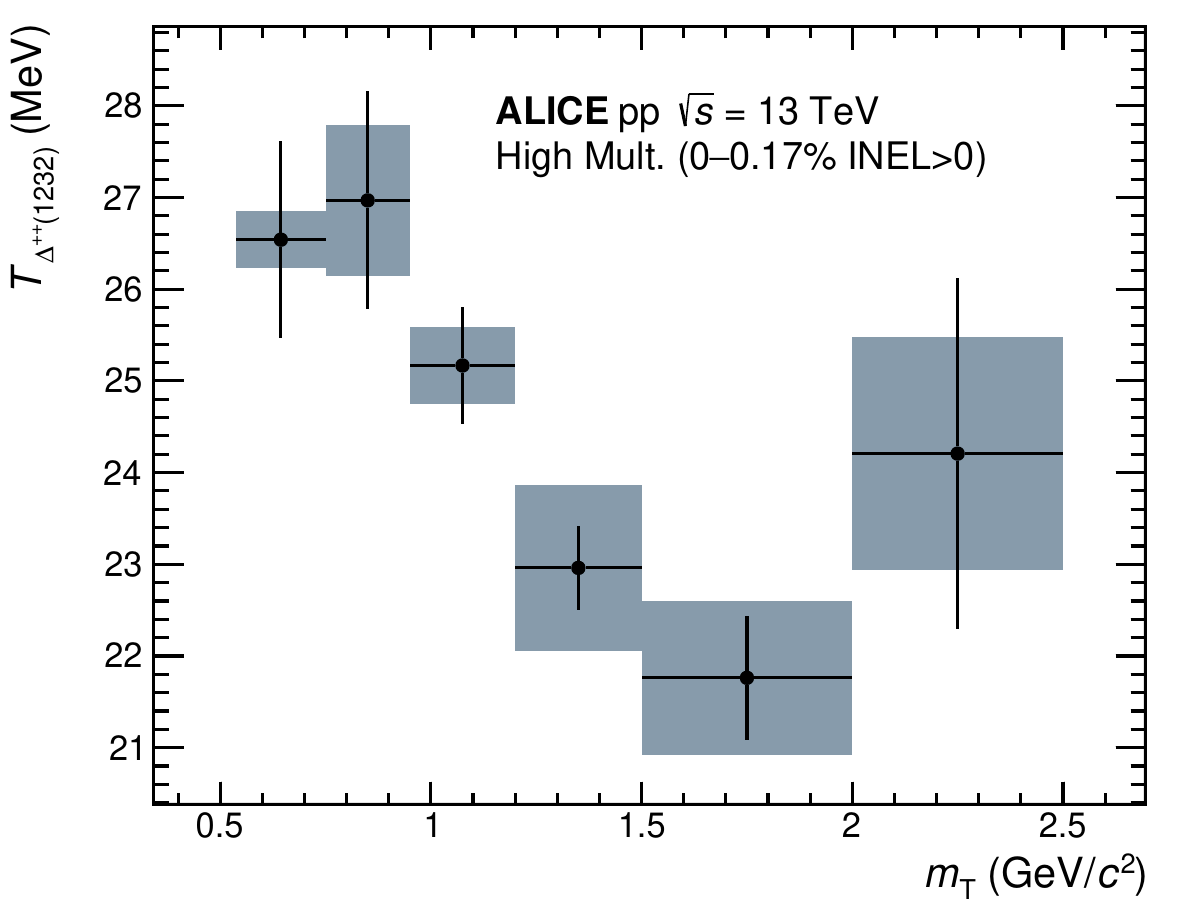}
    \label{Fig. Delta++ Temp}
  }
\subfigure[]{
    \includegraphics[trim={0.1cm 0 0.4cm 0cm},clip,width=0.48\textwidth]{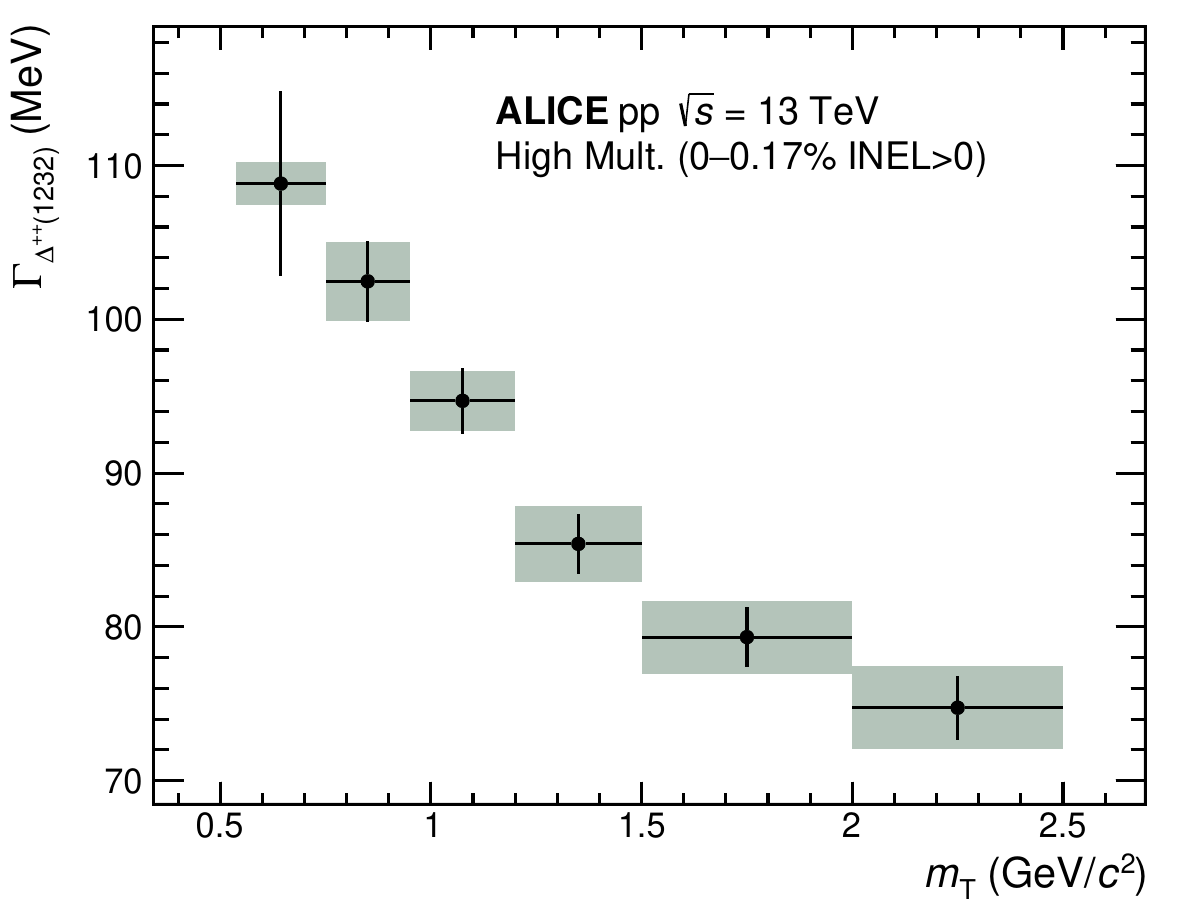}
    \label{Fig. Delta++ Width}
  }
  \caption{Results for the (a) $\Delta^{++}$ spectral temperature and (b) width as a function of the pair transverse mass \mtText{} for the $\Delta^{++}(1232)$ with the statistical (lines) and systematic (boxes) uncertainties. }
  \label{Fig. Delta++ Width Temperature}
\end{figure}
The extracted spectral temperature and resonance width are shown as a function of \mtText{} in Figs.~\ref{Fig. Delta++ Temp} and~\ref{Fig. Delta++ Width}, respectively. The width $\Gamma_{\Delta^{++}}$ shows a decreasing trend from the lowest to the largest \mtText{} range with values centered around $\Gamma_{\Delta^{++}} = 90$~MeV. This is in agreement with the results in Ref.~\cite{GiacosaSill} using the Sill distribution.  The spectral temperature of the $\Delta^{++}(1232)$ shows a similar, slightly decreasing trend with growing \mtText{}. However, it is rather consistent with about 25~MeV in all \mtText{} intervals within the present uncertainties. The lifetime of the $\Delta^{++}(1232)$ is about $1.7~\text{ fm}/c$~\cite{PDG} and thus in the order of the particle-emitting source size. Therefore, during its lifetime, the $\Delta^{++}(1232)$ is surrounded by the particles emitted at short distances at the fm scale. In this range, strong interaction plays a significant role, and thus, rescattering and regeneration effects can occur. This should be more significant for resonances with low \ptText, while resonances with larger transverse momenta can escape the collision environment faster. This is indeed seen in the extracted shape of the $\Delta^{++}(1232)$ as the distortions from a typical Breit--Wigner-like shape are stronger for low pair \mtText, which relates to low \ptText{} of the resonance. The resulting temperature of about 25 MeV is significantly lower than the kinetic freeze-out temperature extracted by blast-wave fits to hadron spectra in pp collisions by ALICE~\cite{ALICE:2020nkc}. The latter method results in values above 100~MeV. Such a salient difference needs to be addressed in future studies, specifically how the values extracted by both methods can be connected to each other. The extracted source size $r_\text{core}$ for primordial $\ppiPlus$ pairs is shown in Fig.~\ref{Fig. ProtonPion Source Size} together with the $r_\text{core}$ extracted from p--p pairs~\cite{ppSourceErratum} as well as $\uppi^\pm$--$\uppi^\pm$ and p--$\text{K}^+$ pairs~\cite{ALICE:2023sjd} in HM pp collisions at $\sqrt{s}=13$~TeV by ALICE. The gray band in Fig.~\ref{Fig. ProtonPion Source Size} shows the extrapolated scaling of p--p pairs~\cite{ppSourceErratum} using a power-law function. The width of the band represents the $3\sigma$ window around the respective experimental data. The source size from $\ppiPlus$ pairs as a function of the pair \mtText{} is compatible within the uncertainties with those of the other analyzed systems. The agreement between the $r_\text{core}$ extracted with $\ppiPlus$ pairs and the extrapolated p--p core radii is evaluated in terms of number of Gaussian standard deviations $n_\sigma$, taking into account the statistical and systematic uncertainties of the data as well as the width of the gray band. It amounts to $n_\sigma = 1.28$ in the interval $0.54 < \nolinebreak \text{\mtText{}} < 2.5 \text{ GeV}/c^2$. The extracted source size of $\ppiPlus$ is also consistent with the $r_\text{core}$ extracted with p--$\Lambda$ pairs by ALICE~\cite{ppSourceErratum}. The latter are not explicitly presented here as they fully overlap with the p--p points. Adding to the results of Ref.~\cite{ALICE:2023sjd}, these data present an additional compelling observation of a common hadron-pair emission source in pp collisions at LHC energies including non-identical pairs. However, if one considers the $\ppiPlus$ source as a convolution of a single-particle $\uppi^+$ and single-particle proton source, such a common scaling is not expected for non-identical particle pairs. The reason is that the source would be dominated by a large pion source, as the \ptText{} of the pion (and hence its single-particle \mtText{}) is usually smaller than the \ptText{} of the proton entering the correlation function at low $k^*$. The observed common source scaling for identical and non-identical hadron-pairs is thus not trivial. The origin of this behavior needs to be addressed in dedicated studies in the future.

%%%%%%%%%%%%%%%%%%%%%%%%%%%%%%%%%%

\begin{figure}[h]
  \centering
  \subfigure[]{
    \includegraphics[trim={0.2cm 0cm 1.5cm 0cm},clip,width=0.48\textwidth]{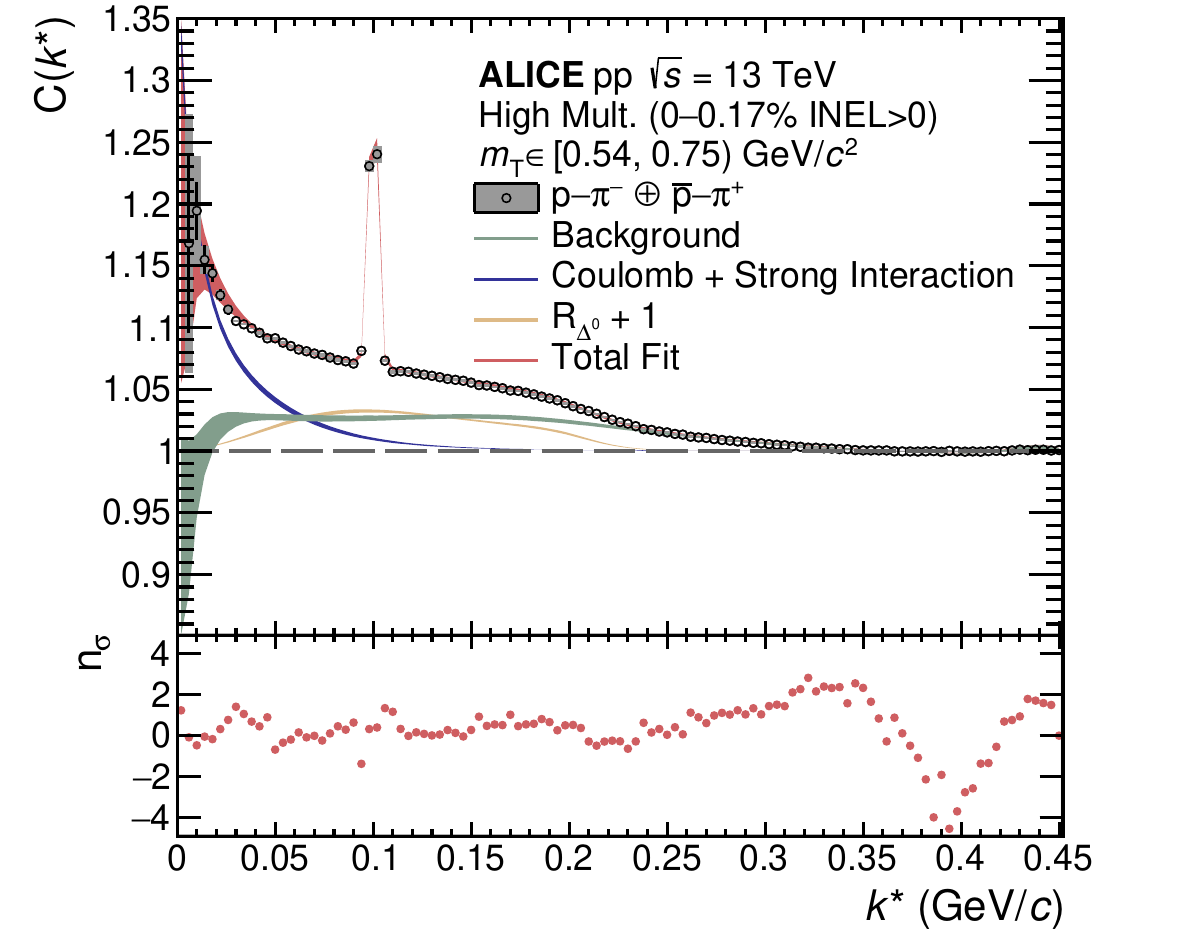}
    \label{Fig. CF ProtonAntiPion mT 1}
  }
  \subfigure[]{
    \includegraphics[trim={0.2cm 0cm 1.5cm 0cm},clip,width=0.48\textwidth]{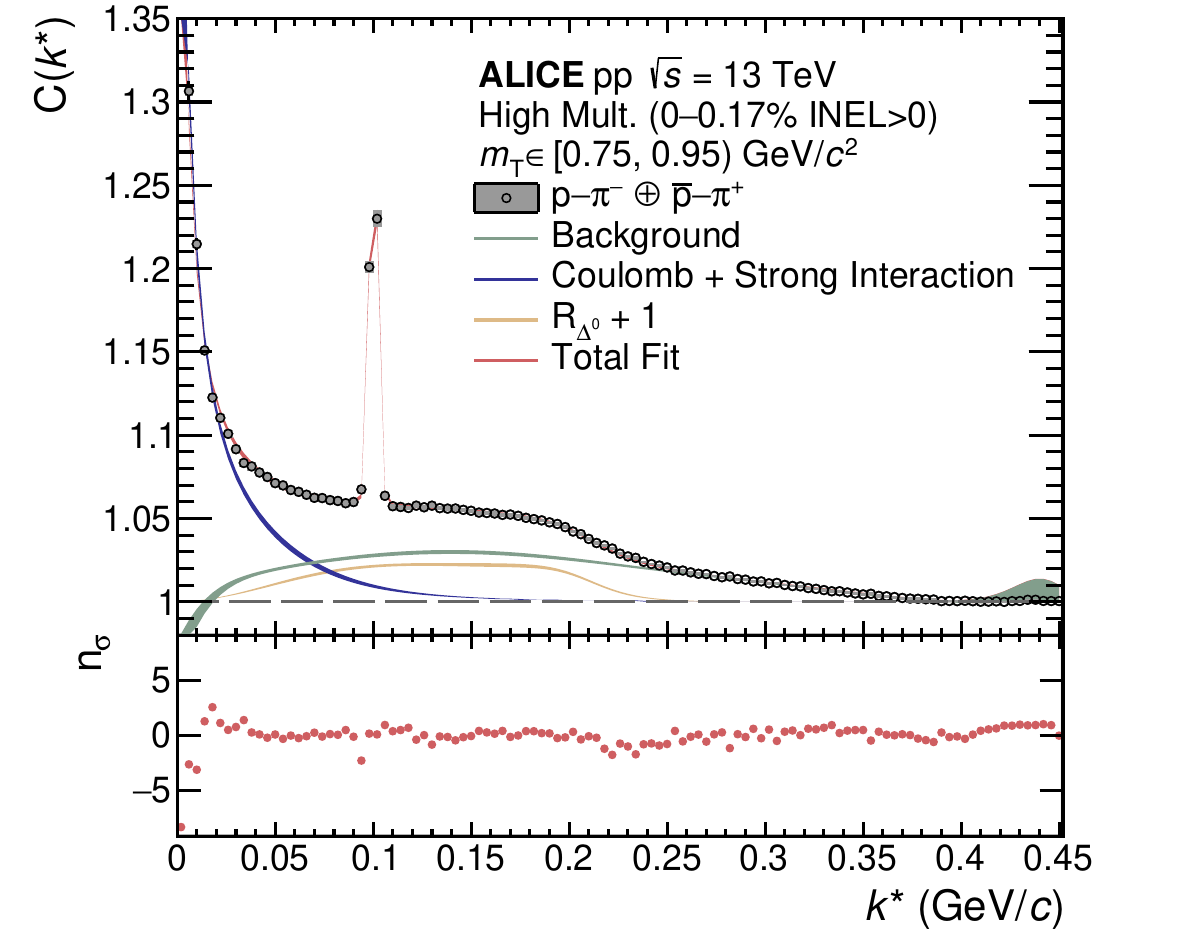}
    \label{Fig. CF ProtonAntiPion mT 2}
  }\\[-0.8em]
    \subfigure[]{
    \includegraphics[trim={0.2cm 0cm 1.5cm 0cm},clip,width=0.48\textwidth]{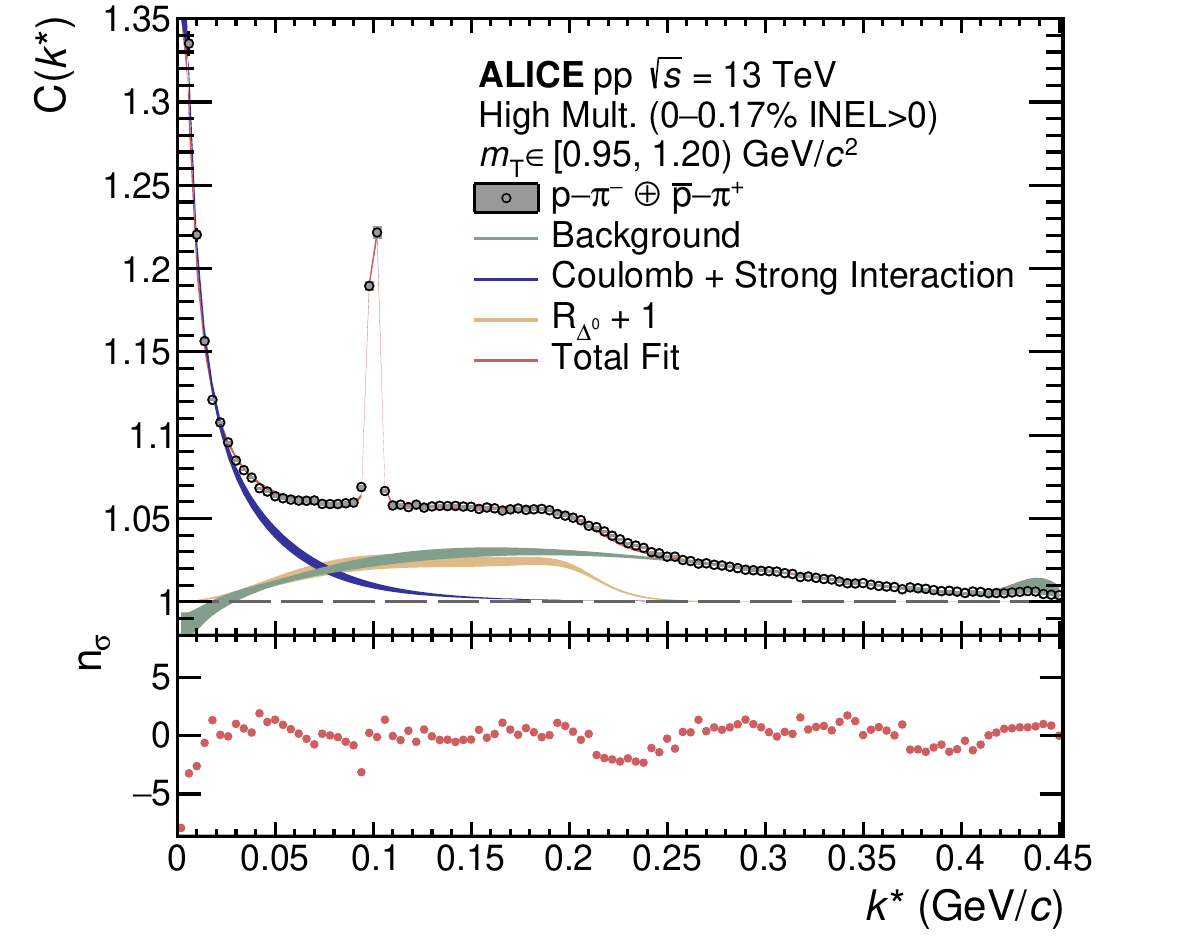}
    \label{Fig. CF ProtonAntiPion mT 3}
  }
  \subfigure[]{
    \includegraphics[trim={0.2cm 0cm 1.5cm 0cm},clip,width=0.48\textwidth]{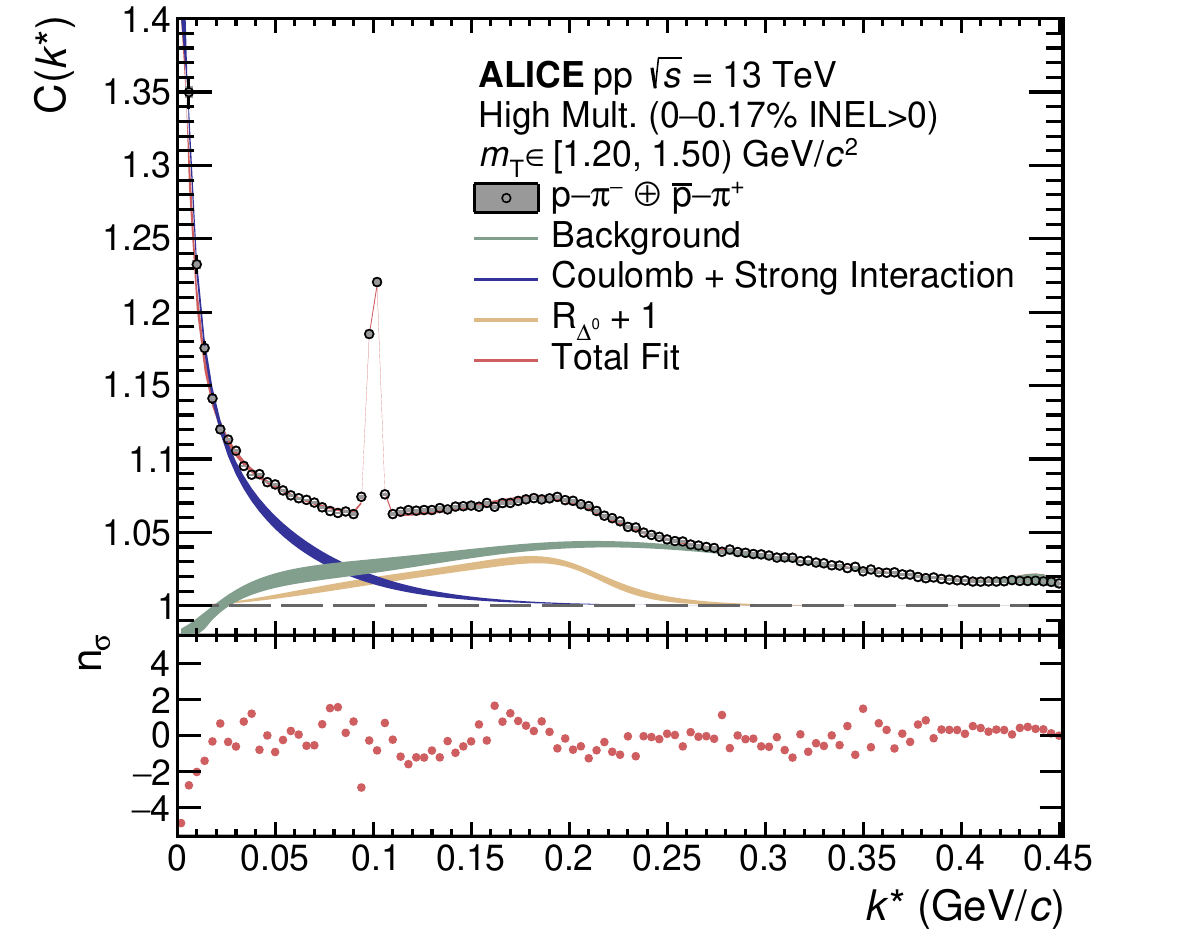}
    \label{Fig. CF ProtonAntiPion mT 4}
  }\\[-0.8em]
    \subfigure[]{
    \includegraphics[trim={0.2cm 0cm 1.5cm 0cm},clip,width=0.48\textwidth]{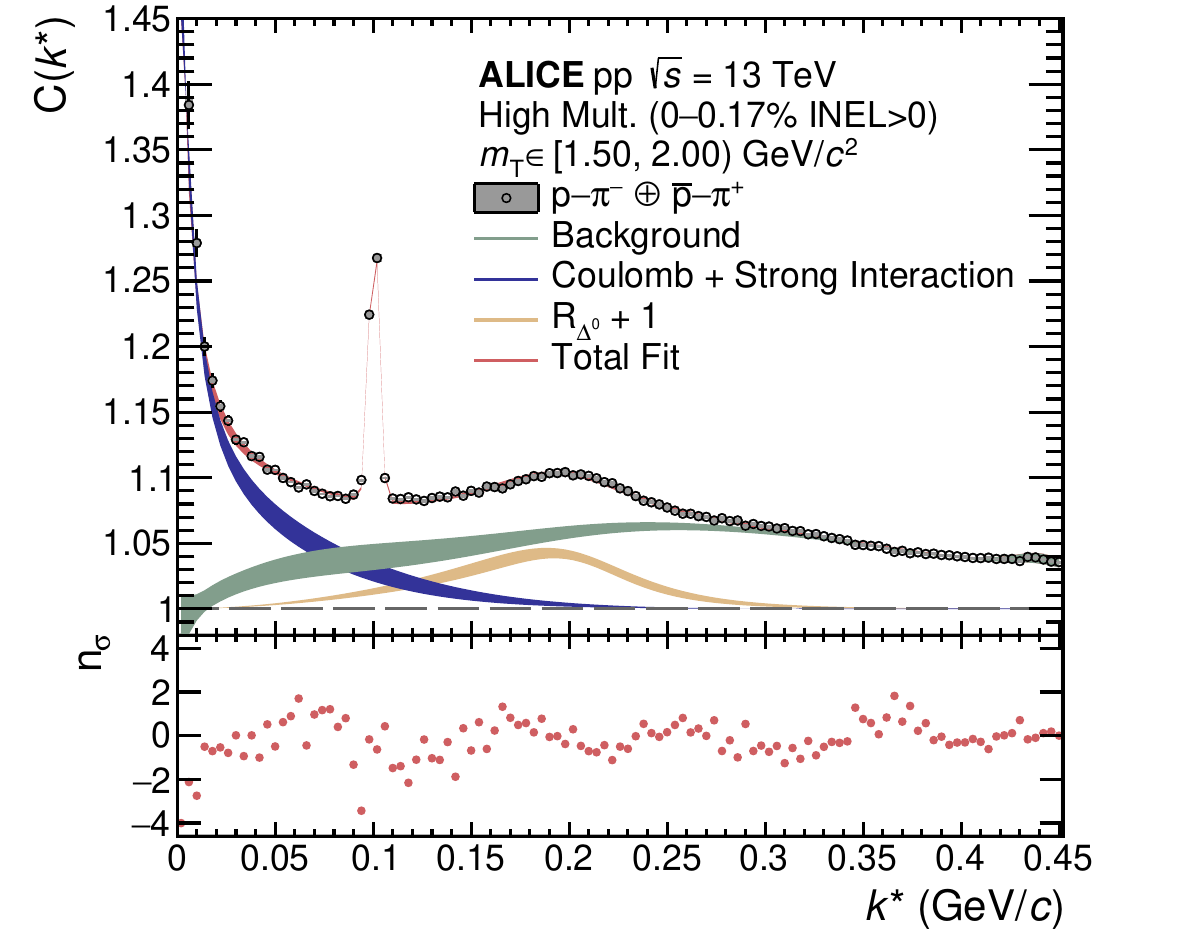}
    \label{Fig. CF ProtonAntiPion mT 5}
  }
  \subfigure[]{
    \includegraphics[trim={0.2cm 0cm 1.5cm 0cm},clip,width=0.48\textwidth]{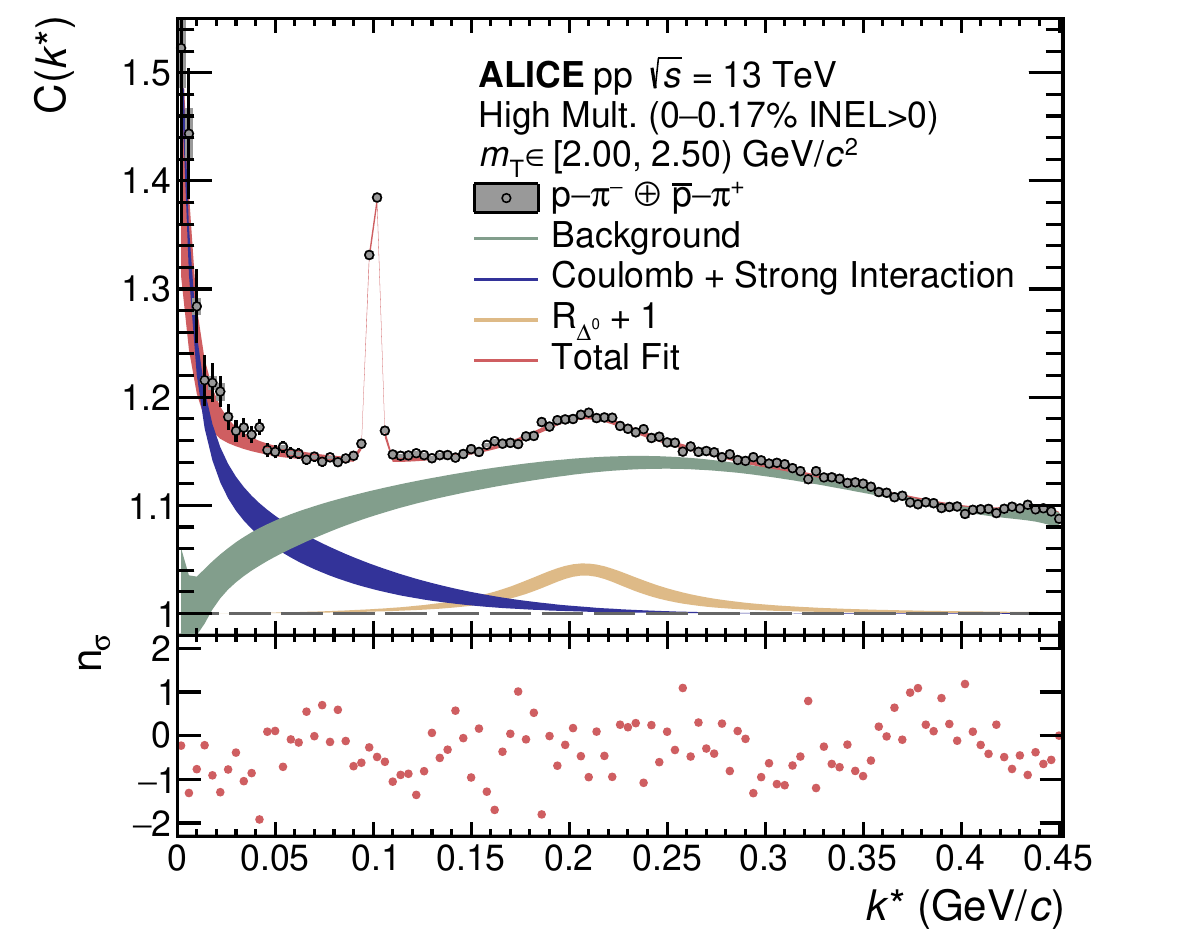}
    \label{Fig. CF ProtonAntiPion mT 6}
  }\\[-0.8em]
  \caption{Upper panel: The $\ppiMinus$ experimental correlation function is shown in black as a function of the pair relative momentum $k^*$ for several intervals of the pair \mtText: (a) $[0.54, 0.75)\text{ GeV}/c^2$, (b) $[0.75, 0.95)\text{ GeV}/c^2$, (c) $[0.95, 1.2)\text{ GeV}/c^2$, (d) $[1.2, 1.5)\text{ GeV}/c^2$, (e) $[1.5, 2.0)\text{ GeV}/c^2$, and (f) $[2.0, 2.5)\text{ GeV}/c^2$. The lines and boxes show the statistical and systematic uncertainties of the experimental data, respectively. The fit results according to Eq.~\eqref{Eq. ProtonAntiPion Exp Fit} are depicted by the red bands. Contributions of correlated background, final-state interaction, and $\Delta^{0}(1232)$ are shown as the green, blue, and yellow bands, respectively. The width of the bands represents the uncertainty from the fitting procedure. Lower panel: point-by-point $n_\sigma$ between the overall fit and the experimental data.}
  \label{Fig. CF ProtonAntiPion}
\end{figure}

\begin{figure} [h]
  \centering
    \subfigure[]{
    \includegraphics[trim={0.1cm 0 0.4cm 0cm},clip,width=0.48\textwidth]{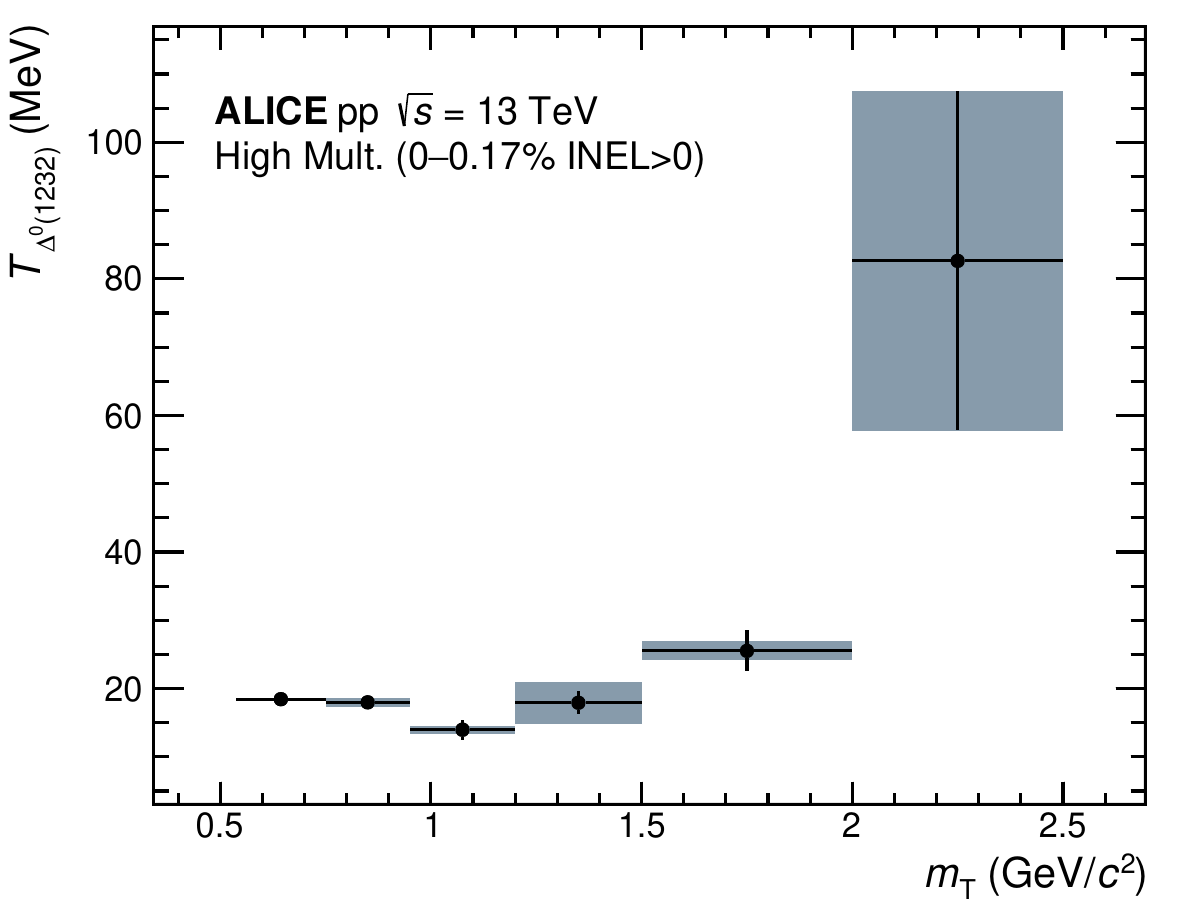}
    \label{Fig. Delta0 Temp}
  }
  \subfigure[]{
    \includegraphics[trim={0.1cm 0 0.4cm 0cm},clip,width=0.48\textwidth]{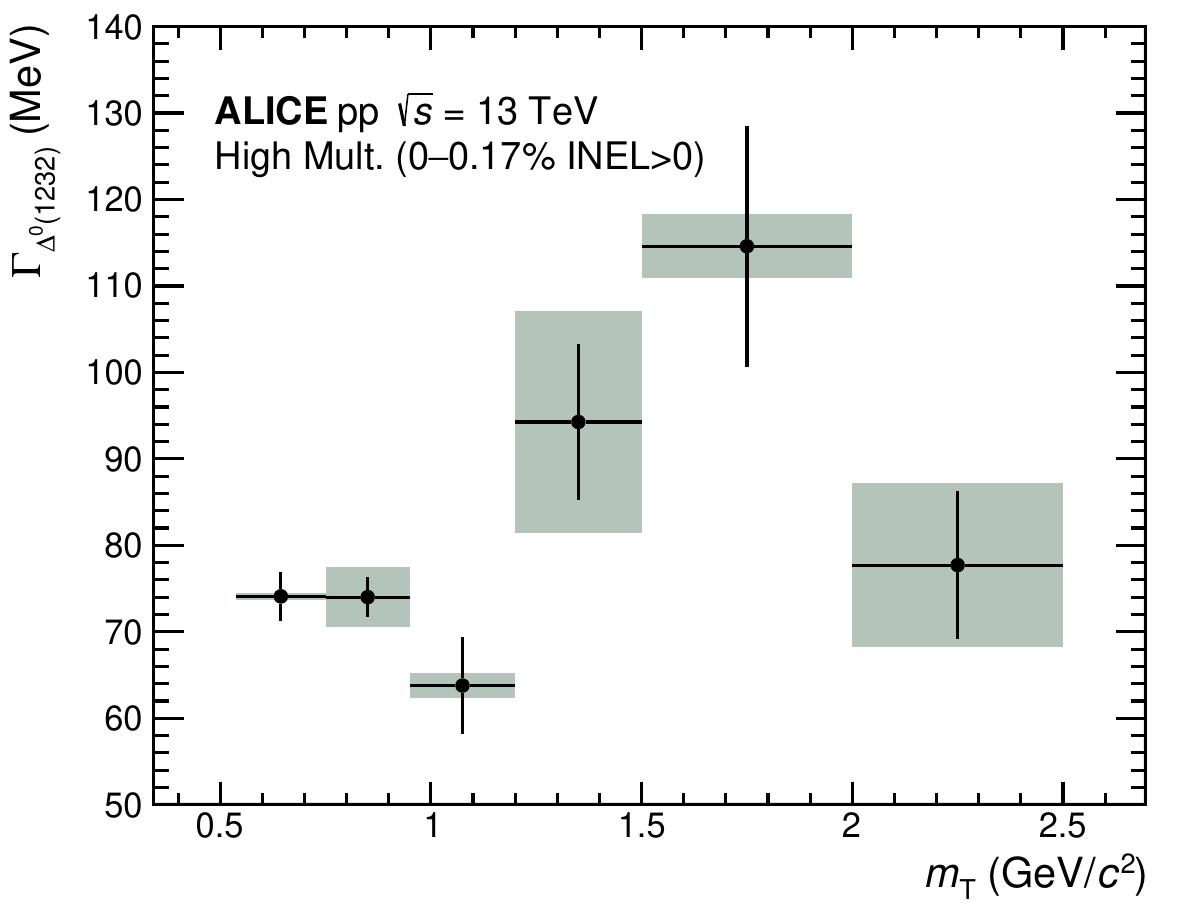}
    \label{Fig. Delta0 Width}
  }

  \caption{Extracted (a) $\Delta^0$ spectral temperature and (b) width as a function of the pair transverse mass \mtText{} for the $\Delta^{0}(1232)$. The lines show the statistical uncertainties, while the boxes represent the systematic uncertainties. }
  \label{Fig. Delta0 Width Temperature}
\end{figure}

\begin{figure}[h]
\centering
\includegraphics[trim={0cm 0cm 0cm 0cm},clip, width=0.7\textwidth]{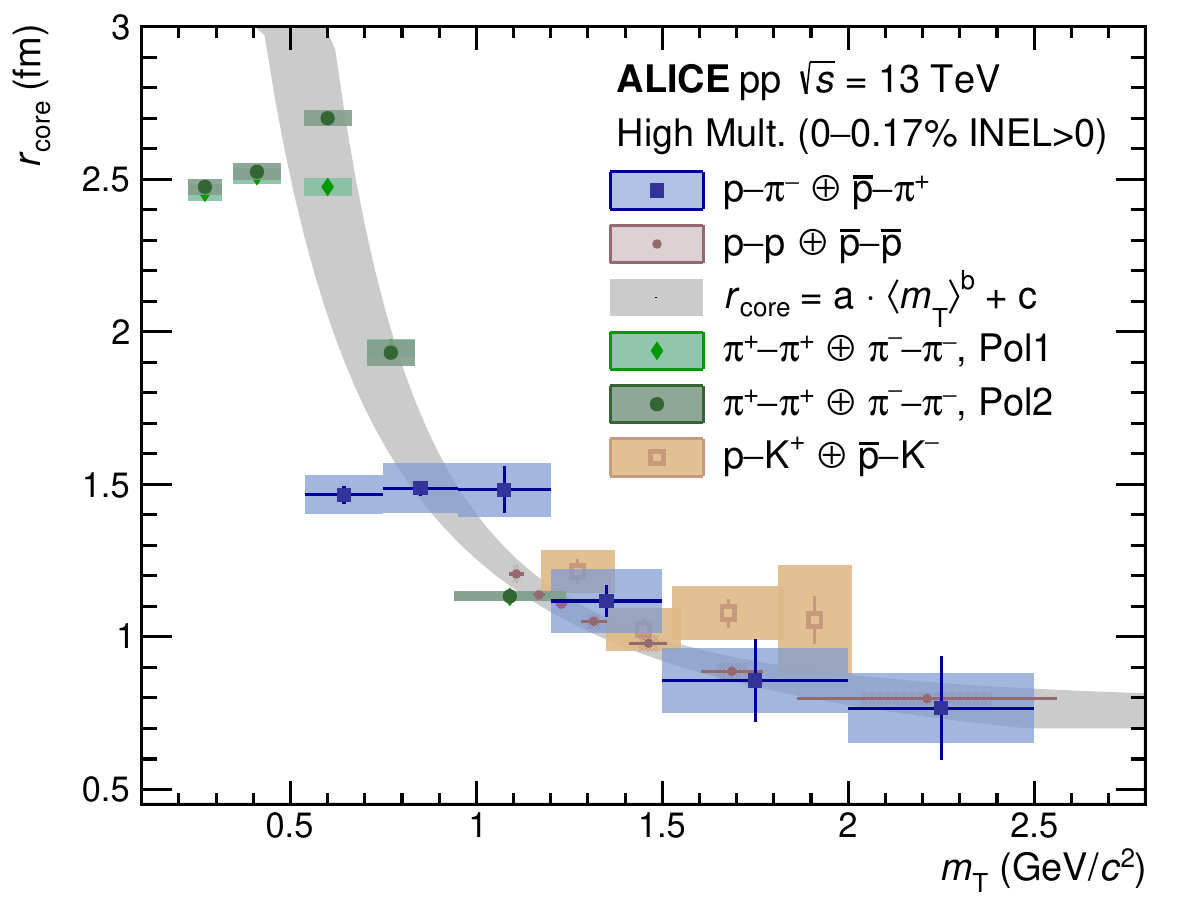}
\caption{Extracted core source size $r_\text{core}$ for primordial $\ppiMinus$ pairs for several \mtText{} ranges. The obtained result is compared with the results from p--p pairs~\cite{ppSourceErratum}, $\uppi^\pm$--$\uppi^\pm$ and $\text{p--}\text{K}^+$ pairs~\cite{ALICE:2023sjd}. The statistical (systematic) uncertainties are shown as lines (boxes). The extrapolated scaling of p--p pairs~\cite{ppSourceErratum} is shown as the gray band.}
\label{Fig. ProtonAntiPion Source Size}
\end{figure}

Figure~\ref{Fig. CF ProtonAntiPion} presents the experimental correlation functions of $\ppiMinus$ pairs obtained in HM pp collisions at $\sqrt{s} = 13$~TeV as well as the fit results using Eq.~\eqref{Eq. ProtonAntiPion Exp Fit}. The correlation functions at low $k^*$ are, in general, above unity due to the attractive strong and Coulomb interactions between the proton and the $\uppi^-$. The compatibility between the overall fit (red band, upper panel) and the experimental data is evaluated in terms of number of standard deviations $n_\sigma$ as a function of $k^*$. The resulting point-by-point $n_\sigma$ is shown in the bottom panel of the respective figure and is generally $\text{n}_\sigma \lesssim 2$. For the lowest \mtText{} interval, a deviation can be seen at $k^* \approx 0.4 \text{ GeV}/c$, as the contribution of the higher mass resonances is not pronounced in the data and, thus, not well constrained in the fit. However, this deviation does not influence the results as it is located far away from the femtoscopic relevant region for this \mtText{} interval.
As in the case of $\ppiPlus$ pairs, a correlated background due to mini-jets is visible. This contribution gets larger with increasing \mtText, which again is related to the contribution of particles with larger \ptText{} at large transverse masses, and hence a larger probability to be related to hard partonic scatterings. However, this background generally is more pronounced in the $\ppiMinus$ system, also at low \mtText. This can be explained as follows. The correlated background originates from a mini-jet background, which is especially dominated by gluon splitting into a quark--antiquark pair. The correlated quark and antiquark then constitute one of the valence quarks in the proton and pion, respectively, which are, in turn, then correlated to each other. In the case of the proton and $\uppi^+$, this correlation can only happen due to correlated d$\overline{\text{d}}$ pairs. For the  proton and $\uppi^-$, this correlation correspondingly stems from u$\overline{\text{u}}$ pairs. As the proton consists of two u valence quarks and only one d valence quark, the correlation from u$\overline{\text{u}}$ is thus more probable. This, in turn, leads to the more pronounced mini-jet background for the $\ppiMinus$ pairs. A clear peak centered at $k^*\approx 0.1 \text{ GeV}/c$ is present for all \mtText{} ranges. This is related to the decay of the $\Lambda$ into a $\ppiMinus$ pair. 
The position of the peak is consistent between all \mtText{} ranges and in agreement with the expected position from the two-body decay kinematics of the $\Lambda$ with mass of 1115~MeV$/c^2$. The contribution of the short-lived $\Delta^0(1232)$ is located within the $k^*$ range of $(0.05-0.3)$~GeV$/c$. As in the case of the $\Delta^{++}(1232)$, distortions from a typical Breit--Wigner-like shape are visible, especially for low pair transverse masses. As in the previous case, this can be related to hadronic rescattering and regeneration phases during the short lifetime of the $\Delta^0(1232)$. This is reflected in the extracted $\Delta^0$ spectral temperature presented in Fig.~\ref{Fig. Delta0 Temp}, which is found to be constant at around $20$~MeV. The resulting width extracted from the fitting procedure is reported in Fig.~\ref{Fig. Delta0 Width}. It shows larger fluctuations than in the case of the $\Delta^{++}(1232)$, which can be related to the larger mini-jet background affecting  the fit results. However, the results are compatible with a width of $90$~MeV extracted in Ref.~\cite{GiacosaSill}.
The resulting scaling of the source size of primordial $\ppiMinus$ pairs as a function of the pair transverse mass is shown in Fig.~\ref{Fig. ProtonAntiPion Source Size}. The results are similar to the previously obtained $r_\text{core}$ in HM pp collisions at $\sqrt{s}=13$~TeV studied with p--p pairs~\cite{ppSourceErratum} in addition to $\uppi^\pm$--$\uppi^\pm$ and $\text{p--}\text{K}^+$ pairs~\cite{ALICE:2023sjd} by ALICE, as well as with the $\ppiPlus$ results shown in Fig.~\ref{Fig. ProtonPion Source Size}. They are also consistent with the source size extracted with p--$\Lambda$ pairs by ALICE~\cite{ppSourceErratum} which are not explicitly shown here as they fully overlap with the p--p points. 
The agreement between the obtained core radii for $\ppiMinus$ pairs and the results from the p--p pairs has been evaluated in terms of number of standard deviations, combining the statistical and systematic uncertainties of the data and the uncertainty of the p--p \mtText{} scaling (gray band). The deviation amounts to $n_\sigma = 2.03$ in the first \mtText{} interval and to $n_\sigma = 1.29$ in the whole range 0.54 $<$ \mtText{} $<$ 2.5 $\text{ GeV}/c^2$. Similar to the case of the $\ppiPlus$ case, this presents further observation of a common hadron-pair source albeit the origin of such a behavior for non-identical pairs is not trivial and needs to be addressed by dedicated studies.
\clearpage

%%%%%%%%%%%%%%%%%%%%%%%%%%%%%%%%%%%%%%%
\section{Analysis of \texorpdfstring{$\mathbf{\pppiPlusMinus}$}{proton--proton--pion} three-particle system} \label{Sec. Three-Body System}
\subsection{Three-particle correlation function}
\label{Sec. Three-particle correlation function}
The systems of $\pppiPlusMinus$ are investigated using three-particle femtoscopy to determine the presence of effects beyond pairwise interactions. The three-particle correlation function is given as 
  \begin{equation}
     C(Q_3) = \mathcal{N} \frac{N_{\text{same}}(Q_3)}{N_{\text{mixed}}(Q_3)}  \,,\label{Eq. CF Three-body triplet}
 \end{equation}
where $Q_3$ is the Lorentz-invariant hyper-momentum~\cite{ALICE:3pi}
\begin{equation}
    Q_3 = \sqrt{-q_{12}^2-q_{23}^2-q_{31}^2}\,. \label{Eq. Q3 Definition}
\end{equation}
The variable $q_{\text{ij}}$~\cite{Lisa:2005dd} is defined as 
\begin{equation}
    q_{\text{ij}}^\mu = (p_\text{i} - p_\text{j})^\mu - \frac{(p_\text{i} - p_\text{j})\cdot P_{\text{ij}}}{P_{\text{ij}}^2}P_{\text{ij}}^\mu, \, \, \, P_{\text{ij}} \equiv p_\text{i} + p_\text{j}
\end{equation}
and represents the relative four-momentum between particles i and j. Hence, $Q_3$ combines all pair-relative momenta within the triplet of particles. The triplet same-event distribution $N_{\text{same}}(Q_3)$ in Eq.~\eqref{Eq. CF Three-body triplet} is obtained by calculating $Q_3$ with all three particles taken from the same collision event. To obtain $N_{\text{mixed}}(Q_3)$, a mixed-event technique is employed, where $Q_3$ is calculated with each of the three particles sampled from another event. As such, no FSI can affect $N_{\text{mixed}}(Q_3)$. The mixing conditions are the same as in the case of the two-particle femtoscopy of $\ppiPlusMinus$ pairs described above. All correlation functions related to the study of the $\pppiPlusMinus$ systems are normalized to unity in the $Q_3$ range of 1.0--1.2~GeV$/c$.

The interpretation of Eq.~\eqref{Eq. CF Three-body triplet} concerning three-body effects cannot be performed directly as the observed signal can be either due to three-body effects or underlying (lower-order) two-body correlations between the pairs in the triplet. To isolate the three-body effects, the cumulant decomposition is employed~\cite{Kubo}, previously utilized in the analysis of p--p--p and p--p--$\Lambda$ systems~\cite{ALICE:ppp-ppL} as well as p--p--$\text{K}^\pm$~\cite{ALICE:ppK}. The three-particle cumulant incorporates information on genuine effects beyond pairwise correlations in the analyzed triplets.
These lower-order correlations are subtracted from the three-body correlation (Eq.~\eqref{Eq. CF Three-body triplet}) to obtain the isolated three-body effects. Concerning three particles X, Y, and Z, the lower-order contributions will be denoted as (X--Y)--Z, where particles X and Y can undergo pairwise interactions, while particle Z is uncorrelated. The three-particle femtoscopic cumulant $c_3(Q_3)$ can then be written as
\begin{equation}
    c_3(Q_3) = C_\text{(X--Y--Z)}(Q_3)  - C_\text{(X--Y)--Z}(Q_3) - C_\text{(X--Z)--Y}(Q_3) - C_\text{(Y--Z)--X}(Q_3) + 2 \,. \label{Eq. Cumulant CF}
\end{equation}
The term $C_{\text{(X-Y-Z)}}(Q_3)$ represents the correlation function where all three particles can undergo two-body and three-body FSI and hence corresponds to Eq.~\eqref{Eq. CF Three-body triplet}. The lower-order contributions of type (X-Y)-Z can be obtained in a data driven-way as 
\begin{equation}
    C_\text{(X--Y)--Z}(Q_3) = \mathcal{N} \frac{N_\text{(X--Y)--Z}(Q_3)}{N_{\text{mixed}}(Q_3)} \,. \label{Eq. Lower-order CF}
\end{equation}
The distribution $N_\text{(X--Y)--Z}(Q_3)$ is obtained by calculating $Q_3$ with particles X and Y from the same event, while particle Z is taken from a another, uncorrelated event. Hence, FSI can only occur between particles X and Y. The term $N_{\text{mixed}}(Q_3)$ is the mixed-event triplet distribution described above, where all three particles are taken from different events. The CPR between protons and pions as well as two protons (see Sec.~\ref{s:DataAnalysis}) is applied to all distributions ($N_{\text{same}}(Q_3)$, $N_{\text{mixed}}(Q_3)$, and any distribution of type $N_\text{(X--Y)--Z}(Q_3)$) to avoid any bias.

Applying this formalism to the $\pppiPlusMinus$ system, 
the cumulant Eq.~\eqref{Eq. Cumulant CF} is written as
\begin{equation}
c_3(Q_3) = C_{(\pppiPlusMinus)}(Q_3) - C_{(\text{p--p})\text{--}\uppi^\pm}(Q_3) - 2\ C_{(\text{p--}\uppi^\pm)\text{--}\text{p}}(Q_3) + 2 \,. \label{Eq. Cumulant nice}
\end{equation}
The lower-order contribution $C_{(\text{p--}\uppi^\pm)\text{--}\text{p}}(Q_3)$ has to be accounted for twice, as the pion can be correlated to either of the two protons. Hence, the cumulant can be expressed with the total lower-order contribution $C_\text{Lower-Order}(Q_3)$  
\begin{equation}
C_\text{Lower-Order}(Q_3) = C_{(\text{p--p})\text{--}\uppi^\pm}(Q_3) + 2\ C_{(\text{p--}\uppi^\pm)\text{--}\text{p}}(Q_3) - 2 \label{Eq. Three-Body Lower-order}
\end{equation}
as $c_3(Q_3) = C_{(\pppiPlusMinus)}(Q_3) - C_\text{Lower-Order}(Q_3)$. The interpretation of cumulants differs from the interpretation of correlation functions.  A signal is considered a three-body effect if the underlying lower-order two-body correlations cannot describe the three-particle correlation function defined in Eq.~\eqref{Eq. CF Three-body triplet}. Hence, an attractive three-body effect results in a cumulant above zero, while repulsive three-body effects lead to a depletion of the cumulant to negative values. In particular, a cumulant compatible with zero indicates an absence of any three-body effects, as the signal can be fully described by the pairwise correlations.

%%%%%%%%%

\subsection{Results} \label{Sec. Three-Body results}
\begin{figure}[h]
  \centering
\subfigure[]{
    \includegraphics[trim={0.5cm 0.45cm 1.5cm 1.1cm},clip,width=0.48\textwidth]{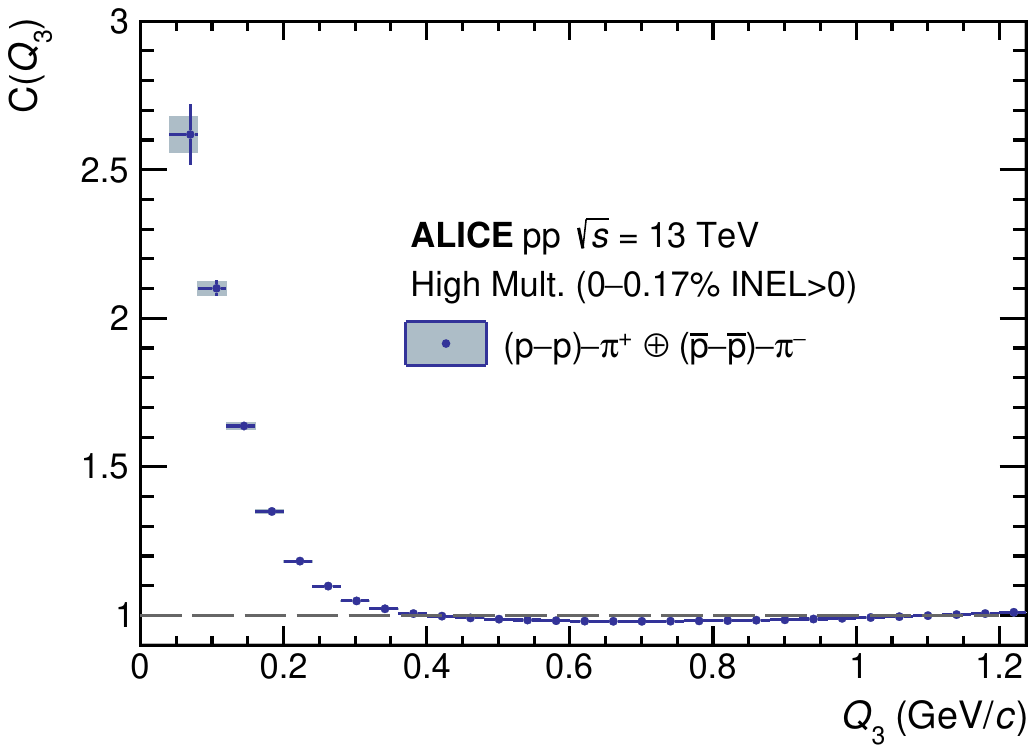}
    \label{Fig. CF PP Same Pion Mixed}
  }
   \subfigure[]{
    \includegraphics[trim={0.5cm 0.45cm 1.5cm 1.1cm},clip,width=0.48\textwidth]{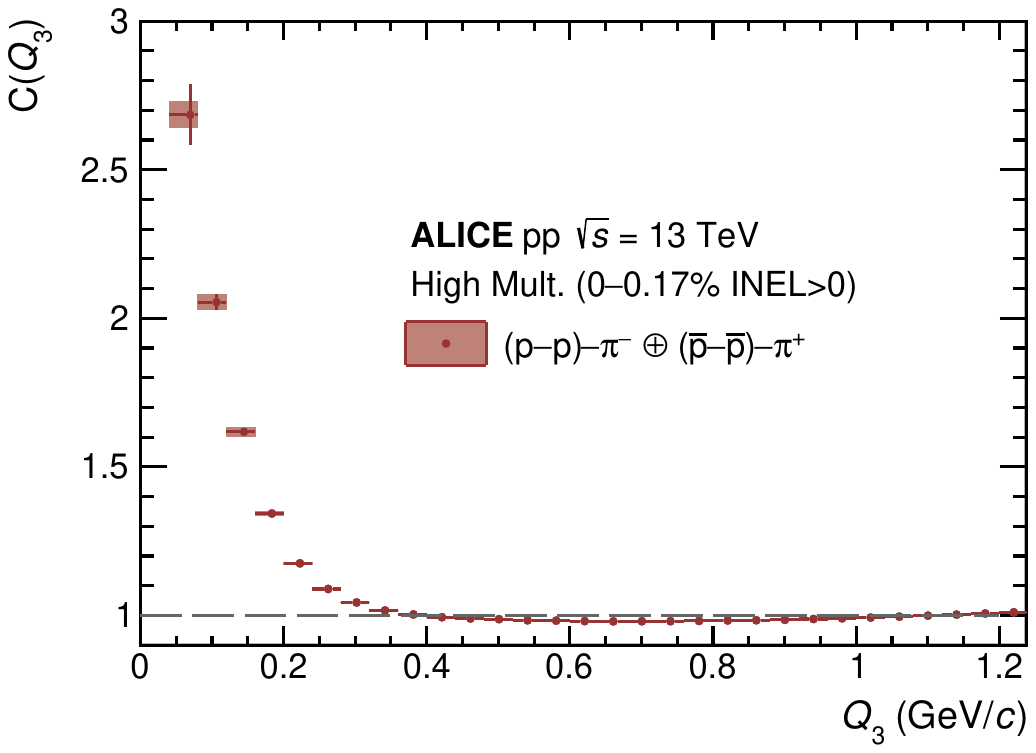}
    \label{Fig. CF PP Same APion Mixed}
  }
  \subfigure[]{
    \includegraphics[trim={0.5cm 0.45cm 1.5cm 1.1cm},clip,width=0.48\textwidth]{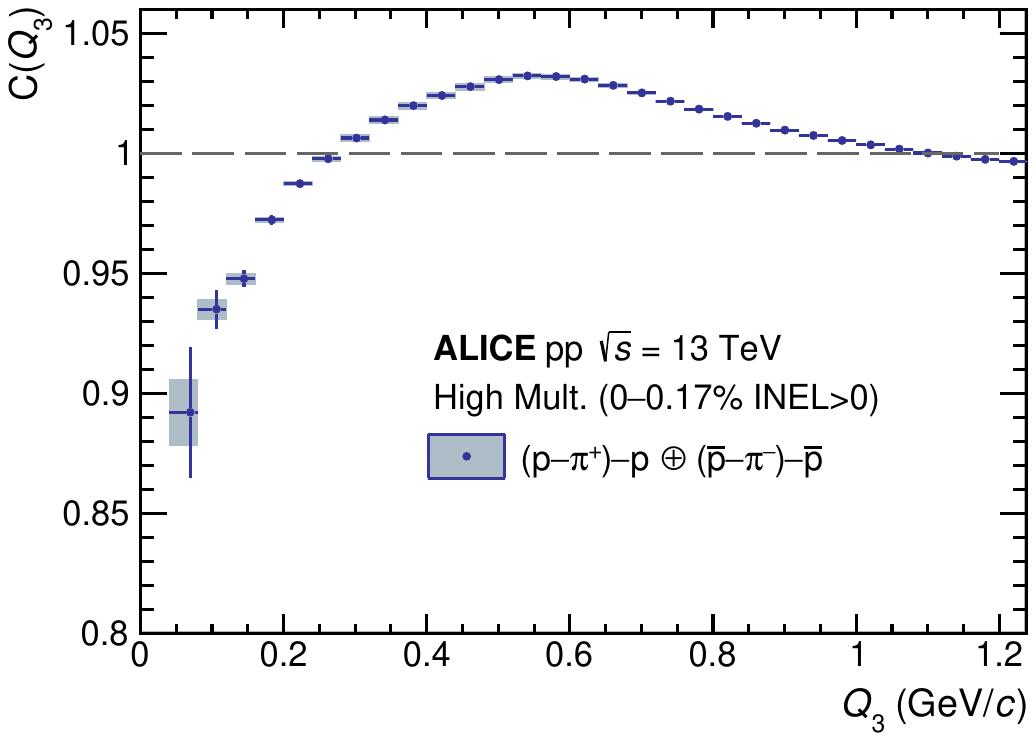}
    \label{Fig. CF PPion Same P Mixed}
  }
  \subfigure[]{
    \includegraphics[trim={0.5cm 0.45cm 1.5cm 1.1cm},clip,width=0.48\textwidth]{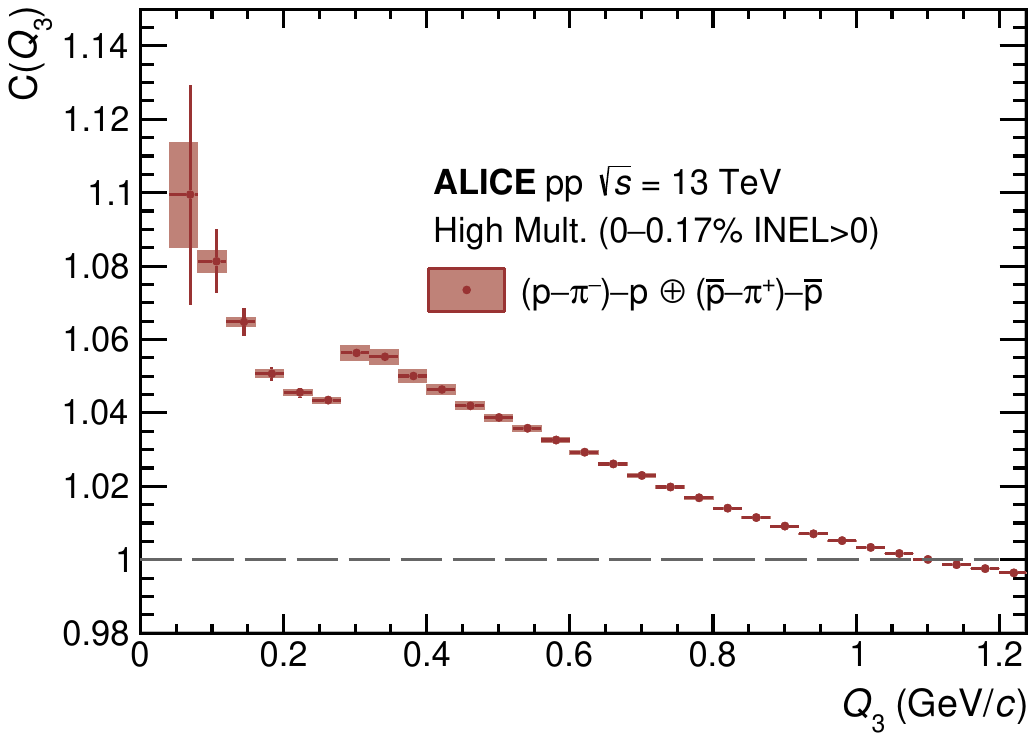}
    \label{Fig. CF PAPion Same P Mixed}
  }
  \caption{Lower-order contributions present in the $\pppiPlusMinus$ systems.  The contributions of (a) $(\text{p--p})\text{--}\uppi^+$ and (b) $(\text{p--p})\text{--}\uppi^-$, as well as (c) $(\text{p--}\uppi^+)\text{--}\text{p}$ and (d) $(\text{p--}\uppi^-)\text{--}\text{p}$ are shown. The statistical (systematic) uncertainties are represented by the lines (boxes). The horizontal lines show the overall bin width, while the respective position of the marker presents the average $Q_3$ within this bin obtained from the mixed event distribution $N_{\text{mixed}}(Q_3)$. }
  \label{Fig. PPPion PPAPion Lower-Order}
\end{figure}

The experimentally obtained lower-order contributions to the $\pppiPlus$ and $\pppiMinus$ systems (defined in Eq.~\eqref{Eq. Lower-order CF}) are presented in Fig.~\ref{Fig. PPPion PPAPion Lower-Order}. Figures~\ref{Fig. CF PP Same Pion Mixed} and ~\ref{Fig. CF PP Same APion Mixed} show the contributions of $(\text{p--p})\text{--}\uppi^+$ and $(\text{p--p})\text{--}\uppi^-$, respectively. Both correlation functions reflect the attractive strong interaction between the two protons (see Ref.~\cite{ALICE:2018ysd} for details) and are in agreement with each other. This is expected, as the effects of the uncorrelated pion are independent of the charge and only depend on the mass. The signal itself is driven by the FSI between the two protons, which is smeared out in the $Q_3$ range due to the mixing with the uncorrelated pion. The correlation function for $(\text{p--}\uppi^+)\text{--}\text{p}$ is presented in Fig.~\ref{Fig. CF PPion Same P Mixed}, while the case of $(\text{p--}\uppi^-)\text{--}\text{p}$ is shown in Fig.~\ref{Fig. CF PAPion Same P Mixed}. In both cases, the presented correlation functions show the same characteristics as the respective two-particle correlation functions of $\ppiPlusMinus$ that were discussed in Sec.~\ref{Sec. Results Two-Body}. In particular, the contribution from the decay of the $\Delta^{++}$ is visible as a bump centered at $Q_3 \approx 0.6~\text{ GeV}/c$ for $(\text{p--}\uppi^+)\text{--}\text{p}$, while the $\Lambda$ is visible as a bump structure for $(\text{p--}\uppi^-)\text{--}\text{p}$ at $Q_3 \approx 0.35~\text{ GeV}/c$. The $\Delta^{++}$ and $\Lambda$ appear broader than in their respective two-particle analyses as the contribution of the uncorrelated proton smears the correlation in $Q_3$. 

\begin{figure}[!h]
  \centering
\subfigure[]{
    \includegraphics[trim={0.7cm 0.42cm 1.5cm 1.1cm},clip,width=0.48\textwidth]{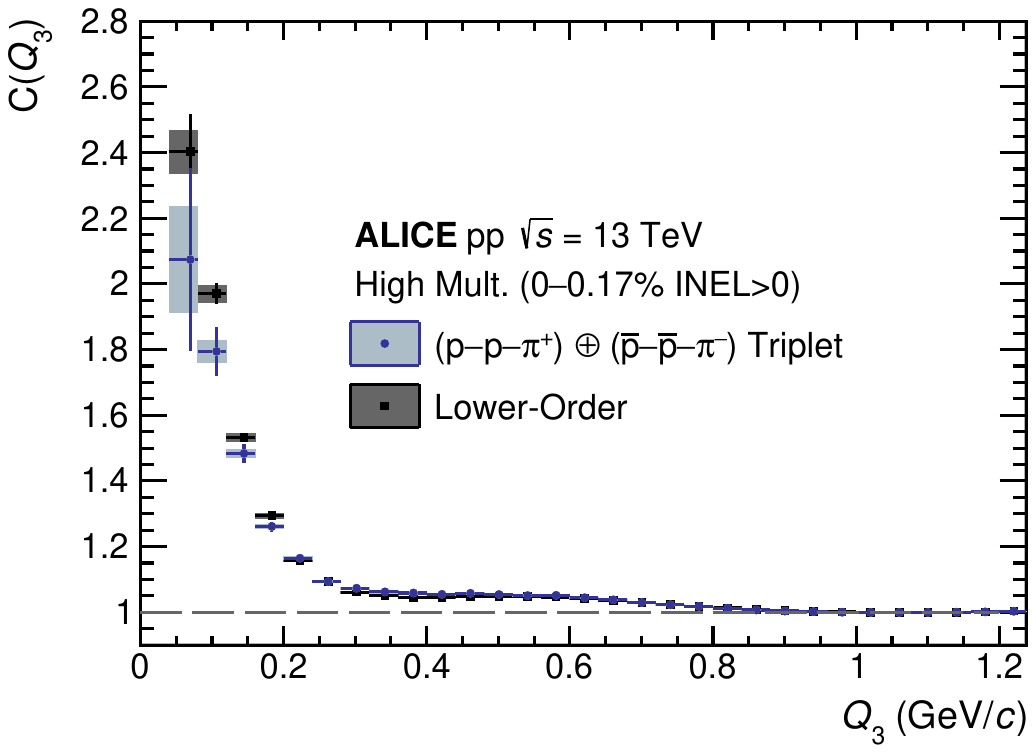}
    \label{Fig. PPPion Lower-Order Triplet Comparison}
  }
   \subfigure[]{
      \includegraphics[trim={0.7cm 0.42cm 1.5cm 1.1cm},clip,width=0.48\textwidth]{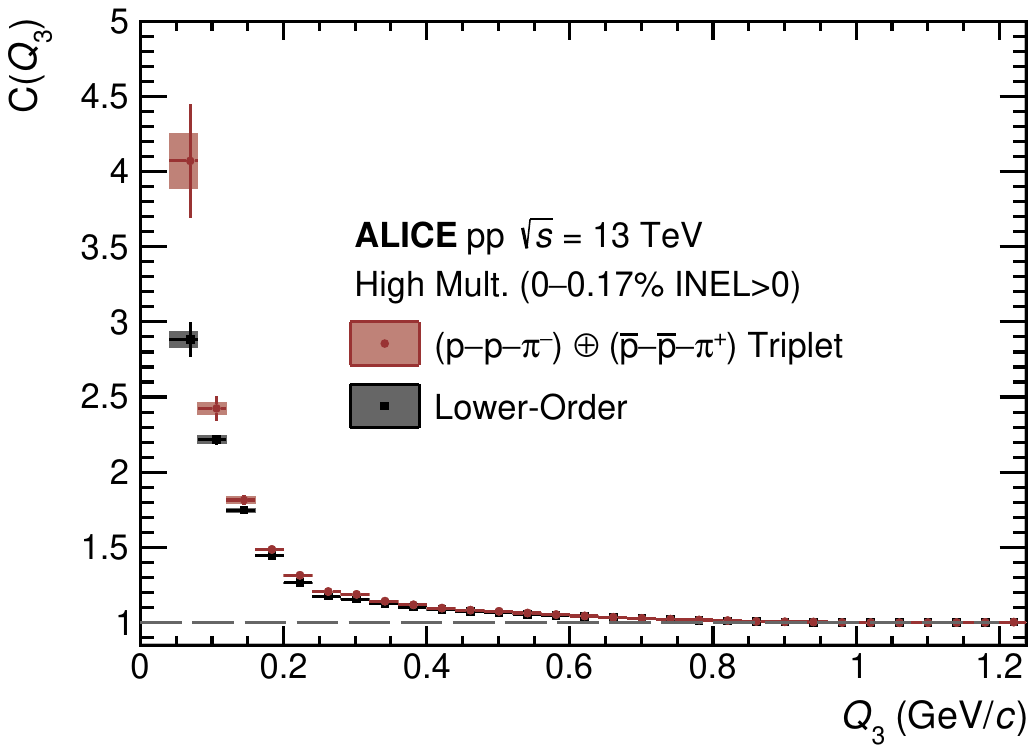}
    \label{Fig. PPAPion Lower-Order Triplet Comparison}
  }

  \caption{Comparison of the three-particle correlation functions of (a) $\pppiPlus$ and (b) $\pppiMinus$ to their respective total lower-order contributions (grey). The statistical and systematic uncertainties are shown by the lines and boxes, respectively. The lower-order contribution is given by Eq.~\eqref{Eq. Three-Body Lower-order}. The horizontal lines represent the overall bin width, while the respective position of the marker show the average $Q_3$ within this bin. The latter is obtained from the mixed event distribution $N_{\text{mixed}}(Q_3)$.}
  \label{Fig. PPPion PPAPion Lower-Order Comp}
\end{figure}

The three-body correlation functions and the comparison to their respective lower-order contributions are presented in Fig.~\ref{Fig. PPPion Lower-Order Triplet Comparison} for the $\pppiPlus$ and in Fig.~\ref{Fig. PPAPion Lower-Order Triplet Comparison} for the $\pppiMinus$ system. In both cases, the three-body correlation function is above unity, indicating overall attractive interactions. However, the $\pppiPlus$ and $\pppiMinus$ systems differ when the effect of the corresponding lower-order correlations is analyzed in detail. In the case of $\pppiPlus$, the lower-order contribution shows a larger signal for $Q_3 < 0.2~\text{ GeV}/c$ than the three-body system. This indicates that the two-body correlations are too attractive to fully describe the three-body system, which correspondingly requires a repulsive three-body interaction in the $\pppiPlus$ system. The opposite effect is visible for $\pppiMinus$, where the lower-order contributions show a weaker correlation signal than the three-body system, indicating an attractive three-body interaction in this system. 

\begin{figure}[h]
\centering
\includegraphics[trim={0cm 0cm 0cm 0cm},clip, width=0.8\textwidth]{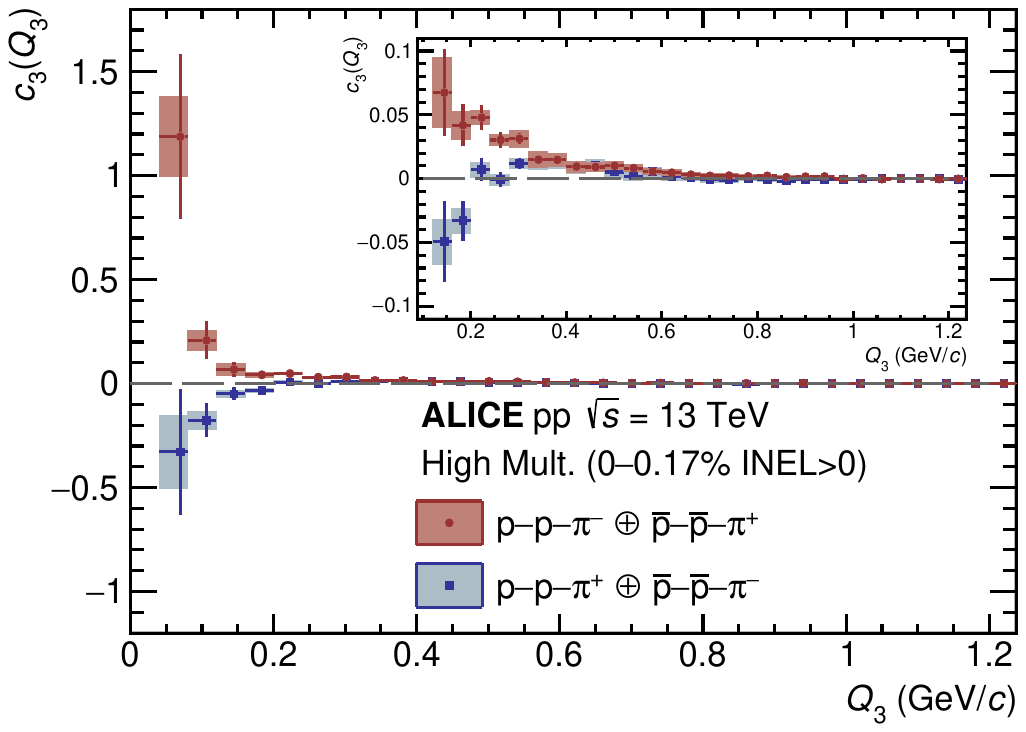}
\caption{Extracted cumulants of $\pppiPlus$ (blue) and $\pppiMinus$ (red) as a function of $Q_3$ after subtracting the lower-order contributions from the triplet correlations. The lines (boxes) represent the statistical (systematic) uncertainties.}
\label{Fig. PPPion PPAPion Cumulants}
\end{figure}

The difference between lower-order and three-body contributions becomes more evident with the cumulant $c_3(Q_3)$, defined in Eq.~\eqref{Eq. Cumulant CF}. The cumulants for $\pppiPlus$ and $\pppiMinus$ are shown in Fig.~\ref{Fig. PPPion PPAPion Cumulants}. In both cases, the cumulant is compatible with zero within uncertainties for $Q_3$ above 0.7~GeV$/c$. In this regime, no three-body effects are present, and the pairwise correlations fully describe the system. Both cumulants are positive in the intermediate $Q_3$ range between  0.2~GeV$/c$ and  0.7~GeV$/c$, indicating attractive three-body effects. The significance of the deviation from zero is estimated by evaluating the p-value of the $\chi^2$ distribution with subsequent conversion into a number $n_\sigma$ of Gaussian standard deviations. In the range $0.2 < Q_3 < 0.7~\text{ GeV}/c $ this significance is $n_\sigma = 2.59$ for the $\pppiPlus$ and $n_\sigma = 8.03$ for the $\pppiMinus$. The significance for the $\pppiPlus$ is lower as the cumulant starts to tend towards zero at $Q_3 \approx 0.3~\text{ GeV}/c$, while the cumulant of $\pppiMinus$ keeps increasing for $Q_3 < 0.3~\text{ GeV}/c$. At $Q_3 \approx 0.2~\text{ GeV}/c$, the cumulant of $\pppiPlus$ exhibits a cross-over to negative values. The two systems then show an opposite behavior for $0.04 < Q_3 < 0.2~\text{ GeV}/c$. While $\pppiMinus$ has a positive cumulant in this regime with a significance of $n_\sigma = 3.24$, the cumulant of the $\pppiPlus$ system is negative ($n_\sigma = 1.99$). Hence, in this low $Q_3$ regime, the cumulants indicate attractive three-body effects for the system involving $\uppi^-$ and repulsive three-body effects for $\pppiPlus$. However, due to its comparably lower significance, the true repulsive nature of the $\pppiPlus$ has to be addressed in future studies with increased amount of data such as the LHC Run 3. The deviation from zero in overall range $0.04 < Q_3 < 0.7~\text{ GeV}/c$ is $n_\sigma = 2.98$ for the $\pppiPlus$ and $n_\sigma = 8.48$ for the $\pppiMinus$. Therefore, the analysis indicates effects beyond pairwise interactions in the $\pppiPlusMinus$ systems. In order to fully interpret the femtoscopic measurements of the $\pppiPlusMinus$ systems, full-fledged three-body calculations of the correlation function are needed. Such techniques have recently been developed for three baryons~\cite{KievskyThreeBody} and will be extended to treat also systems with mesons in the near future.

\section{Summary}
\label{Sec. Summary}
%%%%%%%%%%%%%%%%%%%%%%%%%%%%%%%%%%

\label{s:Conclusion}
This study presented access to the $\ppiPlusMinus$ dynamics in high-multiplicity pp collisions at $\sqrt{s} = 13~$~TeV via measurements of two-particle momentum correlations with the ALICE detector. The size of the particle-emitting source of primordial $\ppiPlusMinus$ pairs has been extracted for several intervals of pair transverse masses \mtText{}  using the Resonance Source Model~\cite{ppSourceErratum}. The results are in agreement with the source sizes extracted with femtoscopic measurements of p--p~\cite{ppSourceErratum}, $\uppi^\pm$--$\uppi^\pm$, and p--$\text{K}^+$ pairs~\cite{ALICE:2023sjd} by ALICE in high-multiplicity pp collisions. This further strengthens the picture of a common emission source of hadron-pairs in small collision systems at LHC energies. The \mtText-dependent scaling is reminiscent of collective effects as seen in the collision of heavy ions~\cite{Lisa:2005dd, Chojnacki:2011hb, Kisiel:2014upa, Shapoval:2014wya}. While signs of collectivity in pp collisions at LHC energies have been reported by various experiments and observables~\cite{CMS:2016fnw, CMS:2019fur, ALICE:2024vzv}, the origin of this behavior in small collision systems has to be addressed by future studies. Further, the observed agreement of the \mtText-dependent source size between different particle pairs, including non-identical particle pairs, has to be studied in the future. Results on the $\Delta^{++}(1232)$ and $\Delta^{0}(1232)$ width and spectral temperature~\cite{ Reichert:2019lny, Reichert:2022uha} have been extracted. For both resonances, the spectral shape for low \mtText{} shows clear deviations from a typical Breit--Wigner shape, indicating a strong influence of hadronic rescattering and regeneration processes of the short-lived resonances. These results provide valuable input to the description of the resonance modification in small collision systems for transport models such as UrQMD~\cite{Bass:1998ca, Bleicher:1999xi} or SMASH~\cite{Petersen:2018jag}. 

Further investigation on the pion dynamics with nucleons has been performed by extending the femtoscopy analysis to $\pppiPlusMinus$ three-particle systems in the same data set. The cumulant decomposition has been used to isolate three-body effects by explicitly removing the pairwise correlations in the system. These lower-order pairwise contributions incorporate all the features of the p--p and $\ppiPlusMinus$ correlations in the triplets, including the effects of the $\Delta$ resonances in the correlation function. The extracted cumulants deviate from zero for hyper-momenta $Q_3 < 0.7$~GeV$/c$, showing a clear indication of the presence of three-body effects in the triplets. Two different regions, corresponding to a change of sign of the cumulants, are discussed. In the intermediate range of $0.2\text{ GeV}/c < Q_3 < 0.7\text{ GeV}/c$, the cumulants of both $\pppiPlus$ and $\pppiMinus$ are positive, which corresponds to attractive three-body effects. The significance of this deviation from zero is $n_\sigma = 2.59$ for the $\pppiPlus$ and $n_\sigma = 8.03$ for the $\pppiMinus$ systems. In the low $Q_3$ region of $0.04\text{ GeV}/c < Q_3 < 0.2\text{ GeV}/c$, the signals for both systems are opposite: the $\pppiPlus$ shows signs of repulsive three-body effects (negative cumulant) with a significance of $n_\sigma = 1.99$ while the $\pppiMinus$ cumulant stays positive ($n_\sigma = 3.24$). The full interpretation of these results calls for a proper theoretical treatment of these three-body systems. The results give the possibility to access potential three-body interaction in the $\pppiPlusMinus$ sytem and hence the coupling of the pion to multiple nucleons. Recent studies show that this coupling is an important input for the studies of the QCD axion at finite baryonic densities and its impact on the equation of state of dense stellar objects such as neutron stars~\cite{Balkin:2020dsr}.

%%%%%%%%%%%%%%%%%%%%%%%%%%%%%%%%
% end main text 
%%%%%%%%%%%%%%%%%%%%%%%%%%%%%%%%

%%%%% acknowledgements - handled by EB chairs 
%\clearpage
\newenvironment{acknowledgement}{\relax}{\relax}
\begin{acknowledgement}

\section*{Acknowledgements}
% add specific acknowledgements here 
% ...but please don't remove the line below: funding agencies
% will be acknowledged with a custom tex file handled by EB chairs after Collab Round 2
% Version: 2024-11-19

The ALICE Collaboration would like to thank all its engineers and technicians for their invaluable contributions to the construction of the experiment and the CERN accelerator teams for the outstanding performance of the LHC complex.
The ALICE Collaboration gratefully acknowledges the resources and support provided by all Grid centres and the Worldwide LHC Computing Grid (WLCG) collaboration.
The ALICE Collaboration acknowledges the following funding agencies for their support in building and running the ALICE detector:
A. I. Alikhanyan National Science Laboratory (Yerevan Physics Institute) Foundation (ANSL), State Committee of Science and World Federation of Scientists (WFS), Armenia;
Austrian Academy of Sciences, Austrian Science Fund (FWF): [M 2467-N36] and Nationalstiftung f\"{u}r Forschung, Technologie und Entwicklung, Austria;
Ministry of Communications and High Technologies, National Nuclear Research Center, Azerbaijan;
Conselho Nacional de Desenvolvimento Cient\'{\i}fico e Tecnol\'{o}gico (CNPq), Financiadora de Estudos e Projetos (Finep), Funda\c{c}\~{a}o de Amparo \`{a} Pesquisa do Estado de S\~{a}o Paulo (FAPESP) and Universidade Federal do Rio Grande do Sul (UFRGS), Brazil;
Bulgarian Ministry of Education and Science, within the National Roadmap for Research Infrastructures 2020-2027 (object CERN), Bulgaria;
Ministry of Education of China (MOEC) , Ministry of Science \& Technology of China (MSTC) and National Natural Science Foundation of China (NSFC), China;
Ministry of Science and Education and Croatian Science Foundation, Croatia;
Centro de Aplicaciones Tecnol\'{o}gicas y Desarrollo Nuclear (CEADEN), Cubaenerg\'{\i}a, Cuba;
Ministry of Education, Youth and Sports of the Czech Republic, Czech Republic;
The Danish Council for Independent Research | Natural Sciences, the VILLUM FONDEN and Danish National Research Foundation (DNRF), Denmark;
Helsinki Institute of Physics (HIP), Finland;
Commissariat \`{a} l'Energie Atomique (CEA) and Institut National de Physique Nucl\'{e}aire et de Physique des Particules (IN2P3) and Centre National de la Recherche Scientifique (CNRS), France;
Bundesministerium f\"{u}r Bildung und Forschung (BMBF) and GSI Helmholtzzentrum f\"{u}r Schwerionenforschung GmbH, Germany;
General Secretariat for Research and Technology, Ministry of Education, Research and Religions, Greece;
National Research, Development and Innovation Office, Hungary;
Department of Atomic Energy Government of India (DAE), Department of Science and Technology, Government of India (DST), University Grants Commission, Government of India (UGC) and Council of Scientific and Industrial Research (CSIR), India;
National Research and Innovation Agency - BRIN, Indonesia;
Istituto Nazionale di Fisica Nucleare (INFN), Italy;
Japanese Ministry of Education, Culture, Sports, Science and Technology (MEXT) and Japan Society for the Promotion of Science (JSPS) KAKENHI, Japan;
Consejo Nacional de Ciencia (CONACYT) y Tecnolog\'{i}a, through Fondo de Cooperaci\'{o}n Internacional en Ciencia y Tecnolog\'{i}a (FONCICYT) and Direcci\'{o}n General de Asuntos del Personal Academico (DGAPA), Mexico;
Nederlandse Organisatie voor Wetenschappelijk Onderzoek (NWO), Netherlands;
The Research Council of Norway, Norway;
Pontificia Universidad Cat\'{o}lica del Per\'{u}, Peru;
Ministry of Science and Higher Education, National Science Centre and WUT ID-UB, Poland;
Korea Institute of Science and Technology Information and National Research Foundation of Korea (NRF), Republic of Korea;
Ministry of Education and Scientific Research, Institute of Atomic Physics, Ministry of Research and Innovation and Institute of Atomic Physics and Universitatea Nationala de Stiinta si Tehnologie Politehnica Bucuresti, Romania;
Ministry of Education, Science, Research and Sport of the Slovak Republic, Slovakia;
National Research Foundation of South Africa, South Africa;
Swedish Research Council (VR) and Knut \& Alice Wallenberg Foundation (KAW), Sweden;
European Organization for Nuclear Research, Switzerland;
Suranaree University of Technology (SUT), National Science and Technology Development Agency (NSTDA) and National Science, Research and Innovation Fund (NSRF via PMU-B B05F650021), Thailand;
Turkish Energy, Nuclear and Mineral Research Agency (TENMAK), Turkey;
National Academy of  Sciences of Ukraine, Ukraine;
Science and Technology Facilities Council (STFC), United Kingdom;
National Science Foundation of the United States of America (NSF) and United States Department of Energy, Office of Nuclear Physics (DOE NP), United States of America.
In addition, individual groups or members have received support from:
Czech Science Foundation (grant no. 23-07499S), Czech Republic;
FORTE project, reg.\ no.\ CZ.02.01.01/00/22\_008/0004632, Czech Republic, co-funded by the European Union, Czech Republic;
European Research Council (grant no. 950692), European Union;
ICSC - Centro Nazionale di Ricerca in High Performance Computing, Big Data and Quantum Computing, European Union - NextGenerationEU;
Academy of Finland (Center of Excellence in Quark Matter) (grant nos. 346327, 346328), Finland;
Deutsche Forschungs Gemeinschaft (DFG, German Research Foundation) ``Neutrinos and Dark Matter in Astro- and Particle Physics'' (grant no. SFB 1258), Germany.

\end{acknowledgement}

%%%%%%%% Bibliography 
\bibliographystyle{utphys}   % Remember we use title in the biblio
\nocite{*}
\bibliography{bibliography.bib}
%\input {bibliography.bib}  

%%%%%%%%%%%%%%%%%%%%%%%%%%%%%%%%
% Appendices: yours (if any) + authorlist
%%%%%%%%%%%%%%%%%%%%%%%%%%%%%%%%
\newpage
\appendix

\section{Two-Particle Correlation Functions}
\label{App. Two-Body CFs}
\begin{figure}[!h]
  \centering
  \subfigure[]{
    \includegraphics[trim={0.8cm 0.55cm 1.5cm 1.2cm},clip,width=0.48\textwidth]{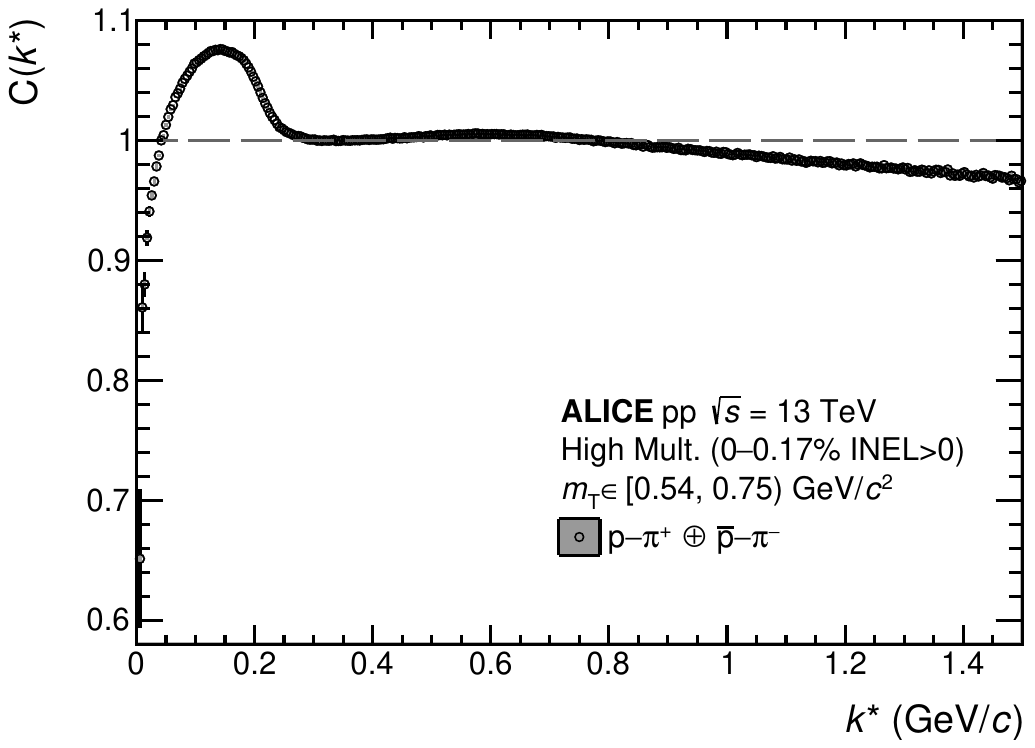}
    \label{Fig. App CF ProtonPion mT 1}
  }
  \subfigure[]{
    \includegraphics[trim={0.8cm 0.55cm 1.5cm 1.2cm},clip,width=0.48\textwidth]{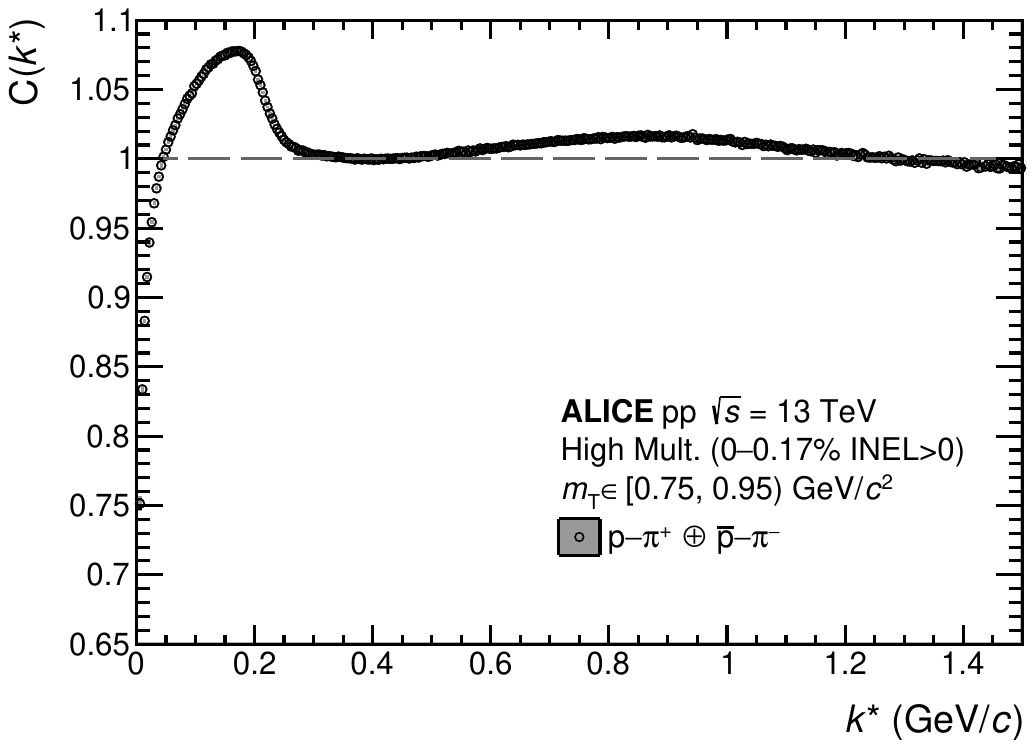}
    \label{Fig. App CF ProtonPion mT 2}
  }
    \subfigure[]{
    \includegraphics[trim={0.8cm 0.55cm 1.5cm 1.2cm},clip,width=0.48\textwidth]{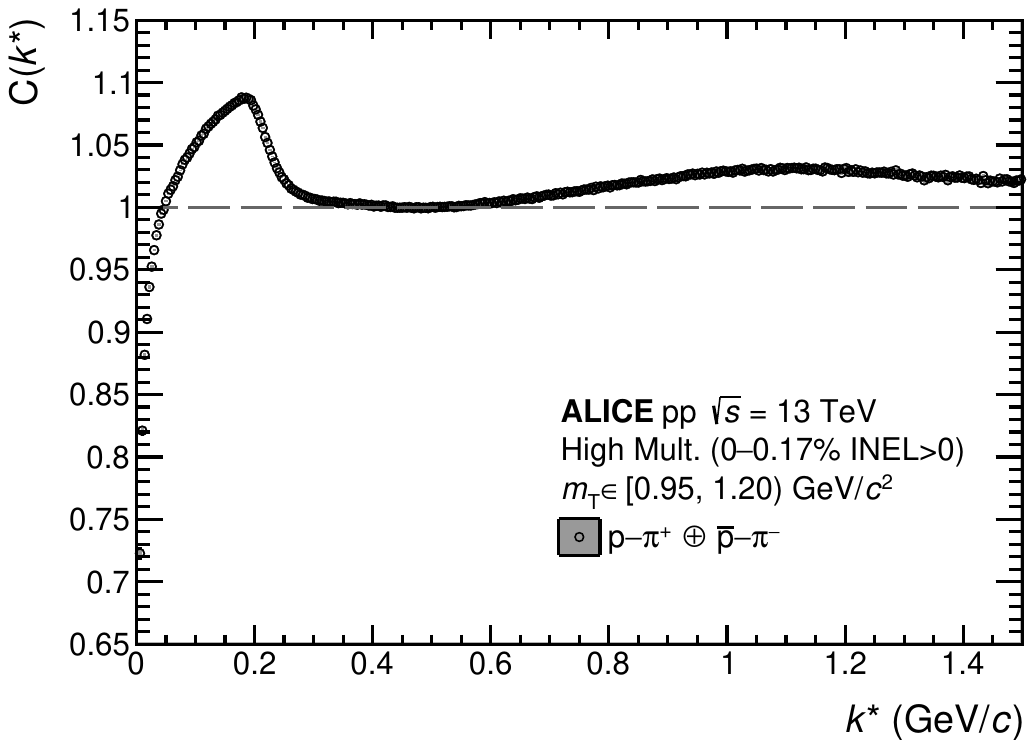}
    \label{Fig. App CF ProtonPion mT 3}
  }
  \subfigure[]{
    \includegraphics[trim={0.8cm 0.55cm 1.5cm 1.2cm},clip,width=0.48\textwidth]{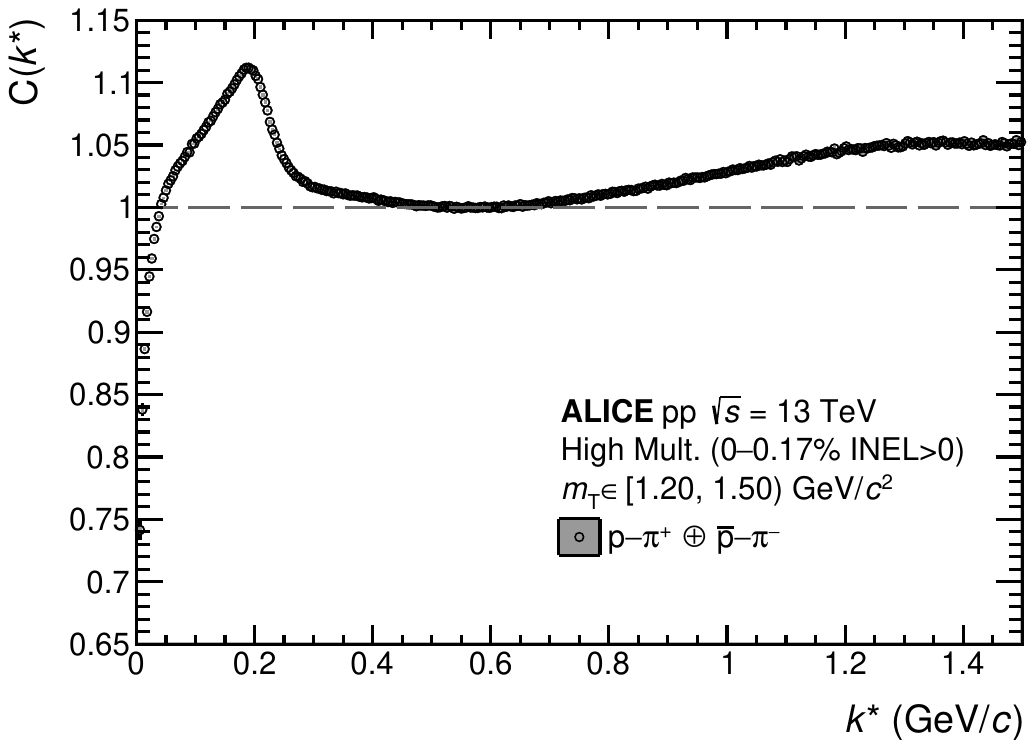}
    \label{Fig. App CF ProtonPion mT 4}
  }
    \subfigure[]{
    \includegraphics[trim={0.8cm 0.55cm 1.5cm 1.2cm},clip,width=0.48\textwidth]{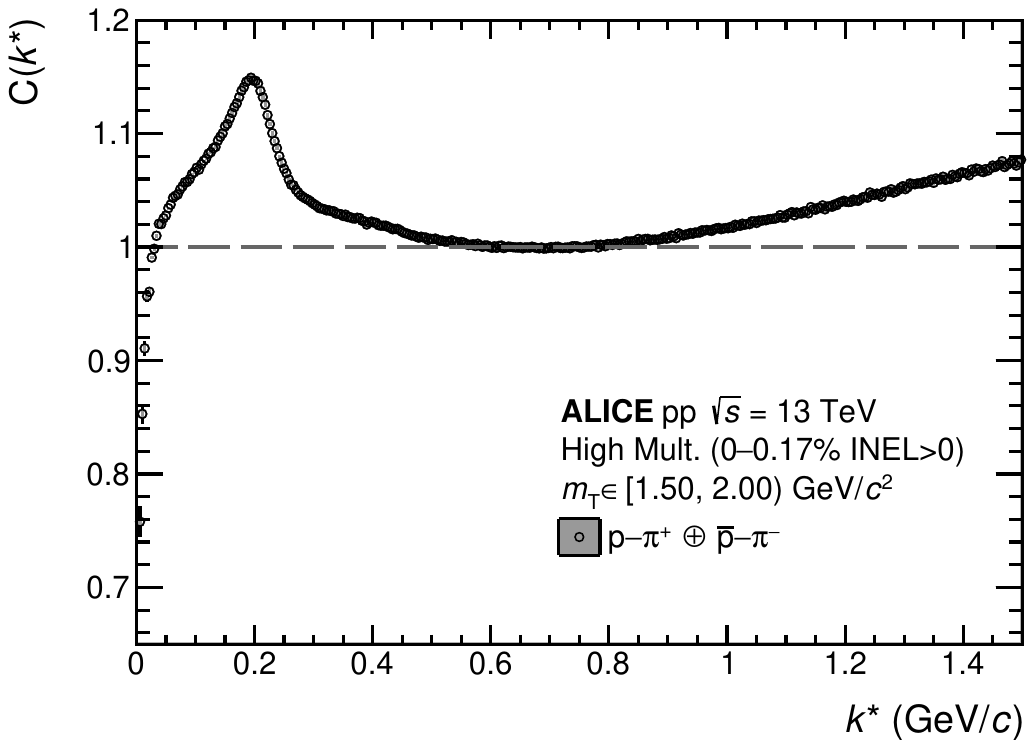}
    \label{Fig. App CF ProtonPion mT 5}
  }
  \subfigure[]{
    \includegraphics[trim={0.8cm 0.55cm 1.5cm 1.2cm},clip,width=0.48\textwidth]{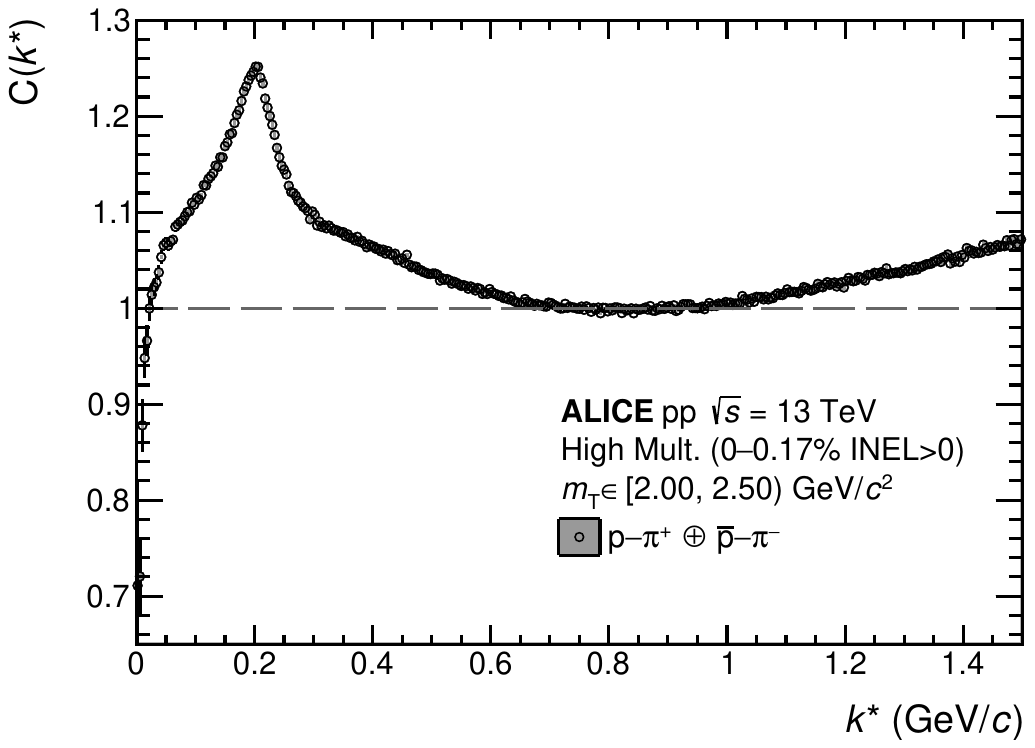}
    \label{Fig. App CF ProtonPion mT 6}
  }
  \caption{The experimental correlation function of $\ppiPlus$ pairs  as a function of the pair relative momentum $k^*$ in several intervals of the pair \mtText: (a) $[0.54, 0.75)\text{ GeV}/c^2$, (b) $[0.75, 0.95)\text{ GeV}/c^2$, (c) $[0.95, 1.2)\text{ GeV}/c^2$, (d) $[1.2, 1.5)\text{ GeV}/c^2$, (e) $[1.5, 2.0)\text{ GeV}/c^2$, and (f) $[2.0, 2.5)\text{ GeV}/c^2$. The lines and boxes show the statistical and systematic uncertainties of the experimental data, respectively.}
  \label{Fig. App CF ProtonPion}
\end{figure}

\begin{figure}[!h]
  \centering
  \subfigure[]{
    \includegraphics[trim={0.8cm 0.55cm 1.5cm 1.2cm},clip,width=0.48\textwidth]{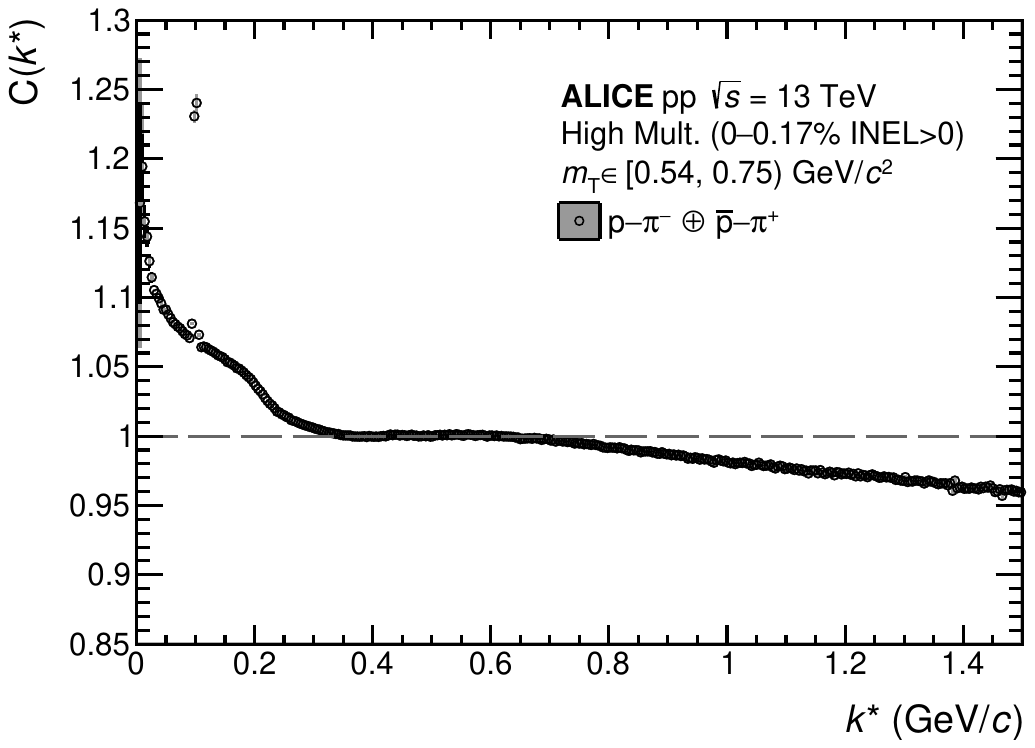}
    \label{Fig. App CF ProtonAntiPion mT 1}
  }
  \subfigure[]{
    \includegraphics[trim={0.8cm 0.55cm 1.5cm 1.2cm},clip,width=0.48\textwidth]{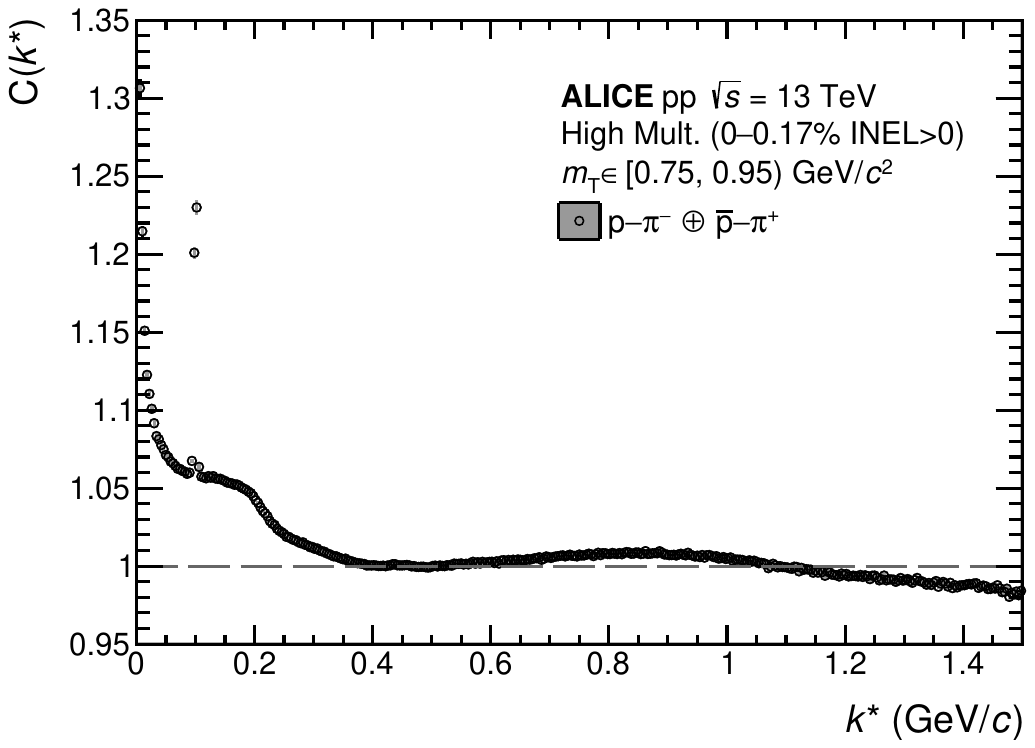}
    \label{Fig. App CF ProtonAntiPion mT 2}
  }
    \subfigure[]{
    \includegraphics[trim={0.8cm 0.55cm 1.5cm 1.2cm},clip,width=0.48\textwidth]{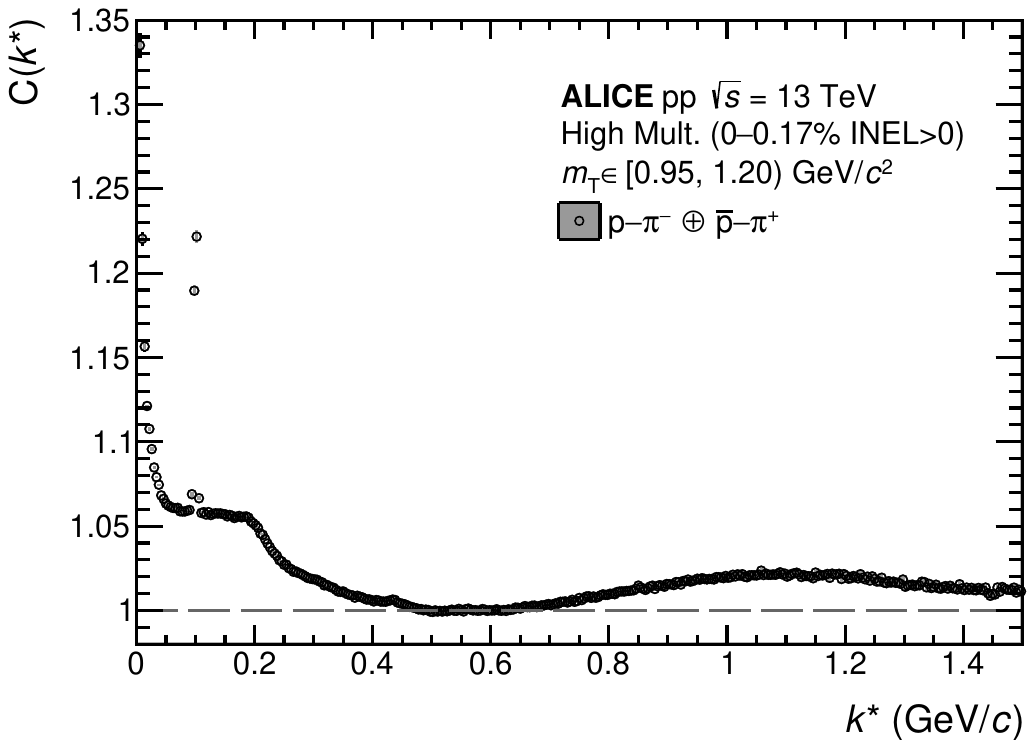}
    \label{Fig. App CF ProtonAntiPion mT 3}
  }
  \subfigure[]{
    \includegraphics[trim={0.8cm 0.55cm 1.5cm 1.2cm},clip,width=0.48\textwidth]{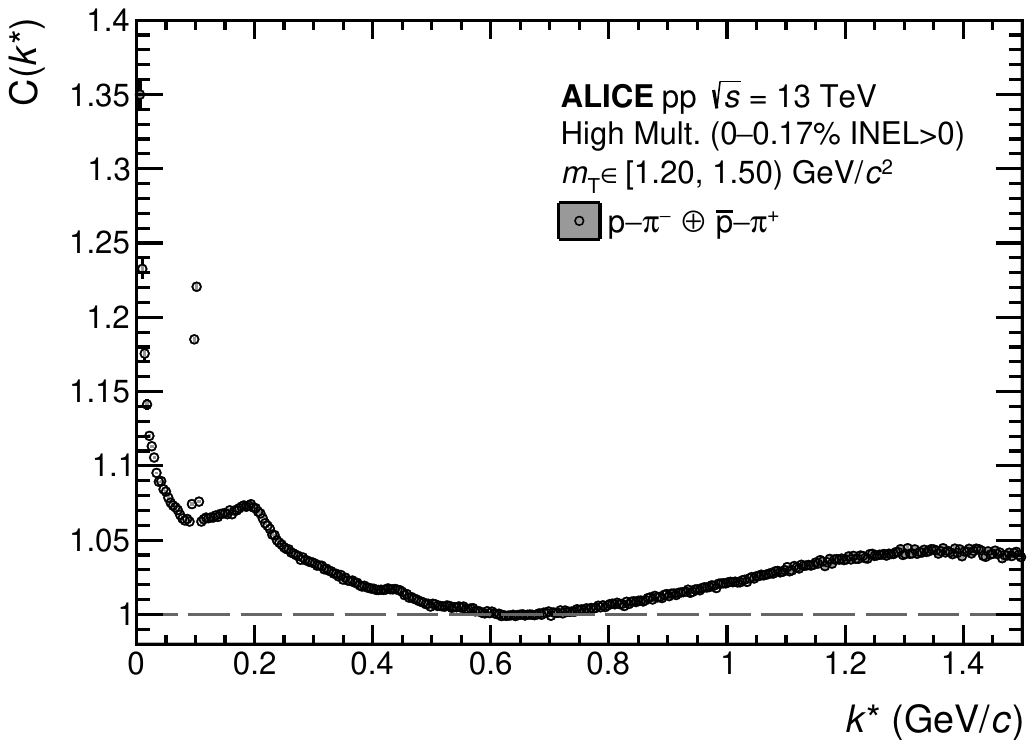}
    \label{Fig. App CF ProtonAntiPion mT 4}
  }
    \subfigure[]{
    \includegraphics[trim={0.8cm 0.55cm 1.5cm 1.2cm},clip,width=0.48\textwidth]{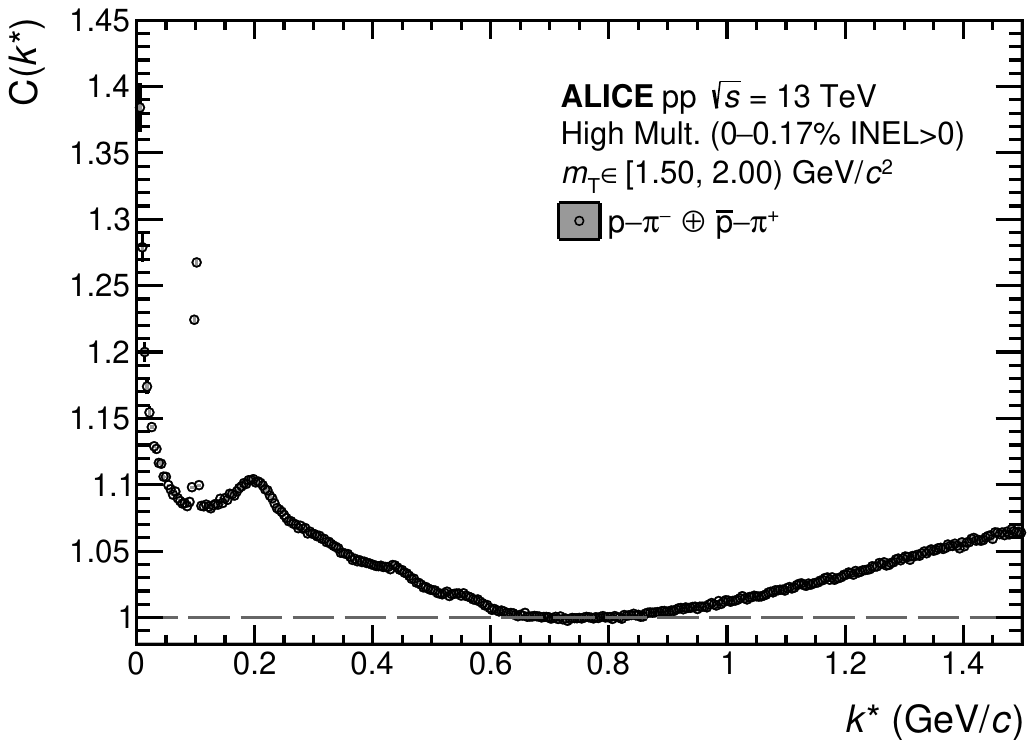}
    \label{Fig. App CF ProtonAntiPion mT 5}
  }
  \subfigure[]{
    \includegraphics[trim={0.8cm 0.55cm 1.5cm 1.2cm},clip,width=0.48\textwidth]{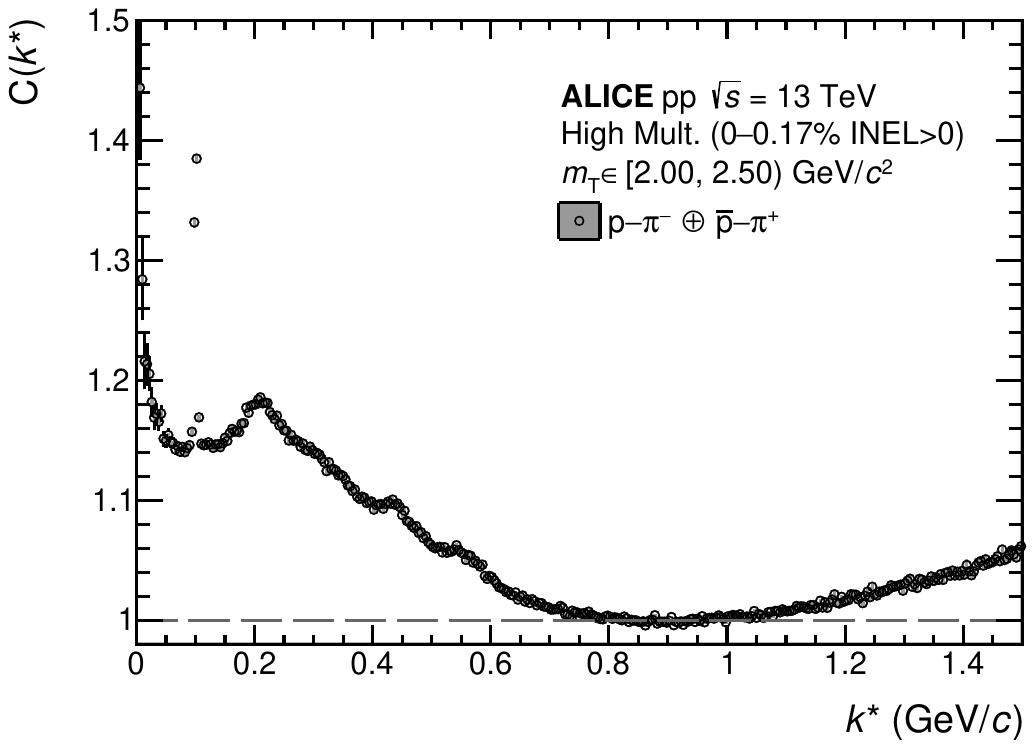}
    \label{Fig. App CF ProtonAntiPion mT 6}
  }
  \caption{The experimental correlation function of $\ppiMinus$ pairs  as a function of the pair relative momentum $k^*$ in several intervals of the pair \mtText: (a) $[0.54, 0.75)\text{ GeV}/c^2$, (b) $[0.75, 0.95)\text{ GeV}/c^2$, (c) $[0.95, 1.2)\text{ GeV}/c^2$, (d) $[1.2, 1.5)\text{ GeV}/c^2$, (e) $[1.5, 2.0)\text{ GeV}/c^2$, and (f) $[2.0, 2.5)\text{ GeV}/c^2$. The lines show the statistical uncertainties of the experimental data, while the respective systematic uncertainties are represented by the boxes.}
  \label{Fig. App CF ProtonAntiPion}
\end{figure}

\begin{figure}[!h]
  \centering
  \subfigure[]{
    \includegraphics[trim={0.8cm 0.55cm 1.5cm 1.2cm},clip,width=0.48\textwidth]{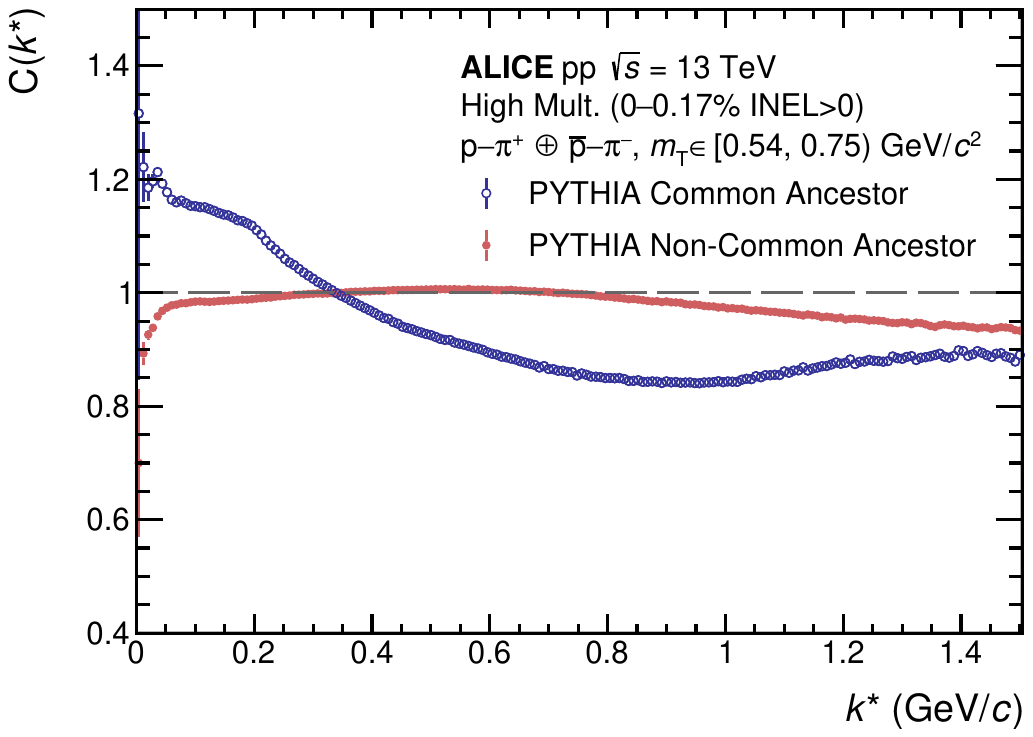}
    \label{Fig. App CF ProtonPion CommonNonCommon mT 1}
  }
  \subfigure[]{
    \includegraphics[trim={0.8cm 0.55cm 1.5cm 1.2cm},clip,width=0.48\textwidth]{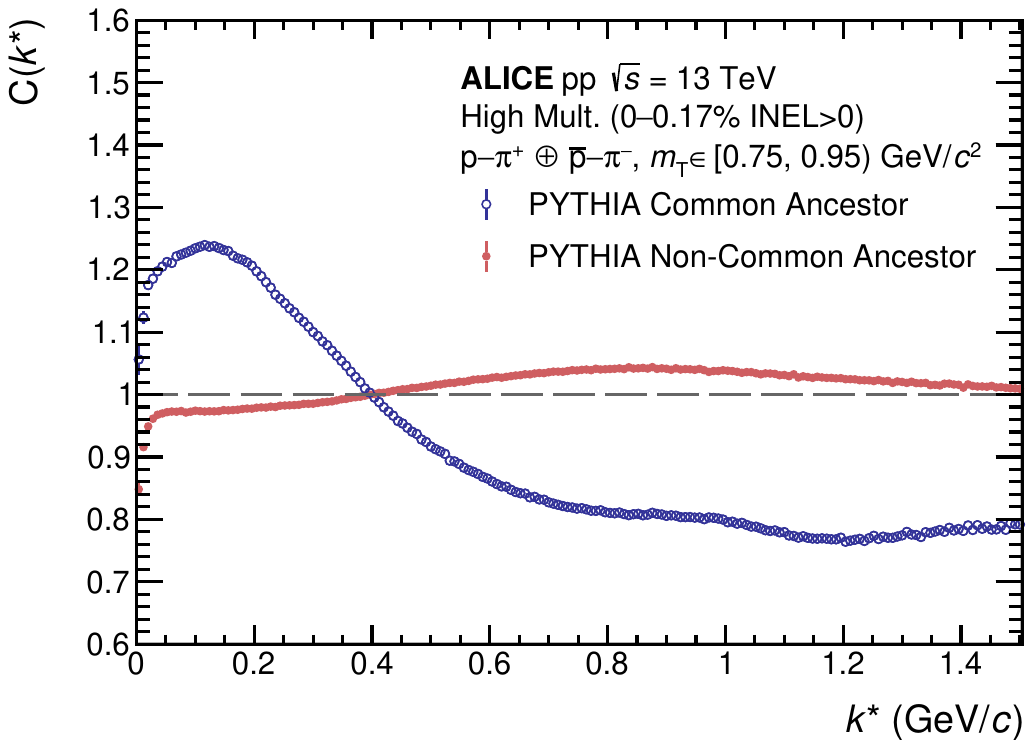}
    \label{Fig. App CF ProtonPion CommonNonCommon mT 2}
  }
    \subfigure[]{
    \includegraphics[trim={0.8cm 0.55cm 1.5cm 1.2cm},clip,width=0.48\textwidth]{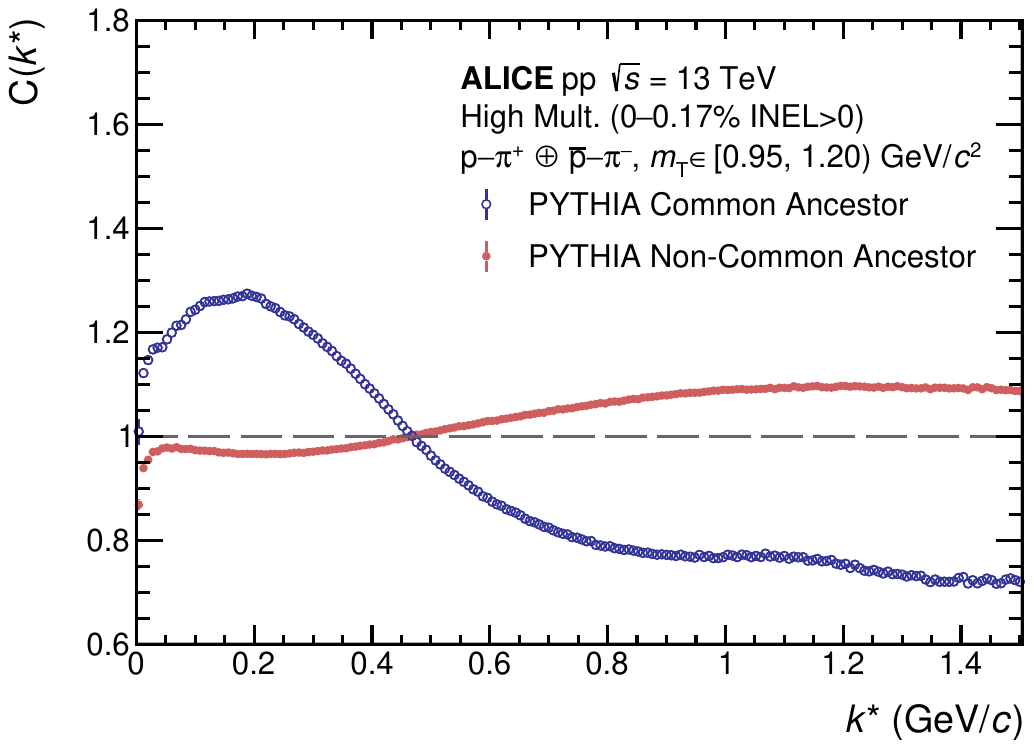}
    \label{Fig. App CF ProtonPion CommonNonCommon mT 3}
  }
  \subfigure[]{
    \includegraphics[trim={0.8cm 0.55cm 1.5cm 1.2cm},clip,width=0.48\textwidth]{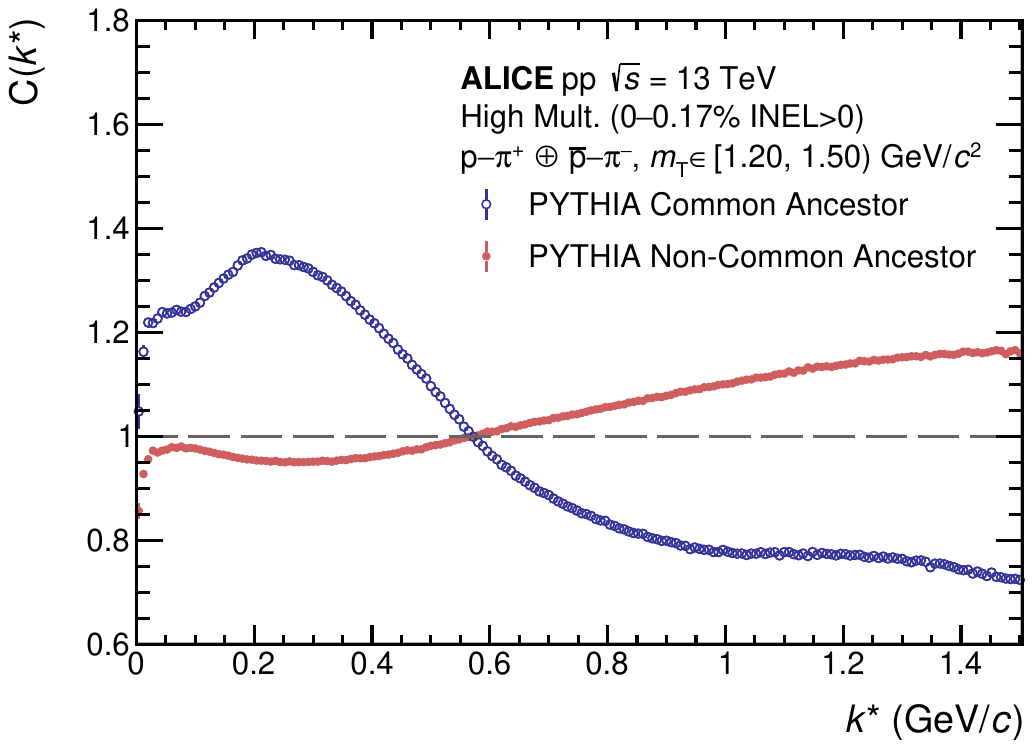}
    \label{Fig. App CF ProtonPion CommonNonCommon mT 4}
  }
    \subfigure[]{
    \includegraphics[trim={0.8cm 0.55cm 1.5cm 1.2cm},clip,width=0.48\textwidth]{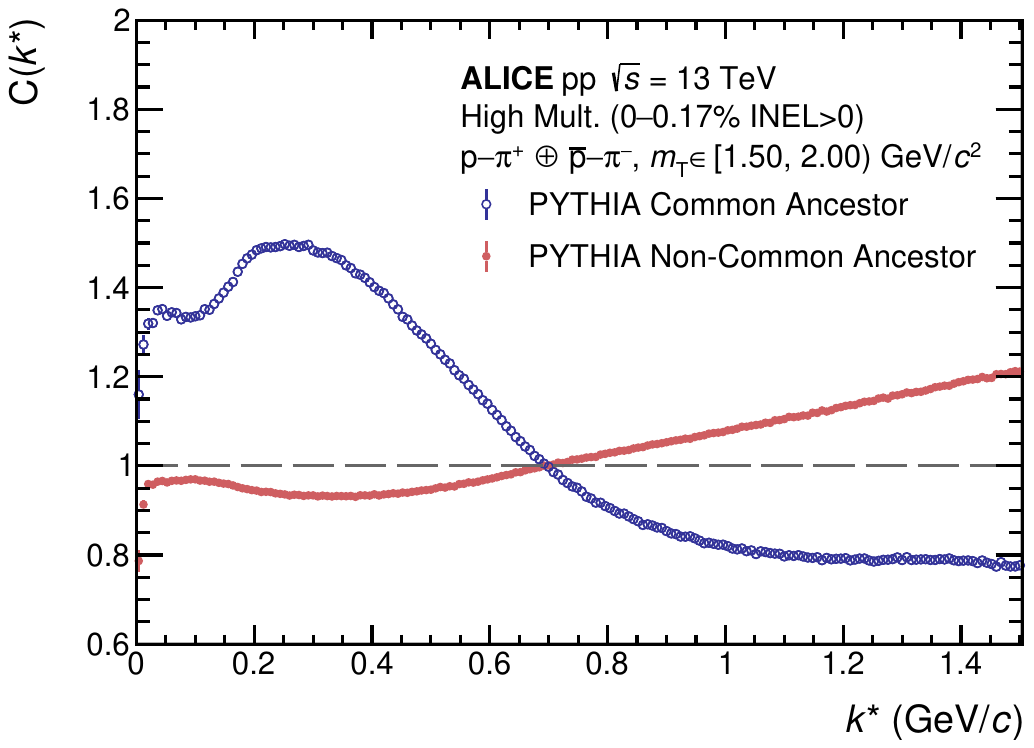}
    \label{Fig. App CF ProtonPion CommonNonCommon mT 5}
  }
  \subfigure[]{
    \includegraphics[trim={0.8cm 0.55cm 1.5cm 1.2cm},clip,width=0.48\textwidth]{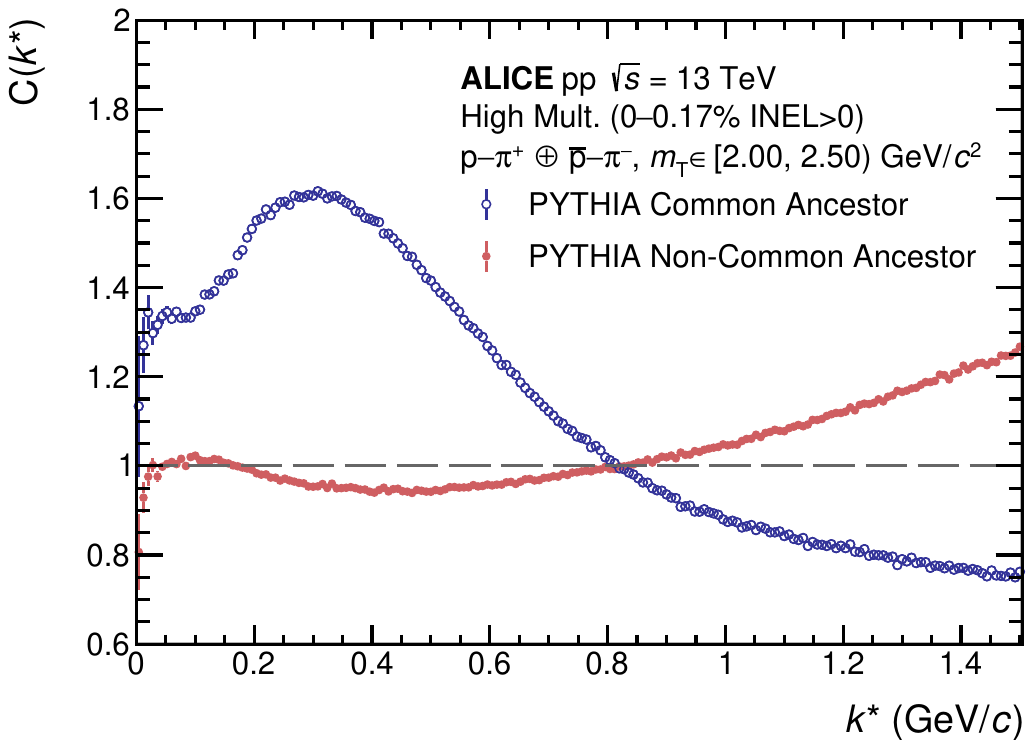}
    \label{Fig. App CF ProtonPion CommonNonCommon mT 6}
  }
  \caption{The correlation function of $\ppiPlus$ pairs for common and non-common ancestor obtained with PYTHIA 8.2 (Monash 2013 Tune)~\cite{Sjostrand:2014zea, Skands:2014pea} together with GEANT3~\cite{GEANT3} as a function of the pair relative momentum $k^*$ in several intervals of the pair \mtText: (a) $[0.54, 0.75)\text{ GeV}/c^2$, (b) $[0.75, 0.95)\text{ GeV}/c^2$, (c) $[0.95, 1.2)\text{ GeV}/c^2$, (d) $[1.2, 1.5)\text{ GeV}/c^2$, (e) $[1.5, 2.0)\text{ GeV}/c^2$, and (f) $[2.0, 2.5)\text{ GeV}/c^2$. The lines represent the statistical uncertainties. Resonance contributions such as the $\Delta^{++}$ have been explicitly removed in the common ancestor correlation functions as these contributions are modeled separately.}
  \label{Fig. App CF ProtonPion CommonNonCommon}
\end{figure}

\begin{figure}[!h]
  \centering
  \subfigure[]{
    \includegraphics[trim={0.8cm 0.55cm 1.5cm 1.2cm},clip,width=0.48\textwidth]{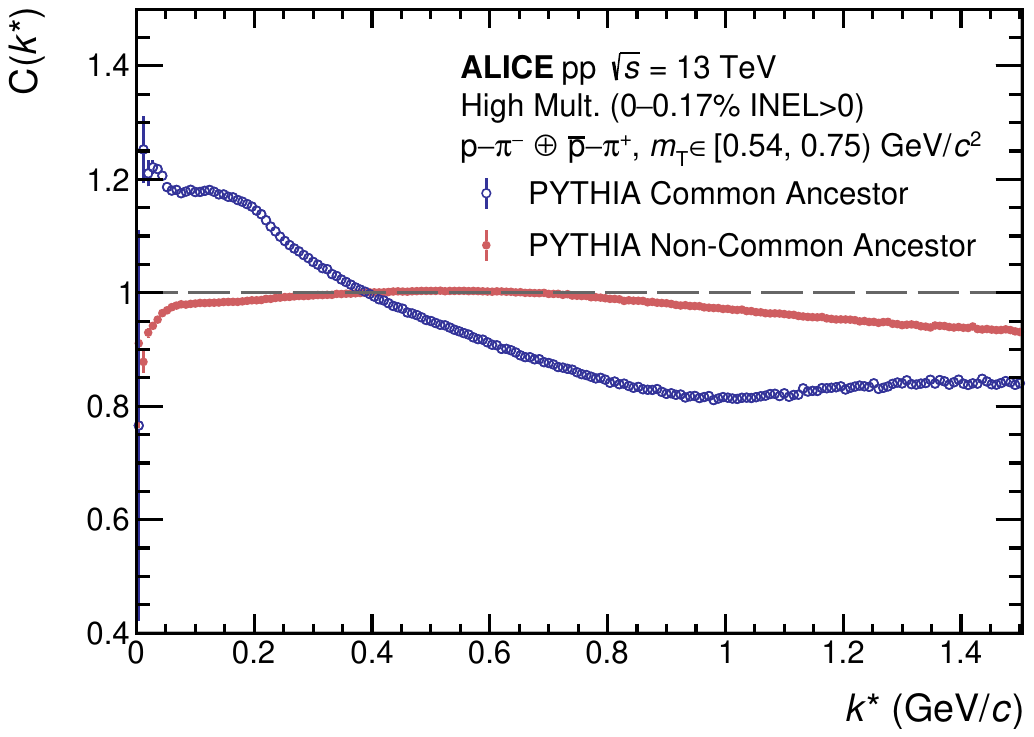}
    \label{Fig. App CF ProtonAntiPion CommonNonCommon mT 1}
  }
  \subfigure[]{
    \includegraphics[trim={0.8cm 0.55cm 1.5cm 1.2cm},clip,width=0.48\textwidth]{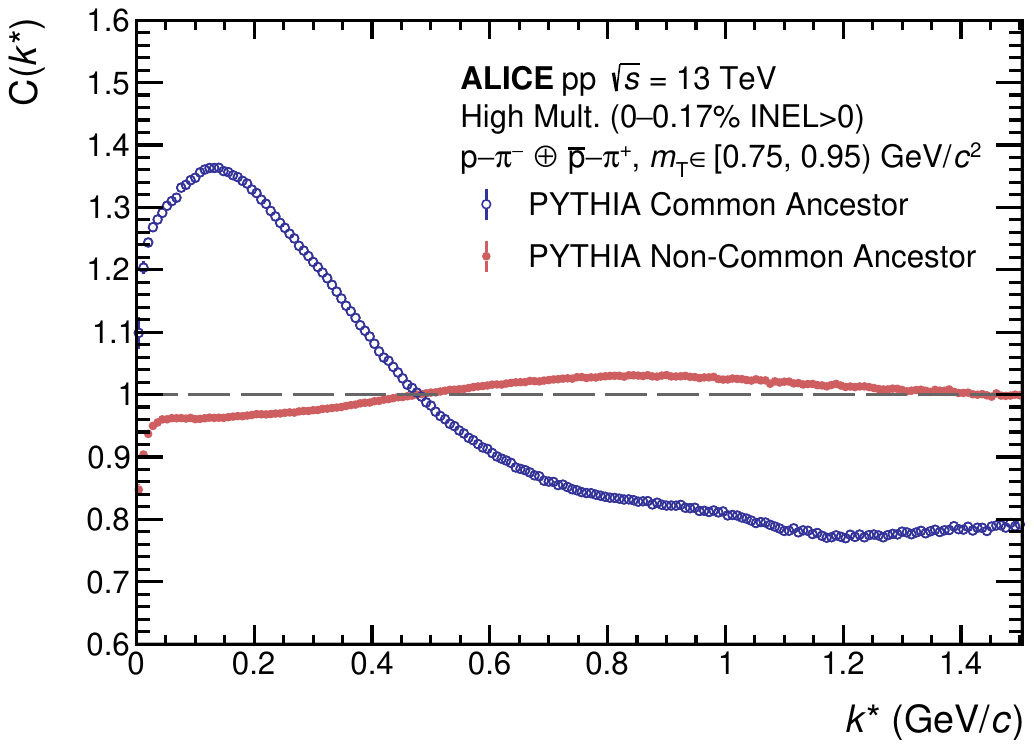}
    \label{Fig. App CF ProtonAntiPion CommonNonCommon mT 2}
  }
    \subfigure[]{
    \includegraphics[trim={0.8cm 0.55cm 1.5cm 1.2cm},clip,width=0.48\textwidth]{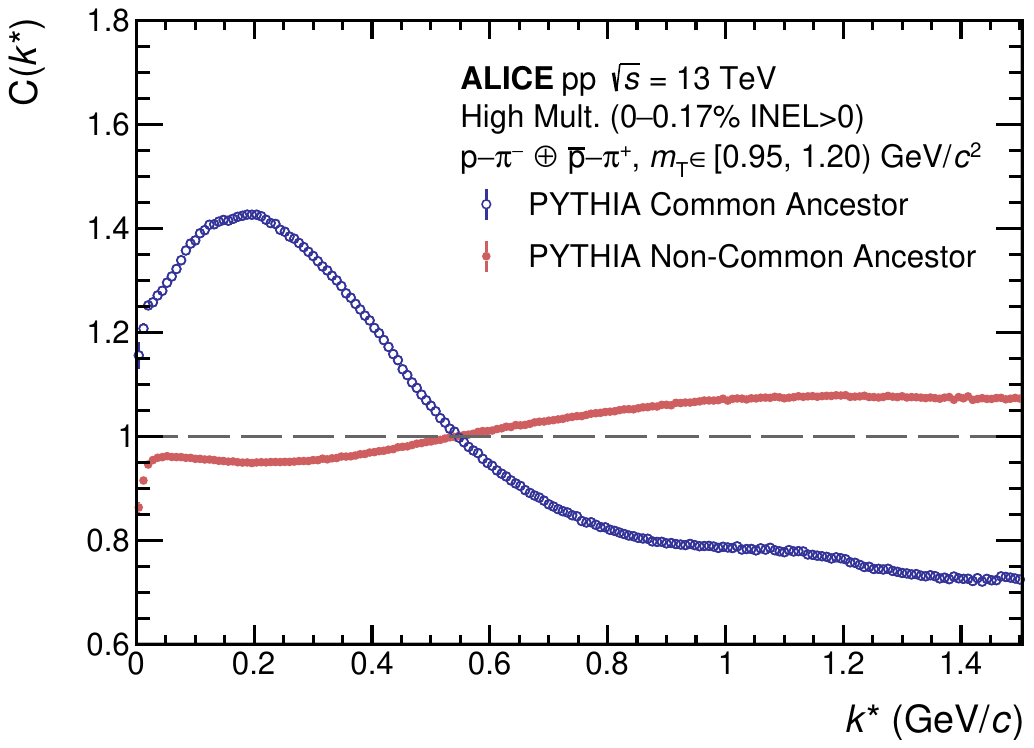}
    \label{Fig. App CF ProtonAntiPion CommonNonCommon mT 3}
  }
  \subfigure[]{
    \includegraphics[trim={0.8cm 0.55cm 1.5cm 1.2cm},clip,width=0.48\textwidth]{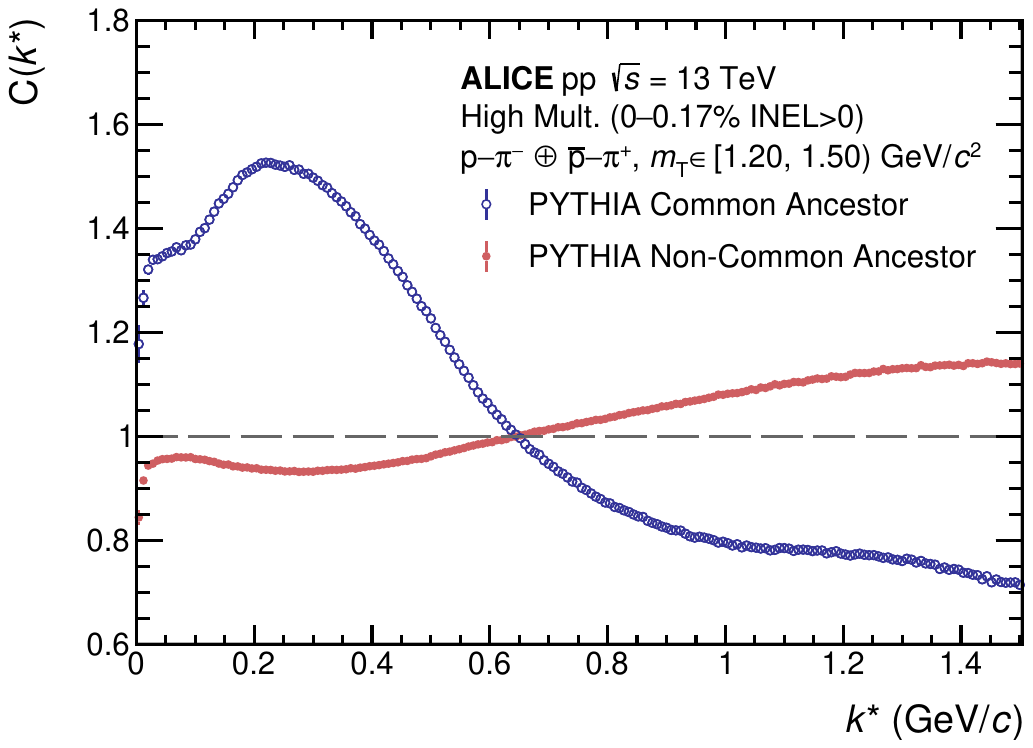}
    \label{Fig. App CF ProtonAntiPion CommonNonCommon mT 4}
  }
    \subfigure[]{
    \includegraphics[trim={0.8cm 0.55cm 1.5cm 1.2cm},clip,width=0.48\textwidth]{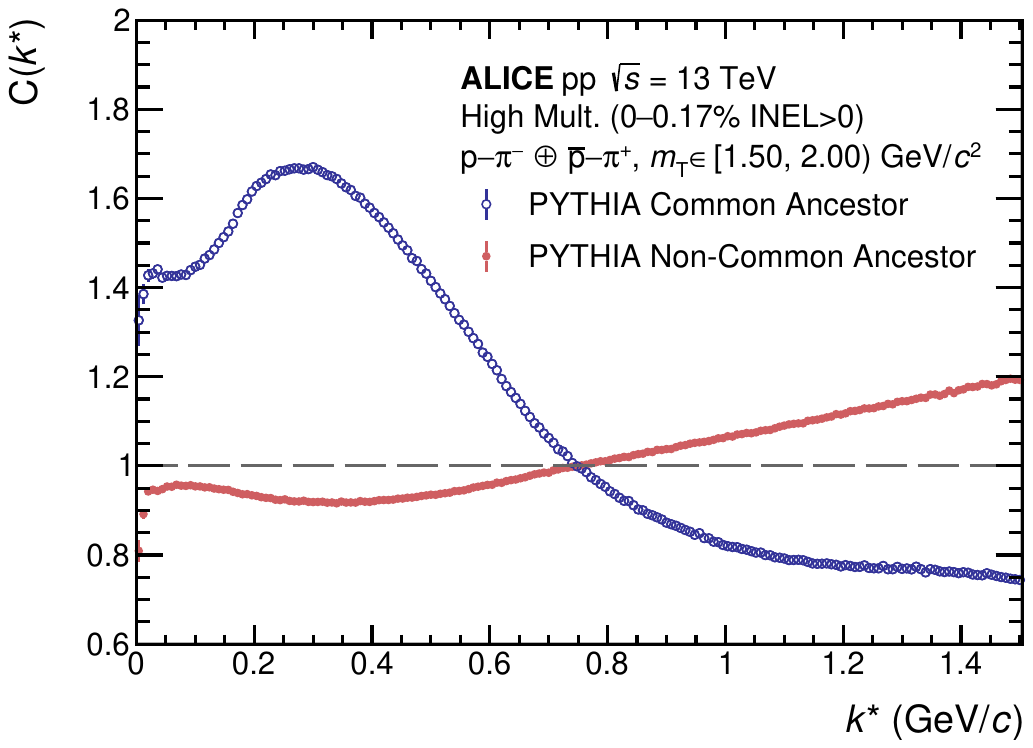}
    \label{Fig. App CF ProtonAntiPion CommonNonCommon mT 5}
  }
  \subfigure[]{
    \includegraphics[trim={0.8cm 0.55cm 1.5cm 1.2cm},clip,width=0.48\textwidth]{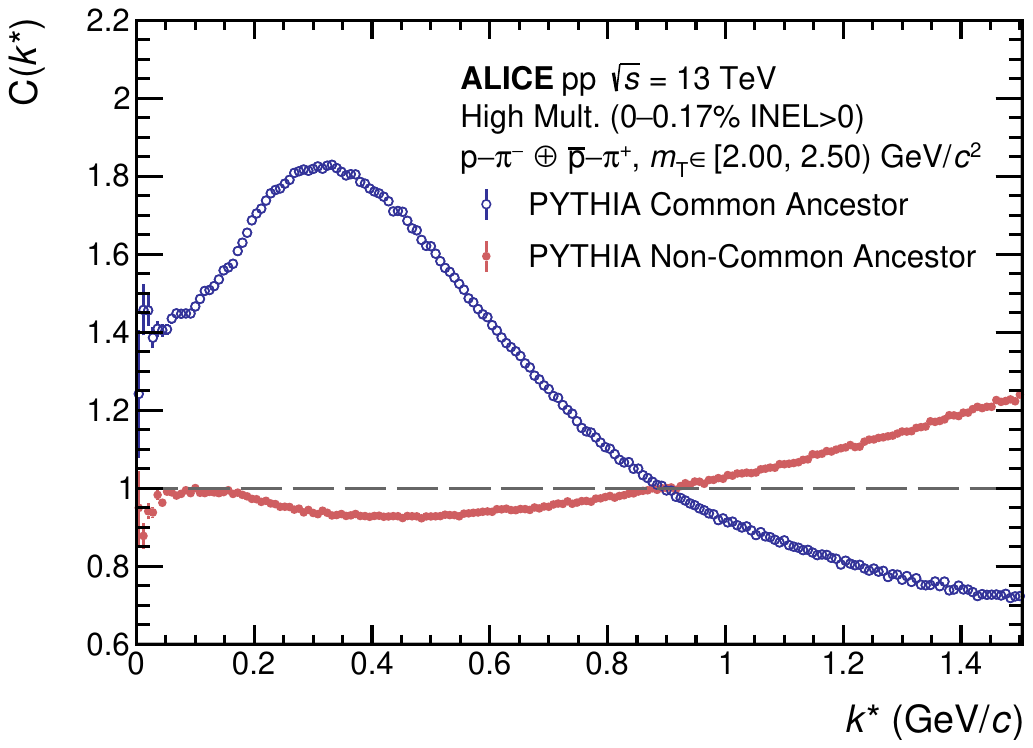}
    \label{Fig. App CF ProtonAntiPion CommonNonCommon mT 6}
  }
  \caption{The correlation function of $\ppiMinus$ pairs for common and non-common ancestor obtained with PYTHIA 8.2 (Monash 2013 Tune)~\cite{Sjostrand:2014zea, Skands:2014pea} together with GEANT3~\cite{GEANT3} as a function of the pair relative momentum $k^*$ in several intervals of the pair \mtText: (a) $[0.54, 0.75)\text{ GeV}/c^2$, (b) $[0.75, 0.95)\text{ GeV}/c^2$, (c) $[0.95, 1.2)\text{ GeV}/c^2$, (d) $[1.2, 1.5)\text{ GeV}/c^2$, (e) $[1.5, 2.0)\text{ GeV}/c^2$, and (f) $[2.0, 2.5)\text{ GeV}/c^2$. The lines represent the statistical uncertainties. Resonance contributions such as the $\Delta^0$ have been explicitly removed in the common ancestor correlation functions as these contributions are modeled separately.}
  \label{Fig. App CF ProtonAntiPion CommonNonCommon}
\end{figure}

The experimental correlation functions of $\ppiPlus$ and $\ppiMinus$ pairs are presented in Figs.~\ref{Fig. App CF ProtonPion} and ~\ref{Fig. App CF ProtonAntiPion}, respectively. All \mtText{} intervals are presented in the $k^*$ range from 0 to 1.5~$\text{ GeV}/c$. For the same \mtText{} and $k^*$ intervals, the common and non-common ancestor correlation functions obtained with PYTHIA 8.2 (Monash 2013 Tune)~\cite{Sjostrand:2014zea, Skands:2014pea} and GEANT3~\cite{GEANT3} are shown in Fig.~\ref{Fig. App CF ProtonPion CommonNonCommon} for $\ppiPlus$ and Fig.~\ref{Fig. App CF ProtonAntiPion CommonNonCommon} for $\ppiMinus$. Resonance contributions such as the $\Delta^0$ for $\ppiMinus$ or the $\Delta^{++}$ for $\ppiPlus$ have been explicitly removed in the common ancestor correlation functions as these contributions are modeled separately (see Sec.~\ref{Sec. Two-Body Modeling}).

%%%%%%%%%%%%%%%%%%%%%%%%%%%%%%% Common and NonCommon 

\clearpage

\section{Details on the modeling of the \texorpdfstring{$\mathbf{\ppiPlusMinus}$}{test} correlation functions}
\label{App. Details Two-Body Modelling}

As described in Sec.~\ref{Sec. Two-Particle Correlation}, the background contribution stemming from mini-jets is modeled using PYTHIA 8.2~\cite{Sjostrand:2014zea} (Monash 2013 Tune~\cite{Skands:2014pea}) MC simulations as $C_\text{Background}(k^*)$ in Eq.~\eqref{Eq. ProtonPion Background}. This $C_\text{Background}(k^*)$ is prefitted to the data in the large $k^*$ regime in order to adjust for differences in the long-range correlations between the MC and the data. The \mtText-dependent prefit ranges can be found in Table~\ref{Tab. ProtonPion PrefitRanges}.
\begin{table}[h]
    \centering
    \begin{tabular}{| c | c |}
    \hline 
        \mtText{} range in GeV$/c^2$ & $k^*$ range of the prefit in  GeV$/c$ \\
        \hline 
        \hline
        $[0.54, 0.75)$ & 0.275 -- 0.7\\
        $[0.75, 0.95)$ & 0.3 -- 0.8\\
        $[0.95, 1.2)$ & 0.3 -- 0.9\\
        $[1.2, 1.5)$ & 0.325 -- 1.0\\
        $[1.5, 2.0)$ & 0.325 -- 1.2\\
        $[2.0, 2.5)$ & 0.325 -- 1.4\\
        \hline
    \end{tabular}
    \caption{Prefit ranges of $C_\text{Background}(k^*)$ of $\ppiPlus$ modeling for all \mtText{} ranges.}
    \label{Tab. ProtonPion PrefitRanges}
\end{table}

The genuine correlation of $\ppiPlusMinus$ due to FSI, $C_\text{Interaction}(k^*)$, is modeled using the Coulomb and strong interactions. The latter is described with a double-Gaussian potential $V_\text{strong}(r)$ of the form
\begin{equation}
    V_\text{strong}(r) = \nu_1 e^{-r^2/\mu_1^2} + \nu_2 e^{-r^2/\mu_2^2}\,, \label{Eq. Double Gauss}
\end{equation}
where each of the Gaussians is parameterized by its strength $\nu$ and width $\mu$. The potential parameters are tuned to reproduce the scattering length~\cite{Hoferichter:2015hva} for $\ppiPlusMinus$ interactions and are given in Table~\ref{Tab. effective potential par}.  The potential $V_\text{strong}(r)$ together with the Coulomb potential is then evaluated with the 
CATS framework~\cite{Mihaylov:2018rva} in order to obtain the two-particle wave function.

\begin{table}[h]
    \centering
    \begin{tabular}{| ccccc | }
    \hline
        System & $\nu_1$ in MeV & $\mu_1$ in fm& $\nu_2$ in MeV & $\mu_2$  in fm\\
        \hline
        \hline
        $\ppiPlus$ & 765.1 & 0.367 & 685.93& 0.331 \\
        $\ppiMinus$ & -32.32 & 1.078 & -222.8 & 0.097 \\
        \hline
    \end{tabular}
    \caption{Parameters for the double-Gaussian potential \eqref{Eq. Double Gauss} for the $\ppiPlusMinus$ strong interaction.}
    \label{Tab. effective potential par}
\end{table}

\section{Two-Particle Momentum Smearing}
\label{App. Two-Body Smearing}
\begin{figure}[!h]
  \centering
\subfigure[]{
    \includegraphics[trim={0.8cm 0.55cm 0cm 0.9cm},clip,width=0.48\textwidth]{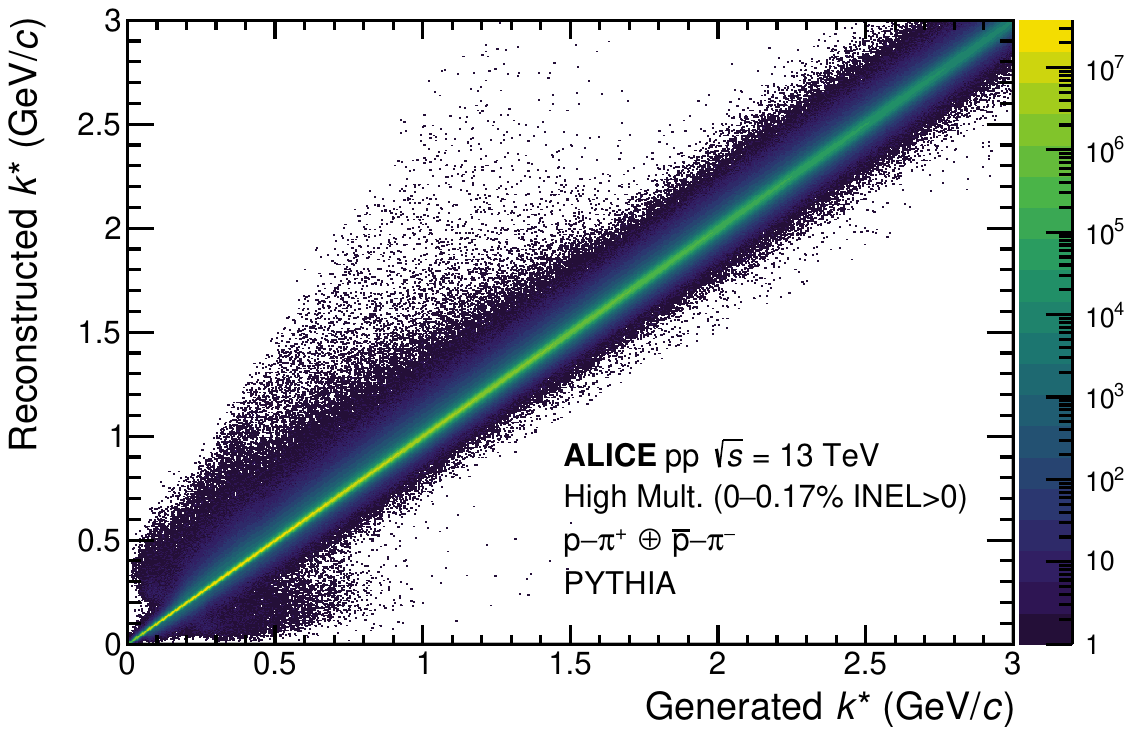}
    \label{Fig. ProtonPion MomentumSmearing}
  }
\subfigure[]{
    \includegraphics[trim={0.8cm 0.55cm 0cm 0.9cm},clip,width=0.48\textwidth]{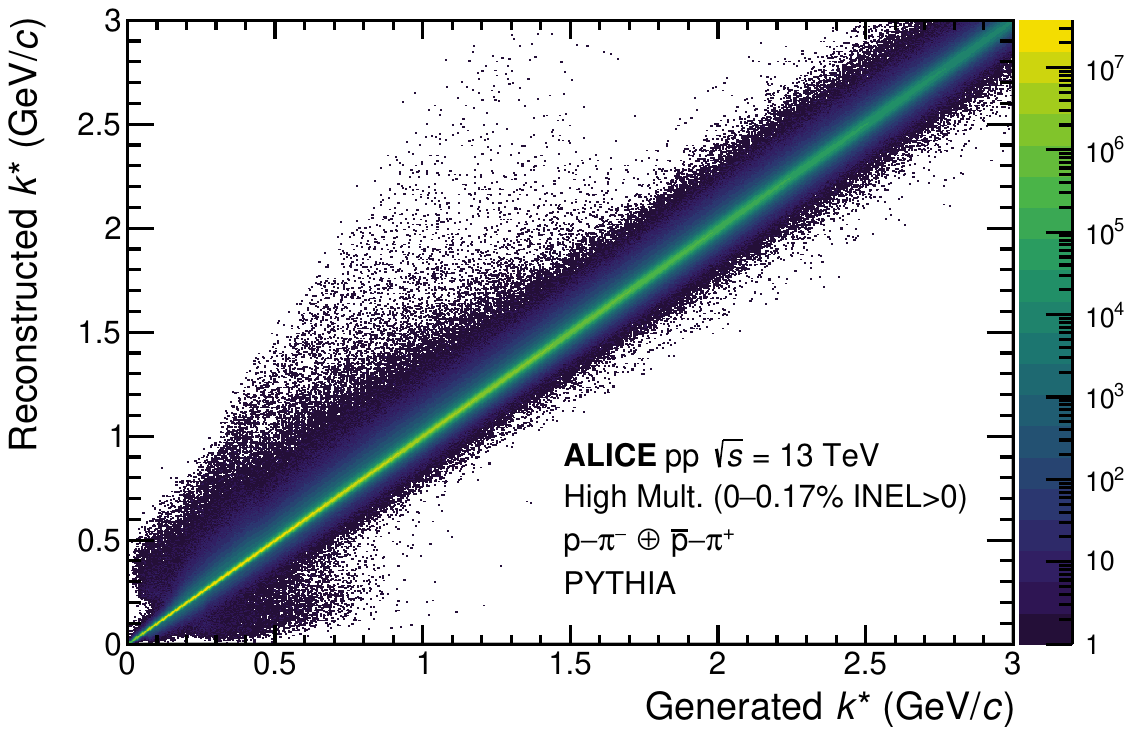}
    \label{Fig. ProtonAntiPion MomentumSmearing}
  } 
  \caption{Momentum smearing matrix of (a) $\ppiPlus$ pairs and (b) $\ppiMinus$ pairs obtained with PYTHIA 8.2 (Monash 2013 Tune)~\cite{Sjostrand:2014zea, Skands:2014pea} together with GEANT3~\cite{GEANT3}.}
  \label{Fig. Smearing Matrices}
\end{figure}

\begin{figure}[!h]
  \centering
  
  \subfigure[]{
      \includegraphics[trim={0.8cm 0.55cm 0cm 0.9cm},clip,width=0.48\textwidth]{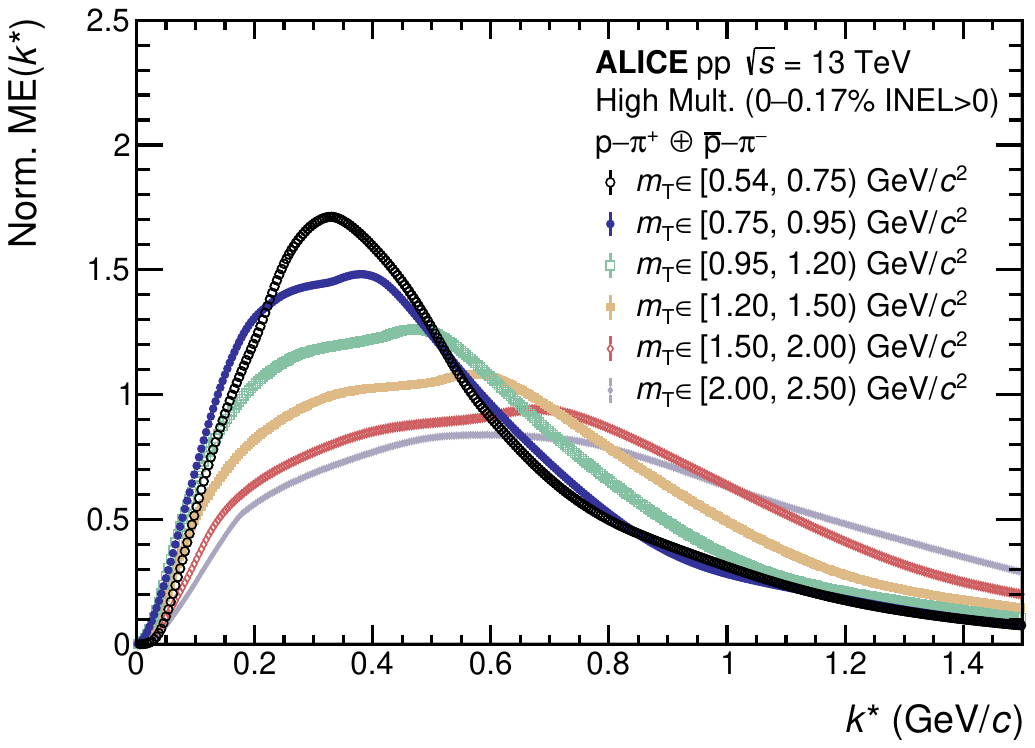}
    \label{Fig. ProtonPion ME}
  }
   \subfigure[]{
      \includegraphics[trim={0.8cm 0.55cm 0cm 0.9cm},clip,width=0.48\textwidth]{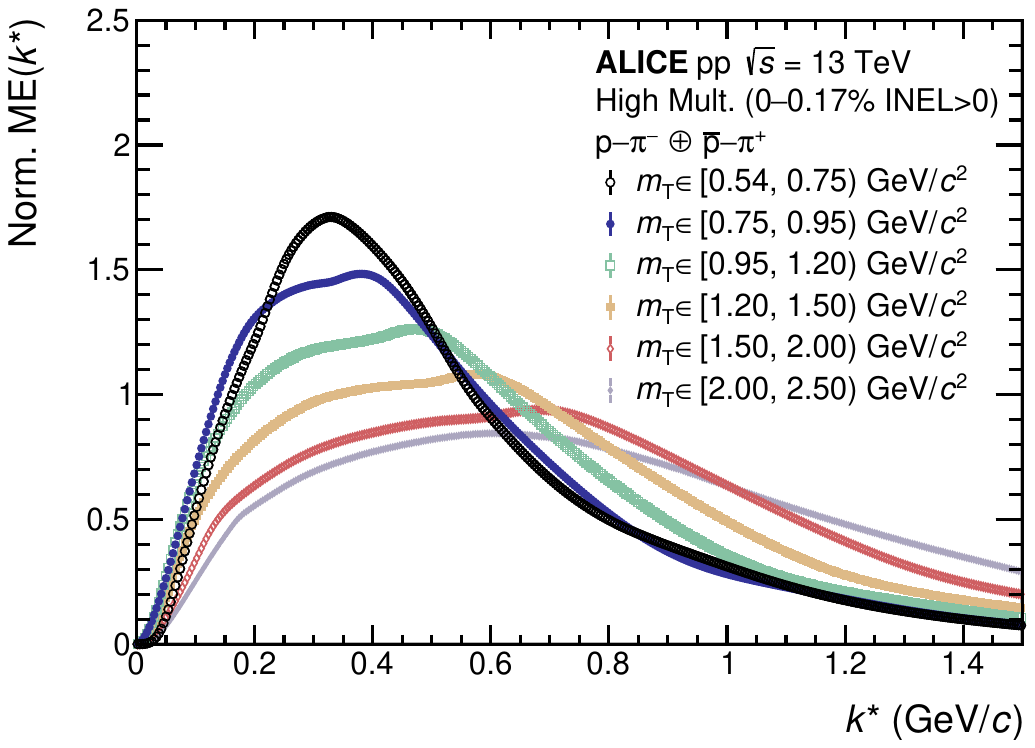}
    \label{Fig. ProtonAntiPion ME}
  }
  \caption{The normalized mixed event distributions of (a) $\ppiPlus$ and (b) $\ppiMinus$ pairs for all relevant \mtText{} intervals.}
  \label{Fig. ME Dists}
\end{figure}

The presented experimental correlation functions are given as a function of the measured $k^*$.  Due to the finite momentum resolution of the detector, this measured (or reconstructed) $k^*$ is not identical to the
true (or generated) relative momentum of the pair. In order to compare the
experimental results with theoretical predictions, the latter can be smeared for these resolution effects. To do so, the momentum smearing matrix, which relates the generated $k^*$ of the particle pair to the reconstructed $k^*$, should be considered. These matrices, obtained with the PYTHIA 8.2~\cite{Sjostrand:2014zea} (Monash 2013 Tune~\cite{Skands:2014pea}) event generator together with the GEANT3~\cite{GEANT3} transport code, are shown in Fig.~\ref{Fig. ProtonPion MomentumSmearing} for $\ppiPlus$ and Fig.~\ref{Fig. ProtonAntiPion MomentumSmearing} for $\ppiMinus$. Additionally, Figs.~\ref{Fig. ProtonPion ME} and ~\ref{Fig. ProtonAntiPion ME} show the mixed event distributions of $\ppiPlus$ and $\ppiMinus$, respectively, for all \mtText{} intervals. They are needed in order to account for the proper phase space during the smearing procedure (see~Ref.~\cite{Acharya:2714920}).

\clearpage
%%%%% Authorlist - please do not touch: handled by EB chairs 
\section{The ALICE Collaboration}
\label{app:collab}
% ALICE Collaboration author list for 2024-11-19
\begin{flushleft} 
\small

S.~Acharya\,\orcidlink{0000-0002-9213-5329}\,$^{\rm 126}$, 
A.~Agarwal$^{\rm 134}$, 
G.~Aglieri Rinella\,\orcidlink{0000-0002-9611-3696}\,$^{\rm 32}$, 
L.~Aglietta\,\orcidlink{0009-0003-0763-6802}\,$^{\rm 24}$, 
M.~Agnello\,\orcidlink{0000-0002-0760-5075}\,$^{\rm 29}$, 
N.~Agrawal\,\orcidlink{0000-0003-0348-9836}\,$^{\rm 25}$, 
Z.~Ahammed\,\orcidlink{0000-0001-5241-7412}\,$^{\rm 134}$, 
S.~Ahmad\,\orcidlink{0000-0003-0497-5705}\,$^{\rm 15}$, 
S.U.~Ahn\,\orcidlink{0000-0001-8847-489X}\,$^{\rm 71}$, 
I.~Ahuja\,\orcidlink{0000-0002-4417-1392}\,$^{\rm 36}$
% ...

\section*{Affiliation Notes}

$^{\rm I}$ Deceased\\
$^{\rm II}$ Also at: Max-Planck-Institut fur Physik, Munich, Germany\\
$^{\rm III}$ Also at: Italian National Agency for New Technologies, Energy and Sustainable Economic Development (ENEA), Bologna, Italy\\
$^{\rm IV}$ Also at: Dipartimento DET del Politecnico di Torino, Turin, Italy\\
$^{\rm V}$ Also at: Department of Applied Physics, Aligarh Muslim University, Aligarh, India\\
$^{\rm VI}$ Also at: Institute of Theoretical Physics, University of Wroclaw, Poland\\
$^{\rm VII}$ Also at: Facultad de Ciencias, Universidad Nacional Aut\'{o}noma de M\'{e}xico, Mexico City, Mexico\\

\section*{Collaboration Institutes}

$^{1}$ A.I. Alikhanyan National Science Laboratory (Yerevan Physics Institute) Foundation, Yerevan, Armenia\\
$^{2}$ AGH University of Krakow, Cracow, Poland\\
$^{3}$ Bogolyubov Institute for Theoretical Physics, National Academy of Sciences of Ukraine, Kiev, Ukraine\\
$^{4}$ Bose Institute, Department of Physics  and Centre for Astroparticle Physics and Space Science (CAPSS), Kolkata, India\\
$^{5}$ California Polytechnic State University, San Luis Obispo, California, United States\\
$^{6}$ Central China Normal University, Wuhan, China\\
$^{7}$ Centro de Aplicaciones Tecnol\'{o}gicas y Desarrollo Nuclear (CEADEN), Havana, Cuba\\
$^{8}$ Centro de Investigaci\'{o}n y de Estudios Avanzados (CINVESTAV), Mexico City and M\'{e}rida, Mexico\\
$^{9}$ Chicago State University, Chicago, Illinois, United States\\
$^{10}$ China Nuclear Data Center, China Institute of Atomic Energy, Beijing, China\\
$^{11}$ China University of Geosciences, Wuhan, China\\
$^{12}$ Chungbuk National University, Cheongju, Republic of Korea\\
$^{13}$ Comenius University Bratislava, Faculty of Mathematics, Physics and Informatics, Bratislava, Slovak Republic\\
$^{14}$ Creighton University, Omaha, Nebraska, United States\\
$^{15}$ Department of Physics, Aligarh Muslim University, Aligarh, India\\
$^{16}$ Department of Physics, Pusan National University, Pusan, Republic of Korea\\
$^{17}$ Department of Physics, Sejong University, Seoul, Republic of Korea\\
$^{18}$ Department of Physics, University of California, Berkeley, California, United States\\
$^{19}$ Department of Physics, University of Oslo, Oslo, Norway\\
$^{20}$ Department of Physics and Technology, University of Bergen, Bergen, Norway\\
$^{21}$ Dipartimento di Fisica, Universit\`{a} di Pavia, Pavia, Italy\\
$^{22}$ Dipartimento di Fisica dell'Universit\`{a} and Sezione INFN, Cagliari, Italy\\
$^{23}$ Dipartimento di Fisica dell'Universit\`{a} and Sezione INFN, Trieste, Italy\\
$^{24}$ Dipartimento di Fisica dell'Universit\`{a} and Sezione INFN, Turin, Italy\\
$^{25}$ Dipartimento di Fisica e Astronomia dell'Universit\`{a} and Sezione INFN, Bologna, Italy\\
$^{26}$ Dipartimento di Fisica e Astronomia dell'Universit\`{a} and Sezione INFN, Catania, Italy\\
$^{27}$ Dipartimento di Fisica e Astronomia dell'Universit\`{a} and Sezione INFN, Padova, Italy\\
$^{28}$ Dipartimento di Fisica `E.R.~Caianiello' dell'Universit\`{a} and Gruppo Collegato INFN, Salerno, Italy\\
$^{29}$ Dipartimento DISAT del Politecnico and Sezione INFN, Turin, Italy\\
$^{30}$ Dipartimento di Scienze MIFT, Universit\`{a} di Messina, Messina, Italy\\
$^{31}$ Dipartimento Interateneo di Fisica `M.~Merlin' and Sezione INFN, Bari, Italy\\
$^{32}$ European Organization for Nuclear Research (CERN), Geneva, Switzerland\\
$^{33}$ Faculty of Electrical Engineering, Mechanical Engineering and Naval Architecture, University of Split, Split, Croatia\\
$^{34}$ Faculty of Nuclear Sciences and Physical Engineering, Czech Technical University in Prague, Prague, Czech Republic\\
$^{35}$ Faculty of Physics, Sofia University, Sofia, Bulgaria\\
$^{36}$ Faculty of Science, P.J.~\v{S}af\'{a}rik University, Ko\v{s}ice, Slovak Republic\\
$^{37}$ Faculty of Technology, Environmental and Social Sciences, Bergen, Norway\\
$^{38}$ Frankfurt Institute for Advanced Studies, Johann Wolfgang Goethe-Universit\"{a}t Frankfurt, Frankfurt, Germany\\
$^{39}$ Fudan University, Shanghai, China\\
$^{40}$ Gangneung-Wonju National University, Gangneung, Republic of Korea\\
$^{41}$ Gauhati University, Department of Physics, Guwahati, India\\
$^{42}$ Helmholtz-Institut f\"{u}r Strahlen- und Kernphysik, Rheinische Friedrich-Wilhelms-Universit\"{a}t Bonn, Bonn, Germany\\
$^{43}$ Helsinki Institute of Physics (HIP), Helsinki, Finland\\
$^{44}$ High Energy Physics Group,  Universidad Aut\'{o}noma de Puebla, Puebla, Mexico\\
$^{45}$ Horia Hulubei National Institute of Physics and Nuclear Engineering, Bucharest, Romania\\
$^{46}$ HUN-REN Wigner Research Centre for Physics, Budapest, Hungary\\
$^{47}$ Indian Institute of Technology Bombay (IIT), Mumbai, India\\
$^{48}$ Indian Institute of Technology Indore, Indore, India\\
$^{49}$ INFN, Laboratori Nazionali di Frascati, Frascati, Italy\\
$^{50}$ INFN, Sezione di Bari, Bari, Italy\\
$^{51}$ INFN, Sezione di Bologna, Bologna, Italy\\
$^{52}$ INFN, Sezione di Cagliari, Cagliari, Italy\\
$^{53}$ INFN, Sezione di Catania, Catania, Italy\\
$^{54}$ INFN, Sezione di Padova, Padova, Italy\\
$^{55}$ INFN, Sezione di Pavia, Pavia, Italy\\
$^{56}$ INFN, Sezione di Torino, Turin, Italy\\
$^{57}$ INFN, Sezione di Trieste, Trieste, Italy\\
$^{58}$ Inha University, Incheon, Republic of Korea\\
$^{59}$ Institute for Gravitational and Subatomic Physics (GRASP), Utrecht University/Nikhef, Utrecht, Netherlands\\
$^{60}$ Institute of Experimental Physics, Slovak Academy of Sciences, Ko\v{s}ice, Slovak Republic\\
$^{61}$ Institute of Physics, Homi Bhabha National Institute, Bhubaneswar, India\\
$^{62}$ Institute of Physics of the Czech Academy of Sciences, Prague, Czech Republic\\
$^{63}$ Institute of Space Science (ISS), Bucharest, Romania\\
$^{64}$ Institut f\"{u}r Kernphysik, Johann Wolfgang Goethe-Universit\"{a}t Frankfurt, Frankfurt, Germany\\
$^{65}$ Instituto de Ciencias Nucleares, Universidad Nacional Aut\'{o}noma de M\'{e}xico, Mexico City, Mexico\\
$^{66}$ Instituto de F\'{i}sica, Universidade Federal do Rio Grande do Sul (UFRGS), Porto Alegre, Brazil\\
$^{67}$ Instituto de F\'{\i}sica, Universidad Nacional Aut\'{o}noma de M\'{e}xico, Mexico City, Mexico\\
$^{68}$ iThemba LABS, National Research Foundation, Somerset West, South Africa\\
$^{69}$ Jeonbuk National University, Jeonju, Republic of Korea\\
$^{70}$ Johann-Wolfgang-Goethe Universit\"{a}t Frankfurt Institut f\"{u}r Informatik, Fachbereich Informatik und Mathematik, Frankfurt, Germany\\
$^{71}$ Korea Institute of Science and Technology Information, Daejeon, Republic of Korea\\
$^{72}$ Laboratoire de Physique Subatomique et de Cosmologie, Universit\'{e} Grenoble-Alpes, CNRS-IN2P3, Grenoble, France\\
$^{73}$ Lawrence Berkeley National Laboratory, Berkeley, California, United States\\
$^{74}$ Lund University Department of Physics, Division of Particle Physics, Lund, Sweden\\
$^{75}$ Nagasaki Institute of Applied Science, Nagasaki, Japan\\
$^{76}$ Nara Women{'}s University (NWU), Nara, Japan\\
$^{77}$ National and Kapodistrian University of Athens, School of Science, Department of Physics , Athens, Greece\\
$^{78}$ National Centre for Nuclear Research, Warsaw, Poland\\
$^{79}$ National Institute of Science Education and Research, Homi Bhabha National Institute, Jatni, India\\
$^{80}$ National Nuclear Research Center, Baku, Azerbaijan\\
$^{81}$ National Research and Innovation Agency - BRIN, Jakarta, Indonesia\\
$^{82}$ Niels Bohr Institute, University of Copenhagen, Copenhagen, Denmark\\
$^{83}$ Nikhef, National institute for subatomic physics, Amsterdam, Netherlands\\
$^{84}$ Nuclear Physics Group, STFC Daresbury Laboratory, Daresbury, United Kingdom\\
$^{85}$ Nuclear Physics Institute of the Czech Academy of Sciences, Husinec-\v{R}e\v{z}, Czech Republic\\
$^{86}$ Oak Ridge National Laboratory, Oak Ridge, Tennessee, United States\\
$^{87}$ Ohio State University, Columbus, Ohio, United States\\
$^{88}$ Physics department, Faculty of science, University of Zagreb, Zagreb, Croatia\\
$^{89}$ Physics Department, Panjab University, Chandigarh, India\\
$^{90}$ Physics Department, University of Jammu, Jammu, India\\
$^{91}$ Physics Program and International Institute for Sustainability with Knotted Chiral Meta Matter (WPI-SKCM$^{2}$), Hiroshima University, Hiroshima, Japan\\
$^{92}$ Physikalisches Institut, Eberhard-Karls-Universit\"{a}t T\"{u}bingen, T\"{u}bingen, Germany\\
$^{93}$ Physikalisches Institut, Ruprecht-Karls-Universit\"{a}t Heidelberg, Heidelberg, Germany\\
$^{94}$ Physik Department, Technische Universit\"{a}t M\"{u}nchen, Munich, Germany\\
$^{95}$ Politecnico di Bari and Sezione INFN, Bari, Italy\\
$^{96}$ Research Division and ExtreMe Matter Institute EMMI, GSI Helmholtzzentrum f\"ur Schwerionenforschung GmbH, Darmstadt, Germany\\
$^{97}$ Saga University, Saga, Japan\\
$^{98}$ Saha Institute of Nuclear Physics, Homi Bhabha National Institute, Kolkata, India\\
$^{99}$ School of Physics and Astronomy, University of Birmingham, Birmingham, United Kingdom\\
$^{100}$ Secci\'{o}n F\'{\i}sica, Departamento de Ciencias, Pontificia Universidad Cat\'{o}lica del Per\'{u}, Lima, Peru\\
$^{101}$ Stefan Meyer Institut f\"{u}r Subatomare Physik (SMI), Vienna, Austria\\
$^{102}$ SUBATECH, IMT Atlantique, Nantes Universit\'{e}, CNRS-IN2P3, Nantes, France\\
$^{103}$ Sungkyunkwan University, Suwon City, Republic of Korea\\
$^{104}$ Suranaree University of Technology, Nakhon Ratchasima, Thailand\\
$^{105}$ Technical University of Ko\v{s}ice, Ko\v{s}ice, Slovak Republic\\
$^{106}$ The Henryk Niewodniczanski Institute of Nuclear Physics, Polish Academy of Sciences, Cracow, Poland\\
$^{107}$ The University of Texas at Austin, Austin, Texas, United States\\
$^{108}$ Universidad Aut\'{o}noma de Sinaloa, Culiac\'{a}n, Mexico\\
$^{109}$ Universidade de S\~{a}o Paulo (USP), S\~{a}o Paulo, Brazil\\
$^{110}$ Universidade Estadual de Campinas (UNICAMP), Campinas, Brazil\\
$^{111}$ Universidade Federal do ABC, Santo Andre, Brazil\\
$^{112}$ Universitatea Nationala de Stiinta si Tehnologie Politehnica Bucuresti, Bucharest, Romania\\
$^{113}$ University of Cape Town, Cape Town, South Africa\\
$^{114}$ University of Derby, Derby, United Kingdom\\
$^{115}$ University of Houston, Houston, Texas, United States\\
$^{116}$ University of Jyv\"{a}skyl\"{a}, Jyv\"{a}skyl\"{a}, Finland\\
$^{117}$ University of Kansas, Lawrence, Kansas, United States\\
$^{118}$ University of Liverpool, Liverpool, United Kingdom\\
$^{119}$ University of Science and Technology of China, Hefei, China\\
$^{120}$ University of South-Eastern Norway, Kongsberg, Norway\\
$^{121}$ University of Tennessee, Knoxville, Tennessee, United States\\
$^{122}$ University of the Witwatersrand, Johannesburg, South Africa\\
$^{123}$ University of Tokyo, Tokyo, Japan\\
$^{124}$ University of Tsukuba, Tsukuba, Japan\\
$^{125}$ Universit\"{a}t M\"{u}nster, Institut f\"{u}r Kernphysik, M\"{u}nster, Germany\\
$^{126}$ Universit\'{e} Clermont Auvergne, CNRS/IN2P3, LPC, Clermont-Ferrand, France\\
$^{127}$ Universit\'{e} de Lyon, CNRS/IN2P3, Institut de Physique des 2 Infinis de Lyon, Lyon, France\\
$^{128}$ Universit\'{e} de Strasbourg, CNRS, IPHC UMR 7178, F-67000 Strasbourg, France, Strasbourg, France\\
$^{129}$ Universit\'{e} Paris-Saclay, Centre d'Etudes de Saclay (CEA), IRFU, D\'{e}partment de Physique Nucl\'{e}aire (DPhN), Saclay, France\\
$^{130}$ Universit\'{e}  Paris-Saclay, CNRS/IN2P3, IJCLab, Orsay, France\\
$^{131}$ Universit\`{a} degli Studi di Foggia, Foggia, Italy\\
$^{132}$ Universit\`{a} del Piemonte Orientale, Vercelli, Italy\\
$^{133}$ Universit\`{a} di Brescia, Brescia, Italy\\
$^{134}$ Variable Energy Cyclotron Centre, Homi Bhabha National Institute, Kolkata, India\\
$^{135}$ Warsaw University of Technology, Warsaw, Poland\\
$^{136}$ Wayne State University, Detroit, Michigan, United States\\
$^{137}$ Yale University, New Haven, Connecticut, United States\\
$^{138}$ Yildiz Technical University, Istanbul, Turkey\\
$^{139}$ Yonsei University, Seoul, Republic of Korea\\
$^{140}$ Affiliated with an institute formerly covered by a cooperation agreement with CERN\\
$^{141}$ Affiliated with an international laboratory covered by a cooperation agreement with CERN.\\

\end{flushleft} 
  
\end{document}